\documentclass[preprint,preprintnumbers,aps,nofootinbib,tightenlines,floatfix,groupedaddress]{revtex4}

\usepackage{hyperref}
\usepackage{amsmath,amssymb}
\usepackage{cleveref}
\usepackage{graphicx}
\usepackage[utf8]{inputenc}
\usepackage{bm}
\usepackage{comment}
\usepackage{color}
\usepackage{qcircuit}
\usepackage{csquotes}
\usepackage{soul}
\usepackage[section]{placeins}
\allowdisplaybreaks
\let\Oldsection\section
\renewcommand{\section}{\FloatBarrier\Oldsection}
\let\Oldsubsection\subsection
\renewcommand{\subsection}{\FloatBarrier\Oldsubsection}

\newcommand{\gsim}{\raisebox{-0.7ex}{$\stackrel{\textstyle >}{\sim}$ }}

\newcount\hour \newcount\hourminute \newcount\minute
\hour=\time \divide \hour by 60
\hourminute=\hour \multiply \hourminute by 60
\minute=\time \advance \minute by -\hourminute
\newcommand{\mydate}{\ \today \ - \number\hour :\number\minute}

\begin{document}

\title{Digitization of
Scalar Fields for Quantum Computing }

\author{Natalie Klco}
\email{klcon@uw.edu}
\affiliation{Institute for Nuclear Theory, University of Washington, Seattle, WA 98195-1550, USA}
\author{Martin J.~Savage}
\email{mjs5@uw.edu}
\affiliation{Institute for Nuclear Theory, University of Washington, Seattle, WA 98195-1550, USA}

\date{\mydate}

\preprint{INT-PUB-18-044}

\begin{figure}[!t]
 \vspace{-1.5cm} \leftline{
 	\includegraphics[width=0.2\textwidth]{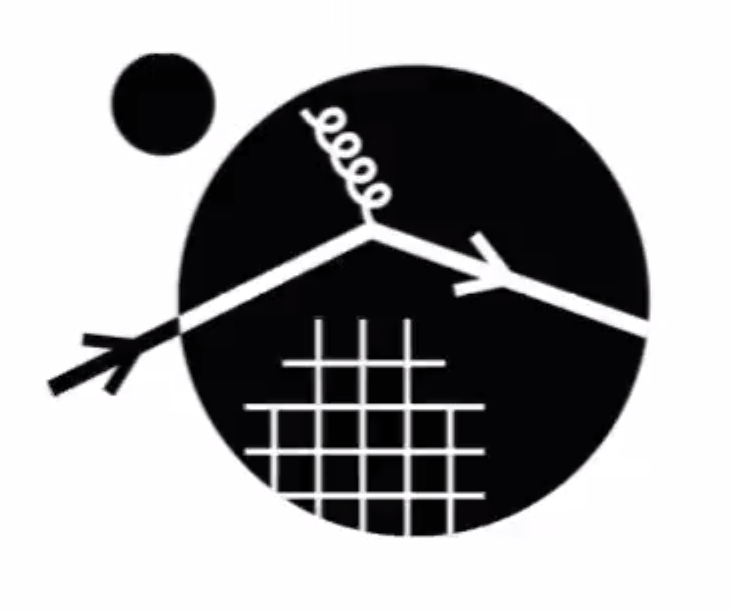}}
\end{figure}

\begin{abstract}
  \noindent
Qubit, operator and gate resources required for the digitization of
lattice $\lambda\phi^4$ scalar field theories onto quantum computers are considered, building upon the foundational work by
Jordan, Lee and Preskill, with a focus towards noisy intermediate-scale quantum (NISQ) devices.
The Nyquist-Shannon sampling theorem,
introduced in this context by Macridin, Spentzouris, Amundson and Harnik building on the work of Somma,
provides a guide with which to evaluate the efficacy of two field-space bases, the eigenstates of the field operator, as used by
Jordan, Lee and Preskill, and eigenstates of a harmonic oscillator, to describe $0+1$- and $d+1$-dimensional scalar field theory.
We show how  techniques associated with improved actions, which are heavily utilized in Lattice QCD calculations
to systematically reduce  lattice-spacing artifacts, can be
used to reduce the impact of the field digitization in  $\lambda\phi^4$,
but are found to be inferior to a complete digitization-improvement of the Hamiltonian using a Quantum Fourier Transform.
When the Nyquist-Shannon sampling theorem is satisfied, digitization errors scale as
$|\log|\log |\epsilon_{\rm dig}|||\sim n_Q$
(number of qubits describing the field at a given spatial site)
for the low-lying states,
leaving the familiar power-law lattice-spacing  and finite-volume effects that
scale as $|\log |\epsilon_{\rm latt}||\sim N_Q$
(total number of qubits in the simulation).
For localized(delocalized) field-space wavefunctions, it is found that $n_Q\sim4(7)$ qubits per spatial lattice site are sufficient to reduce theoretical digitization errors below error contributions associated with
approximation of the time-evolution operator and noisy implementation on near-term quantum devices. \footnote{
Only classical computing resources have been used to obtain the results presented in this work.}
\end{abstract}
\maketitle

\tableofcontents
\vfill\eject

\section{Introduction}

While offering the potential to greatly refine calculations that can be performed through classical computation,
Quantum Computing (QC) also
holds the potential to enable calculations of quantities in quantum field theories and other quantum many-body systems that are not possible with classical techniques~\cite{Lloyd1996,OrtizPhysRevA2001,SommaPhysRevA2002,Byrnes:2005qx,OvrumHjorthJensen,Jordan:2011ci,Jordan:2011ne,Jordan:2014tma,Jordan:2017lea,Zohar:2012xf,Zohar:2012ay,Banerjee:2012pg,Banerjee:2012xg,Wiese:2013uua,Marcos:2014lda,Wiese:2014rla,Zohar:2016iic,Bermudez:2017yrq,Banuls:2017ena,Kaplan:2018vnj}.
In particular,  real-time dynamics, such as the fragmentation of quarks into hadrons at
particle accelerators, the dynamics of non-equilibrium systems,
and the nature of  finite-density systems for which sampling with classical computation is limited by sign problems,
are key areas for which a \emph{quantum advantage}  is anticipated to be achieved.
Quantum devices with a range of underlying qubit architectures without error correction
are now becoming available for domain scientists to seek inroads into these problems and other
important scientific applications,
and to envisage attributes of quantum devices necessary to outperform classical computations of scientific significance.
The performance of present day quantum devices is limited by a number of basic attributes, including
 coherence times and
the  number of gates (specifically entangling gates) that can be applied prior to decoherence,
the accuracy and precision of applied gates,
the number and interconnectivity of qubits, and
the lack of error correction.
While significant efforts are in progress to reduce or eliminate these deficiencies,
and remarkable progress is being made,
these  limitations are expected to persist in near-term quantum devices.  This has led John Preskill to name the present and upcoming time period the
``NISQ era'' (Noisy Intermediate-Scale Quantum era)~\cite{Preskill2018quantumcomputingin}.  While formidable in its destruction of pure quantum states, quantum noise has been recently suppressed sufficiently for a number of small quantum simulations of physical systems  \cite{Martinez2016,OMalleyPhysRevX,Kandala2017,PhysRevLett.120.210501,PhysRevA.98.032331,Hempel2018}, encouraging the expectation of meaningful scientific applications of NISQ-era devices.

Scalar field theories are ubiquitous in physics, from
describing  densities in condensed matter systems,
to  fundamental fields  in the electroweak sector
from which the Higgs Boson emerges after spontaneous symmetry breaking.
The quantum theory
describing the dynamics of a self-interacting, real scalar field represents, perhaps,  the simplest quantum field theory  (QFT)
that can be explored through direct digitization of the field with a quantum computer.
Such studies are anticipated to provide important insights into how quantum devices can be used to
simulate  gauge field theories,
such as quantum electrodynamics (QED) and quantum chromodynamics (QCD)
that describe the interactions in electronic systems and
between quarks
and gluons responsible for the nuclear forces and the
structure and dynamics of strongly interacting matter.
It is exciting to observe the  advances that are being made in developing~\cite{Jordan:2011ci,Jordan:2011ne,Jordan:2014tma,Jordan:2017lea,Wiese:2013uua,Wiese:2014rla,GarciaAlvarezFermionFermionPRL2015,Bazavov:2015kka,zohar2016,Pichler2016,Bermudez:2017yrq,Banuls:2017ena,Moosavian:2017tkv,Gonzalez-Cuadra:2017lvz,PhysRevLett.121.110504,PhysRevA.98.032331,Zhang:2018ufj,Macridin:2018oli,Kaplan:2018vnj,Bender:2018rdp,Zohar:2018cwb,Zohar:2018nvl,Jefferson:2017sdb,Alsup:2016fii,Yeter-Aydeniz:2017ubh,Marshall:2015mna,Bradler:2016fyk}
and implementing~\cite{Zohar:2012xf,Zohar:2012ay,Banerjee:2012xg,Banerjee:2012pg,Marcos:2014lda,Marshall:2015mna,Zohar:2016iic,Martinez2016,Muschik:2016tws,PhysRevA.98.032331}
algorithms for both Abelian and non-Abelian gauge theories and scalar field theories
that may be useful for QFT calculations with quantum computers.

It is not expected that NISQ-era devices will surpass the computational capabilities of classical devices for the evolution of scalar fields discussed in this paper.
While an advantage may be found in a contrived endeavour for increased precision in the ground state energy of a non-local spatial wavefunction (see section \ref{subsubsec:delocalized}), the more-likely regimes of \emph{quantum advantage} in simulation are those highlighted at the beginning of this introduction.
Creating a computational framework making these systems accessible, taking advantage of superpositions and interference while remaining robust to quantum noise, is arduous and has become the focus of many current avenues of research; porting the knowledge and understanding of high-performance classical computation will be vital but insufficient to achieve this goal.
As is the case in classical computing, performance and scaling of quantum devices for scientific application cannot be completely understood before computations are implemented at scale on hardware.
While quantum calculations at scale are unreasonable today and the currently-available hardware is likely far from future fault-tolerant devices, this document examines scalar quantum field theory calculations on near-term quantum devices in preparation for a future in which substantial quantum resources allow exploration of classically-unattainable states of matter.

In a series of foundational papers, Jordan, Lee and Preskill (JLP) formulated and analyzed scalar field theories
for quantum computers~\cite{Jordan:2011ci,Jordan:2011ne,Jordan:2014tma,Jordan:2017lea}
and estimated the resource-requirement
scaling of calculations
of static properties and  of elastic and inelastic particle scattering processes
determined through direct time evolution.
A real scalar field, $\phi({\bf x})$,
is discretized on a spatial lattice using techniques that are standard in lattice QCD (LQCD)
calculations using classical computers.
The spacing between lattice sites along a cartesian axis is denoted by $a$
and the extent of each spatial direction is denoted by $L$, and
$\phi({\bf x})$ is subject to, for example,  periodic boundary conditions (PBCs)
or twisted boundary conditions
(e.g. Refs.~\cite{Lin:2001zz,Sachrajda:2004mi,Bedaque:2004kc,Briceno:2013hya})
in each direction.
However, in NISQ-era  quantum computations,
$\phi({\bf x})$ can only assume values from a modest-sized set of possibilities,
with extreme values of $|\phi({\bf x})|\le \phi_{\rm max}$ and a digitization  $\delta_{\phi({\bf x})}$.
Therefore, the computational layout of these JLP simulations is that
a number of qubits, $n_Q$,  describe the value of
$\phi({\bf x})$ at each position ${\bf x}$,
with a total number of qubits of $N_Q=n_Q \left(L/a\right)^d$ for spatial dimension, $d$.
This system is evolved under the action of the time-evolution operator,
$\hat U(t) = e^{-i \hat H t}$ where $\hat H$ is the Hamiltonian operator, to evolve
isolated wave packets forward in time to determine scattering amplitudes.

In nice work by Macridin, Spentzouris, Amundson and Harnik (MSAH)~\cite{PhysRevLett.121.110504,Macridin:2018oli},
focused on phonon-electron interactions
and building upon work by Somma~\cite{Somma:2016:QSO:3179430.3179434},
it was emphasized that the Nyquist-Shannon (NS) sampling theorem
should be considered in
 the  architecture of a quantum computer, the mapping of $\phi({\bf x})$
and the implementation of the Hamiltonian to achieve the desired accuracies in quantum simulations.
The localization of the $\phi({\bf x})$ wavefunction in $\phi$-space and its curvature
determine the extent and interval of sampling in
$\phi$-space, i.e. $\phi_{\rm max}$ and  $\delta_{\phi({\bf x})}$ (which dictate $n_Q$),
required to reproduce the $\phi$-space wavefunction with exponential precision,
scaling as $|\log|\log |\epsilon_{\rm dig} |||\sim n_Q$,
where $\epsilon_{\rm dig}$ is the error introduced through digitization,
thereby removing inaccuracies due to  field digitization.
These studies of the NS sampling theorem
determined the minimum number of qubits per phonon field required to accurately describe
harmonic oscillator (HO) wavefunctions  up to a given excitation level of the phonon field~\cite{Somma:2016:QSO:3179430.3179434,PhysRevLett.121.110504,Macridin:2018oli}.
The digitization errors make contributions that are
parametrically smaller than spatial lattice-spacing artifacts and spatial finite-volume effects,
which typically scale as
$|\log |\epsilon_{\rm latt}||\sim N_Q$,
where $\epsilon_{\rm latt}$ is the error introduced by the non-zero spatial lattice spacing.

\begin{figure}[!ht]
  \includegraphics[width = 0.9\textwidth]{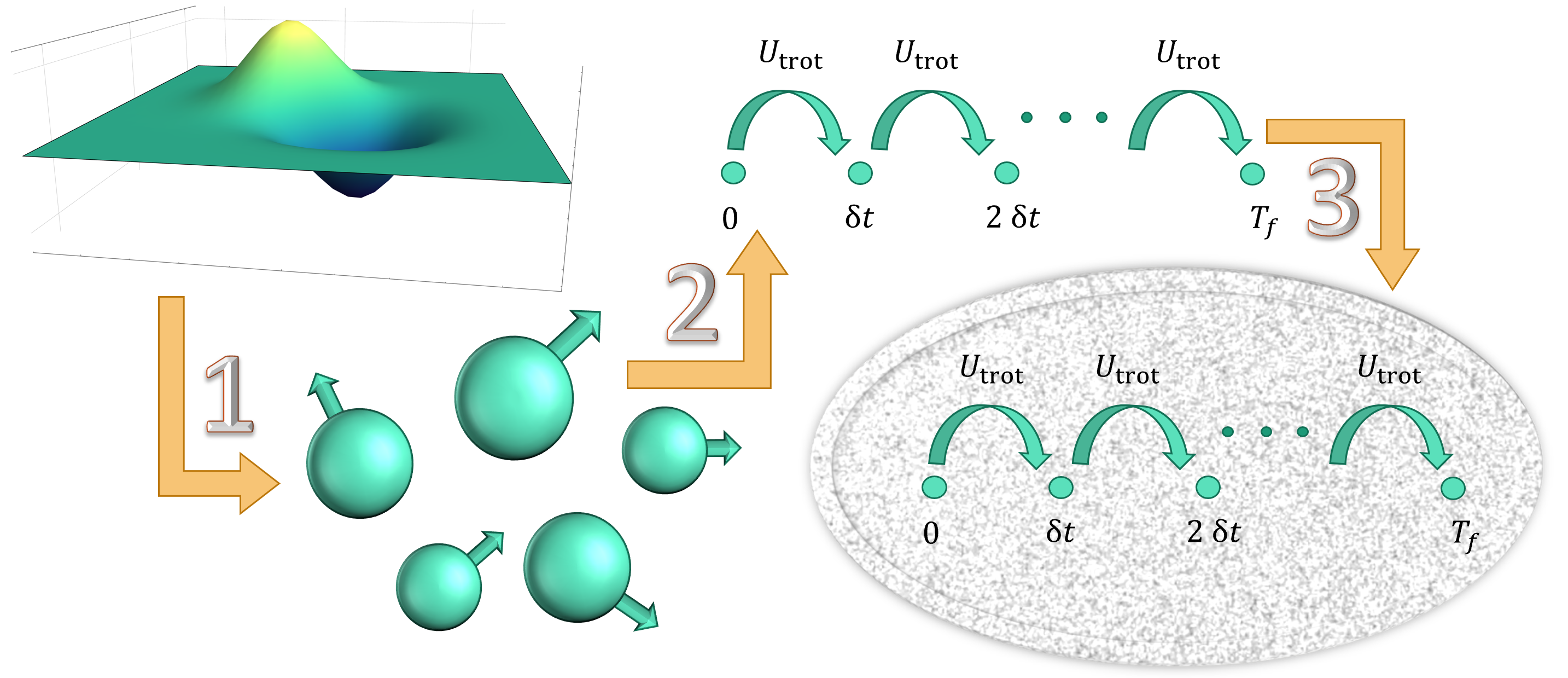}
  \caption{
  Identifying three distinct sources of error in quantum simulations of scalar field theory:
  (1) the error of digitizing and latticizing the continuous field onto qubit degrees of freedom;
  (2) the error of simulation due to the use of approximations to the exact digitized propagator;
  (3) the error due to noise in the quantum device's implementation of the approximate digitized propagator.
  The first two sources of error are independent of the quantum hardware. The first is the main focus of this paper.
  }
  \label{fig:errorsourcediagram}
\end{figure}
The use of $\epsilon$ in this paper indicates the precision to which the ground state energy of the continuous scalar field can be reproduced by a Hamiltonian digitized with qubit degrees of freedom (step (1) of
Fig.~\ref{fig:errorsourcediagram}).
This is a theoretical source of systematic error accompanying the formulation of the Hamiltonian
before it is implemented on hardware or employed  in a simulation algorithm.
Most importantly, this is \emph{not} the $\epsilon$ commonly used in the quantum simulation literature to express the precision with which properties of a given Hamiltonian can be extracted on quantum hardware (steps (2,3)
of Fig.~\ref{fig:errorsourcediagram}).
Thus, $\epsilon$ here characterizes the physics of field-digitization necessary to map the system onto a qubit Hamiltonian, and does not include precision reductions entering from the Hamiltonian simulation (e.g., Trotterization) or phase estimation
that may be implemented to extract features of this system on quantum hardware.
Examples of progress in bounding these latter sources of error can be found in  Refs.~\cite{PhysRevA.90.022305,Reiher201619152,Poulin:2015:TSS:2871401.2871402,Babbush2015ChemicalBasis,LowChuang2016}.
The distinction between these three sources of error are depicted in Fig.~\ref{fig:errorsourcediagram}.
The scalar field begins in a continuous representation in a formally-infinite-dimensional Hilbert space.
With the reduction in step (1), this infinite dimensional Hilbert space is truncated, digitized, and formulated on qubit degrees of freedom.
Step (2) designs a quantum simulation algorithm to approximate the time evolution of the quantum state
e.g., Trotterization, which introduces errors scaling polynomially in the temporal digitization step size $\delta t$.
Step (3) implements this approximate time evolution on quantum hardware susceptible to noise and
(likely) without quantum error correction in the NISQ-era.

The digitization errors represented in step (1) of Fig.~\ref{fig:errorsourcediagram} are the only errors considered in the main text of this paper.
The results presented in this work establish the precision of low-energy calculations that could be
obtained on an ideal quantum device with exact implementation of the time evolution operator based upon formal considerations of digitizations and discretizations that must be performed when formulating the field theories onto quantum devices with a modest number of qubits.
These are presented in order to determine the best-case scenarios for modest-sized devices of the near term, and intentionally neglect the errors of steps (2,3) necessary to accurately reflect the precision attainable with realistic NISQ-era devices.
Simulations of the effects of first-order Trotterization (2) and simple unitary gate noise (3) can be found in Appendix~\ref{app:noisySimulation}.
Together, Appendix~\ref{app:noisySimulation} and the main text indicate that digitization errors (1) may be controlled to high precision with a number of qubits reasonable for the NISQ era, leaving the steps (2,3) to dominate the current scalar simulation error budget.

In this work, we consider implications of the digitization of scalar fields when mapped to qubit degrees of freedom, with a focus on the associated limits in accuracy of calculations on NISQ-era scale devices.
In particular, we examine
digitizing
0+1 and 1+1 dimensional $\lambda\phi^4$ scalar field theory describing a single real scalar field,
including estimating qubit requirements, estimating the number of operators and number of gates required for such simulations, and extrapolating these estimates to $d$+1-dimensional simulations~\footnote{The necessary ingredients for this extrapolation are detailed resource requirements for
implementation of 1.) the 0+1 dimensional self-interacting scalar field and 2.) the nearest-neighbor finite-difference gradient operator.
We analyze these two pieces in depth and subsequently discuss the compilation procedure for applying the analysis to
scalar lattices of arbitrary size and dimensionality.}.
As the sign of the mass-squared term in the Hamiltonian determines whether the ground state of these theories are localized around $\phi = 0$ or are delocalized around two minima of the potential, estimates are provided for both situations.  Making a connection with classical calculations of  lattice QFTs,
we discuss Hamiltonian improvement that can be included to parametrically reduce the impact
of the field digitization by powers of $\delta_{\phi({\bf x})}^2$.
However, as used in
Refs.~\cite{Jordan:2011ci,Jordan:2011ne,Jordan:2014tma,Jordan:2017lea}  and emphasized in
Refs.~\cite{Somma:2016:QSO:3179430.3179434,PhysRevLett.121.110504,Macridin:2018oli},
the use of the Quantum Fourier Transform (QuFoTr)
on the $n_Q$ qubits at each spatial site
to evaluate the action of the conjugate-momentum term in the Hamiltonian,
and the freedom it provides in applying phases in field conjugate-momentum space,
provides the opportunity to arbitrarily improve the digitized Hamiltonian,
removing all polynomials in $\delta_{\phi({\bf x})}$ and rendering digitization effects
to be exponentially small (once the conditions imposed by the NS sampling theorem are satisfied).
Analogous implementations have been utilized previously in Monte-Carlo calculations of non-relativistic systems~\cite{Bulgac:2008zz,Endres:2012cw}.
We present the complete operator structure required to implement simulations with $n_Q=3,4,5$ qubits per spatial site, along with associated quantum circuits for $n_Q = 3$.
As different bases in $\phi$-space can be used to span the Hilbert space at each  point in space, we examine
the JLP implementation using the eigenstates of the $\phi$-operator and a basis defined by the eigenstates
of a harmonic oscillator (HO), that is distinct from the frameworks developed in Refs.~\cite{Somma:2016:QSO:3179430.3179434,Macridin:2018oli,PhysRevLett.121.110504}.
From our analysis, we conclude that the properties and dynamics of interacting scalar field theories may be simulated with
only a modest number of qubits per site required to render digitization artifacts negligible
compared to other expected systematic errors \footnote{see processes (2,3) of Fig.~\ref{fig:errorsourcediagram} and Appendix~\ref{app:noisySimulation}} in the NISQ-era.

\section{Lattice Scalar Field Theory with Qubits}
\label{sec:SFT}
\noindent
The continuum Lagrange density describing the dynamics of a scalar field with self interactions,  retaining
only renormalizable terms in 3+1 dimensions, is
\begin{eqnarray}
{\cal L} & = & {1\over 2} (\partial_\mu\phi)^2 - {1\over 2} m^2 \phi^2 - {\lambda\over 4!} \phi^4
\ \ \ ,
\label{eq:Llamphi4}
\end{eqnarray}
with a Hamiltonian density of
\begin{eqnarray}
{\cal H} & = & {1\over 2} \Pi^2 + {1\over 2} ({\bm\nabla}\phi)^2 + {1\over 2} m^2 \phi^2 + {\lambda\over 4!} \phi^4
\ \ \ .
\label{eq:Hlamphi4}
\end{eqnarray}
The conjugate momentum operator, $\Pi({\bf x})$ has the standard equal-time commutation relation with the field operator,
$\left[\ \phi({\bf x}) , \Pi({\bf y})\ \right] = i\delta^3 ({\bf x}-{\bf y})$.
Numerical evaluation of observables resulting from this Hamiltonian density can be
accomplished by discretizing space with a cubic grid with a distance between adjacent lattice sites on the
Cartesian axes of $a$ (the lattice spacing) and extent $L$ in each direction,
as previously defined.   The number of sites in each spatial direction is $L/a$.
The discretized Hamiltonian on a $d$-dimensional spatial lattice is
\begin{eqnarray}
H & = &  a^d\ \sum_{\bf x}\  {1\over 2} \Pi^2 - {1\over 2} \phi {\bm\nabla}^2_a\phi + {1\over 2} m_0^2 \phi^2 + {\lambda_0\over 4!} \phi^4
\ \ \ ,
\label{eq:Hlatt}
\end{eqnarray}
where the discretized Laplacian operator is defined as
$ {\bm\nabla}^2_a \phi ({\bf x}) = \sum\limits_{j=1}^d
\left(  \phi ({\bf x} + a \hat \mu_j)  +  \phi ({\bf x} -  a \hat \mu_j)  -  2 \phi ({\bf x})  \right)/a^2 $
where $\hat\mu_j$ is the unit vector in the $j^{\rm th}$ direction.
The quantities $m_0$ and $\lambda_0$ are bare parameters that are tuned to recover,
for example,
correct values of the $\phi$ mass, $M_\phi$, and the $4\phi$ scattering amplitude.
The conjugate momentum is required to satisfy
\begin{eqnarray}
\left[\ \phi({\bf x}) ,  \Pi({\bf y})\ \right] & = & {i\over a^d} \delta^d_{{\bf x},{\bf y}}\ \hat I
\ \ \ ,
\label{eq:phiPicomm}
\end{eqnarray}
where $\hat I$ is the identity operator in field space.
Redefining the fields, Hamiltonian and mass as
$\hat\phi = a^{(d-1)/2} \phi$,
$\hat\Pi = a^{(d+1)/2} \Pi$,
$\hat H = a H$,
$\hat m_0 = a m_0$,
$\hat \lambda_0  = a^{3-d} \lambda_0$,
Eq.~(\ref{eq:Hlatt}) can be written
in terms of dimensionless quantities,
\begin{eqnarray}
\hat H & = &  \sum_{\bf x}\  {1\over 2} \hat\Pi^2 - {1\over 2} \hat\phi \hat{\bm\nabla}^2_a \hat\phi
+ {1\over 2} \hat m_0^2 \hat \phi^2 + {\hat \lambda_0\over 4!} \hat\phi^4
\ \ \ ,
\label{eq:Hhatlatt}
\end{eqnarray}
with an equal-time commutator of
\begin{eqnarray}
\left[\ \hat \phi({\bf x}) ,  \hat \Pi({\bf y})\ \right] & = & i \delta^d_{{\bf x},{\bf y}}\ \hat I
\ \ \ .
\label{eq:hatphiPicomm}
\end{eqnarray}
The eigenstates of the momentum operator, $|{\bf k}\rangle$,
satisfy $\hat {\cal K} |{\bf k}\rangle = {\bf k} |{\bf k}\rangle$
where ${\bf k}$ is quantized by
the boundary conditions,
and, for example,  takes the values ${\bf k}= {\bf n} {2\pi\over L}$ for PBCs,
where the integer-triplets ${\bf n}$ are constrained to lie within the first Brillouin zone $| n_{x,y,z} | < {L\over 2 a}$.
The finite-difference operator that is used to define the latticized $\hat {\bf K}=\hat{\bm\nabla}^2_a \hat\phi$ has eigenvalues such that
$\hat {\bf K} |{\bf k}\rangle = \hat{\bf k} |{\bf k}\rangle$  with
$ \hat k_j = {2\over a} \sin \left({k_j a\over 2}\right)$.

The construction of the latticized Hamiltonian in Eq.~(\ref{eq:Hhatlatt}) is such that the long-distance, or low-energy,
quantities (compared to $\pi/a$) will be faithfully reproduced in numerical evaluations up to corrections that are
polynomial in the lattice spacing,
$\sim (a E)^n$, or exponential in the volume, $\sim e^{n M_\phi L}$ (for spatially localized states).
Therefore,  such lattice frameworks should be considered as low-energy effective field theories (EFTs),
with an ultra-violet (UV) cut-off set by the inverse lattice spacing.
Considerable effort by the LQCD community has been put in to construct improved actions
in which
additional terms are added to the Lagrange density that are parametrically suppressed by
powers of the lattice spacing and consistent with the underlying (hyper-)cubic symmetry of the spacetime lattice.
The additional terms in the QCD action are termed the Symanzik action~\cite{Symanzik:1983dc,Symanzik:1983gh}.
Coefficients of the operators in the Symanzik action depend upon the lattice spacing and the discretized action, and are determined both by tree-level matching and nonperturbatively through tuning for higher precision.
As an example, the Wilson discretization of the light-quark field in LQCD calculations
leads to spatial finite-difference
discretization errors that scale
linearly with the lattice spacing, ${\cal O}(a)$.
By adding one dimension-5 operator to the lattice action, the Sheikholeslami-Wohlert term~\cite{Sheikholeslami:1985ij},
and tuning its coefficient, this improved action produces low-energy and
long-distance observables that have errors at ${\cal O}(a^2)$.
In principle, an arbitrary number of operators in the Symanzik action can be included in
numerical computations to improve the action to high orders.
However, the  requirements for such calculations that include,
for instance, four-quark operators, make this impractical.
We will apply similar considerations when proposing improvements for the digitization of the scalar field onto quantum degrees of freedom.

\section{Implications of the Nyquist-Shannon Sampling Theorem}
\noindent
The work of MSAH~\cite{PhysRevLett.121.110504,Macridin:2018oli}
stressed the importance of the NS Sampling Theorem,
implicit in the work of Somma~\cite{Somma:2016:QSO:3179430.3179434},
which is central to signal processing, communications and data compression,
 to quantum computations.
It is worth reminding the reader of its main elements and  implications.
While the results of this theorem are used implicitly in the formulation and analysis of LQCD calculations,
connections between the two are typically not dwelt upon.

Consider the reconstruction of a real function, $C(x)$, that has support only between $x=0$ and $x=x_{\rm max}$ in position space and
between $k=-k_{\rm max}$ and $k=+k_{\rm max}$ in momentum space,
from discrete sampling.
If $C(x)$ is sampled over the interval $x\in \left[0,L\right]$ with $L>x_{\rm max}$
and at intervals of $\delta x < {\pi \over k_{\rm max}}$ then
the NS Sampling Theorem ensures that
$C(x)$ can be reconstructed up to corrections that are exponentially small.
The Poisson resummation formula is at the heart of this result,
which is also used extensively in deriving, for example,  finite-volume effects in LQCD calculations.
The implications of this theorem are clear.
As long as the function is sampled in both position-space and momentum-space over the entire region
where the function has support, then it can be reconstructed with only exponentially small errors introduced by the discretization.
In quantum simulations of field theories,
and in particular the computation of the low-lying eigenstates and eigenvalues,
this imposes constraints for both the spatial discretization and the digitization of the field at any given spatial site.
From the viewpoint of lattice calculations,
this dictates that the lattice spacing must be small enough to include all
spatial-momentum states that contribute  (to the level of precision to which the calculation is being performed),
and the volume large enough to contain the eigenstates of interest, in order for deviations
between the calculated eigenstates and eigenvalues  and the true eigenstates and eigenvalues  to be exponentially small.
For LQCD calculations, this underpins L{\"u}scher's finite-volume analysis of QCD observables~\cite{Luscher:1985dn,Luscher:1986pf,Luscher:1990ux}, which is used extensively to both quantify uncertainties
and to extract S-matrix elements.

The NS theorem does not specify how to ``cover'' the region of support in position-space and momentum-space,
i.e. what basis should be used to span the spaces, and some bases will be better than others for any given function.
For smooth functions that fall exponentially (or as a Gaussian) at large distances, the plane-wave basis is efficient,
defined over the spatial interval where the function has support and with a discretization that encompasses
its highest frequency component.
For a more localized function, such as those that fall as  a Gaussian at large distance,
eigenstates of the HO that are approximately tuned to the function can also be efficient.

For quantum computations of a  field theory using a given set of basis functions to define the
spatial discretization and the field digitization, including plane waves or eigenstates of the HO,
this theorem dictates the number of qubits required  to achieve a desired accuracy.
The number of qubits and the
number and complexity of operators required to execute the computation are  basis dependent.
Identifying the optimal basis with which to perform the quantum computation requires examining both the number of qubits
and the number of gates required to perform the computation with the desired precision.

It is worth commenting that the NS sampling bounds are likely satisfied in LQCD calculations
of localized quantities, such as hadron masses and nuclear bound states.
Therefore, the eigenvalues and eigenstates obtained in such calculations would be exponentially close to the
values associated with the lattice Hamiltonian if infinite statistics were accumulated
in the stochastic sampling of the quantum fields.
The power-law deviations that scale as $\sim (a E)^n$ result from
deviations in the lattice Hamiltonian from the continuum Hamiltonian
and are not due to under-sampling in the NS sense.
We  are
unaware of the NS theorem being implemented in classical quantum Monte-Carlo calculations, and consider
the possibility
worthwhile to explore.

\section{0+1 Dimensional Scalar Field Theory}
\noindent
In order to demonstrate some important features of the construction presented in the previous section,
we examine a
0+1 dimensional non-interacting scalar field theory, which is simply a HO.
After a further field and Hamiltonian redefinition,
$\hat \phi = {1\over \sqrt{\hat m_0}} \bar\phi$,
$\hat \Pi = \sqrt{\hat m_0} \bar\Pi$,
$\hat H=\hat m_0  \bar H$,
the HO is described by the Hamiltonian,
\begin{eqnarray}
\bar H & = &  {1\over 2} \bar\Pi^2 + {1\over 2}  \bar \phi^2
\ \ \ ,
\label{eq:Hhatlatt01free}
\end{eqnarray}
with a commutation relation
$\left[\ \bar \phi ,  \bar \Pi \ \right] = i \hat I$.
It is the digitization of this system that was studied by
Somma~\cite{Somma:2016:QSO:3179430.3179434} and by
MSAH~\cite{PhysRevLett.121.110504,Macridin:2018oli} with the identification
$\bar\phi\rightarrow X$, $\bar\Pi\rightarrow P$ and $\bar H\rightarrow H_h$.
Without field digitization, $\delta_{\bar\phi}=0$, this is simply the Hamiltonian describing a HO without self-interactions,
with energy eigenstates $|\psi_n\rangle$ and energy eigenvalues $E_n= n+{1\over 2}$.
The conjugate momentum operator can be identified with a derivative in field space,
$\bar\Pi = -i{d\over d\bar\phi}$,
to satisfy the equal-time commutation relation.

\subsection{Jordan-Lee-Preskill Basis}
\begin{figure}[!ht]
	\centering
	\includegraphics[width=0.85\columnwidth]{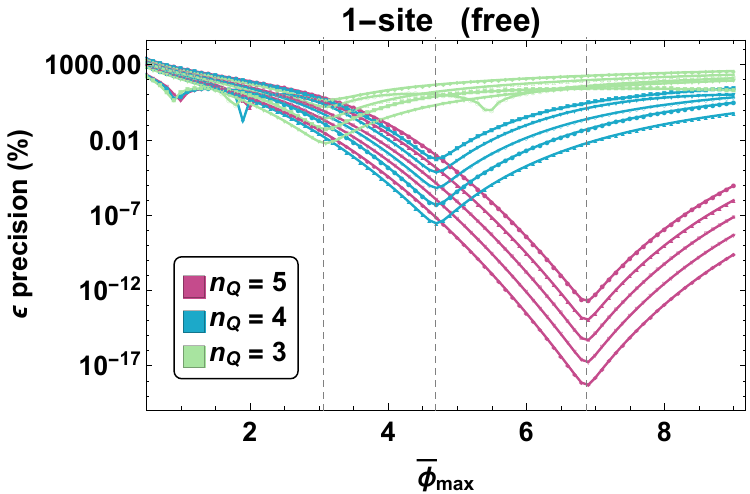}
	\caption{
	The precision  of the energies of the lowest five states of the HO in Eq.~(\ref{eq:Hhatlatt01free})
	expected from calculations on an ideal quantum computer with JLP digitizations over a range of values of $\bar\phi_{\rm max}$
	for a system digitized on
	$n_Q=3$, $n_Q=4$ and $n_Q=5$ qubits (minimized at $\bar{\phi}_{\rm max}= $3.1, 4.7, and 6.9, respectively) .
	The vertical gray-dashed lines correspond to saturation of the NS sampling bound.
		}
		\label{fig:HOphiMnq3}
\end{figure}

When the field is digitized,
$\bar\phi\rightarrow\tilde\phi$ (using the notation of MSAH),
and sampled at regular intervals $\delta_{\tilde\phi}\ne 0$, the conjugate momentum operator can be
replaced by a finite difference operator in field space,
in analogy with lattice field theory spatial discretization.
It has a matrix representation in $\tilde\phi$-space of
\begin{eqnarray}
\langle \tilde\phi^\prime|\
\tilde\Pi^2
 \ |\tilde\phi\rangle
 & = &
 {1\over \delta_{\bar\phi}^2}\
 \left(
 \begin{array}{cccccc}
 2 & -1 & 0 & 0 & \cdots & -1\\
-1 & 2 & -1 &   0 & 0 & \cdots  \\
 \vdots &  \vdots &  \vdots &  \vdots &  \vdots &  \vdots  \\
 0 & 0 &  \cdots & -1 & 2 & -1\\
-1 &  0 & 0 &  \cdots & -1 & 2 \\
 \end{array}
 \right)
\ \ \ ,
\label{eq:PiFD}
\end{eqnarray}
and acts in the space defined by field values
$ -\bar\phi_{\rm max},  -\bar\phi_{\rm max}+\delta_{\tilde\phi},
\cdots , -\frac{\delta_{\tilde{\phi}}}{2} , \frac{\delta_{\tilde{\phi}}}{2}, \cdots, \bar\phi_{\rm max} -\delta_{\tilde\phi} , \bar\phi_{\rm max}$.
For a space spanned by $n_s = 2^{n_Q}$ basis states in field space, the field takes values
\begin{eqnarray}
\bar\phi & = & -\bar\phi_{\rm max}\ +\ \delta_{\tilde\phi}\ \beta_\phi
\ \ ,\ \
 \delta_{\tilde\phi}\ =\ {2 \bar\phi_{\rm max}\ \over n_s-1}
\ \ \ ,
\end{eqnarray}
where
$\beta_\phi \ =\ 0, 1,  ... , n_s-1$.  Note that this formulation allows the field operator to be decomposed as $\phi = \frac{\bar{\phi}_{\rm max}}{n_s -1} \sum\limits_{j = 0}^{n_Q-1} 2^{j} \sigma^z_j$ (with qubits labeled right to left in tensor product spaces) and thus requires only single-qubit Pauli-Z operators for its implementation.
As is familiar from classical lattice simulations,
the  momentum modes of this system satisfying PBCs are,
\begin{eqnarray}
k_{\tilde\phi} & = &
-k_{\tilde\phi}^{\rm max} + \beta_k\  \delta k_{\tilde\phi}
\ \ ,\ \
k_{\tilde\phi}^{\rm max}\ =\ {\pi\over \delta_{\tilde\phi} }
\ \ ,\ \
\delta k_{\tilde\phi}
\ =\ {2\pi\over  \delta_{\tilde\phi} n_s}
\ \ \ ,
\label{eq:kphi}
\end{eqnarray}
with $\beta_k \ =\ 1, 2,  ... , n_s$.
It is interesting to note that this conjugate momentum-space basis may not be optimal in terms of the number
of gates in a quantum circuit required to apply the Hamiltonian to any given state.
Satisfying the NS theorem does not require any particular momentum components to be present in the conjugate momentum-space basis set and, as such, there is freedom to shift each momentum state by the same constant momentum.
It is convenient to shift each basis state in conjugate momentum space by $\Delta k_{\tilde\phi} = -\delta k_{\tilde\phi}/2$, so that
\begin{eqnarray}
k_{\tilde\phi}^\Delta & = &
-k_{\tilde\phi}^{\rm max} + \left(\beta_k - {1\over 2}\right)\  \delta k_{\tilde\phi}
\ \ \ ,
\label{eq:kDeldef}
\end{eqnarray}
which is equivalent to imposing twisted boundary conditions in field space~\cite{Lin:2001zz,Sachrajda:2004mi,Bedaque:2004kc,Briceno:2013hya}, resulting in $+1$'s in the off-diagonal corners of Eq.~\eqref{eq:PiFD} and momentum states that are symmetrically distributed within the edges of the first Brillouin zone between values of $\pm \frac{\pi}{\delta_{\tilde{\phi}}} \frac{n_s-1}{n_s}$.
For any choice of basis states spanning conjugate momentum space,
the finite-difference operator has matrix elements
\begin{eqnarray}
\langle k_{\tilde\phi}^\prime|\
 \tilde\Pi^2
 \ | k_{\tilde\phi}\rangle
 & = &
 \hat  k_{\tilde\phi}^2 \ \delta_{ k_{\tilde\phi}, k_{\tilde\phi}^\prime}
 \ \ \ ,\ \ \ \
  \hat  k_{\tilde\phi}^2\ =\ {4\over \delta_{\tilde\phi} ^2}\ \sin^2\left( {k_{\tilde\phi}\  \delta_{\tilde\phi}  \over 2}\right)
 \ \ \ .
\label{eq:kPi}
\end{eqnarray}
The Hamiltonian resulting from this field digitization is denoted by $\bar H\rightarrow\tilde H$.  The precision expected from computations on an ideal quantum computer
for a range of values of $\bar\phi_{\rm max}$ is shown in Fig.~\ref{fig:HOphiMnq3}.  Encouragingly, this calculation indicates that a $\bar{\phi}_{\rm max}$ of 4.7 for a 4-qubit representation of the scalar field can achieve a precision of better than $10^{-3}$\% on the energies of the lowest 5 eigenstates of the HO with an ideal quantum simulation.
For explicit examples of this field digitization implementation with three, four, and five qubits per site, see Appendix~\ref{sec:HOnQ3}.

With any finite computing device, classical or quantum, only a finite representation of a continuous quantity is possible.
In the JLP formulation, $|\phi|$ is bounded by $\phi_{\rm max}$ and sampled at intervals dictated by the number of qubits per site.
Focusing on the $\phi_{\rm max}$ truncation of the scalar field and allowing an infinite momentum-space coverage, formal quantum field theory studies \cite{PhysRevLett.88.141601,PhysRevD.69.045014} have shown that the asymptotic perturbative series becomes convergent.  For a sufficiently large $\phi_{\rm max}$, results for low-lying quantities are exponentially close to those obtained with unbounded values of the field.

\begin{figure}[!ht]
	\centering \includegraphics[,angle=0,width=0.8\columnwidth]{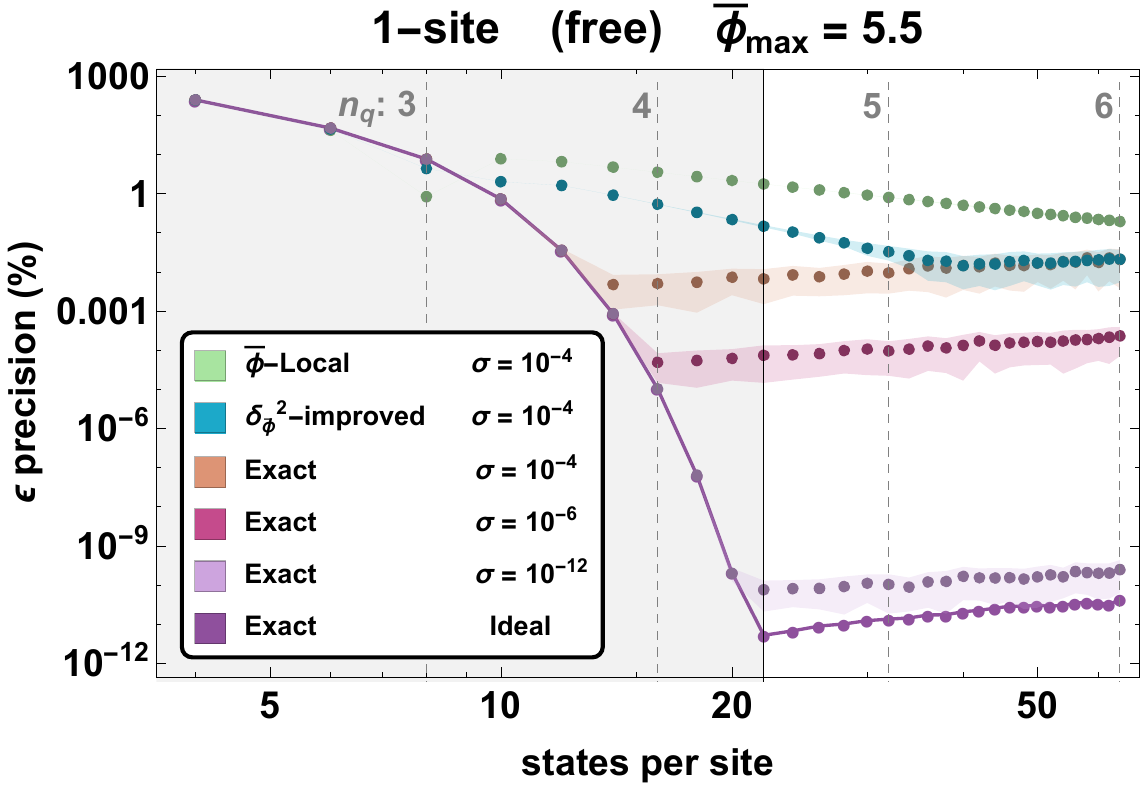}
	\caption{
	The precision of the ground state energy of the HO in Eq.~(\ref{eq:Hhatlatt01free})
	for unimproved, improved and exact conjugate-momentum operators,
	over a range of digitizations of $\bar\phi$ 	with different levels of gate noise.
	The light-green points correspond to implementing the finite-difference  conjugate momentum operator,
	the light-blue corresponds to the
	${\cal O}(\delta_{\tilde\phi}^2)$-improved conjugate momentum operator,
	and the purple-points correspond to the exact conjugate momentum operator.
	Gaussian noise with a width $\sigma$ is  added to the diagonal elements of the
	eigenvalues of the conjugate momentum operators (resembling a simplified version of process (3) in Fig.~\ref{fig:errorsourcediagram}).
	The maximum value of the field is fixed to be $\bar\phi_{\rm max} = 5.5$,
	enabling a precision of $\sim 10^{-12}$ for an ideal quantum computer.
	The vertical light-gray dashed lines correspond to the number of qubits associated with the number of states,
	while the solid darker-gray line corresponds to the na\"ive estimate of saturation of the NS sampling bound
	based upon the properties of the HO ground state wavefunction (the six calculations cross the latter of these lines top-to-bottom in the order of the legend).
For the computational errors anticipated in the NISQ-era, 4 qubits are seen to be sufficient to eliminate the digitization of the scalar field as a source of important error.
		}
		\label{fig:HOa}
\end{figure}

In a quantum simulation of this HO,
the JLP framework using the eigenstates of $\tilde\phi$ and its conjugate momentum can be used, as discussed above.
By tuning $ \bar\phi_{\rm max}$ to be larger than the spatial support of the $n^{\rm th}$ state of the HO at some level of precision,
the NS sampling bound will be satisfied for these levels as long as the largest value of $|k_{\tilde\phi}|$
in Eq.~(\ref{eq:kphi}) is greater than the region of support in conjugate momentum-space of the $n^{\rm th}$ state.
The action of the Hamiltonian on this set of qubits is most easily accomplished in two parts, as prescribed by JLP.
First, the $\tilde\phi^2$ operator, represented by a diagonal matrix in this basis, is
directly evaluated.
Second, a QuFoTr is performed to render the matrix representation of $ \tilde\Pi^2 $ diagonal and thus easily evaluated.
The ability to move back and forth between representations in which  $\tilde\phi$ or $\tilde \Pi $ is diagonal is
typically
not practical in classical field theory
computations and permits more freedom in choosing the operators that can be
applied in either representation.
Using the $\tilde\Pi$ operator in momentum space that is conjugate to the finite-difference operator,
Eq.~(\ref{eq:kPi}),
yields  exponentially-converged
eigenvalues and eigenvectors for the lowest $n$ states (by the NS sampling theorem).
However,
these quantities  differ from the corresponding HO quantities by even powers of $\delta_{\tilde\phi}$
because of the difference between $\hat  k_{\tilde\phi}^2$ and  $k_{\tilde\phi}^2$ in Eq.~(\ref{eq:kPi}),
as shown in Fig.~\ref{fig:HOa}.
However, if instead,  the $k_{\tilde\phi}^2$ eigenvalues in $\tilde\Pi^2$ are used in the quantum computation,
corresponding to  using $\bar\Pi^2$ and not $\tilde\Pi^2$, the eigenvalues and eigenvectors of
the lowest $n$ states are exponentially close to the
$\delta_{\tilde\phi} =0$ undigitized HO
quantities~\cite{Somma:2016:QSO:3179430.3179434,PhysRevLett.121.110504,Macridin:2018oli},
as can be observed in Fig.~\ref{fig:HOa}.
In performing the quantum simulations discussed in this paragraph, as the number of qubits is increased from
being insufficient to satisfy the NS sampling bound to exceeding the bound for a given state,
the deviation between the true   and   calculated energies will reduce as a polynomial in
$\delta_{\tilde\phi}^2$ until the NS sampling bound is satisfied, from which point on the gains will
become exponentially small.
It would appear  that working at this saturation point is an effective way to perform such computations.

\subsubsection{Perturbatively Improved Hamiltonian}
\label{sec:PertImprovHam}
\noindent
It is interesting to note that terms can be added  to the finite-difference conjugate-momentum operator
$\tilde\Pi$  in Eq.~(\ref{eq:PiFD})
to systematically improve it by powers of $\delta_{\tilde\phi}^2$.
Finding the improvement term is straightforward in conjugate-momentum space, which can then be transformed into $\tilde\phi$ space.
By including appropriate terms to systematically cancel deviations from the true conjugate-momentum operator,
\begin{eqnarray}
  \hat  k_{\tilde\phi}^2
  & =  & {4\over \delta_{\tilde\phi}^2}\ \sin^2\left( { k_{\tilde\phi}\  \delta_{\tilde\phi}  \over 2}\right)
  \ \rightarrow\
  k_{\tilde\phi}^2\ -\ { k_{\tilde\phi}^4\  \delta_{\tilde\phi}^2 \over 12}\ +\ ...
  \nonumber\\
    \hat  k_{\tilde\phi}^{\prime 2}
  & =  & {4\over \delta_{\tilde\phi}^2}\ \sin^2\left( { k_{\tilde\phi}\  \delta_{\tilde\phi}  \over 2}\right)
\ +\
{4\over 3 \delta_{\tilde\phi}^2}\ \sin^4\left( { k_{\tilde\phi}\  \delta_{\tilde\phi}  \over 2}\right)
  \ \rightarrow\
  k_{\tilde\phi}^2\ - \ { k_{\tilde\phi}^6\  \delta_{\tilde\phi}^4 \over 90}\ +\ ...
 \ \ \ .
\label{eq:kPiImprove}
\end{eqnarray}
and the corresponding effective action can be derived that is parametrically improved.
In $\tilde\phi$-space, the first term in this improvement is reproduced by an additional term in the Hamiltonian of the form,
\begin{eqnarray}
\delta \tilde H & = & {1\over 24} \ \delta_{\tilde\phi}^2\  \tilde\Pi^4
 \ \ \ .
\label{eq:LOdH}
\end{eqnarray}
The quadratic improvement in the energy of the ground state of the HO due to the inclusion of this improvement term in the Hamiltonian is shown in Fig.~\ref{fig:HOa}.
Numerical improvements on the order of one to two orders of magnitude in the accuracy of the improved calculations versus the unimproved calculations are found,
and that the residual dependence on $ \delta_{\tilde\phi}$ becomes ${\cal O}\left(\delta_{\tilde\phi}^4\right)$.

For systematic errors arising from approximation of the conjugate-momentum operator with a finite difference operator, the exact form of errors introduced into the Hamiltonian are well known.
If the situation was not so fortunate, the polynomial digitization errors could still be systematically removed.
Through a series of modest-sized calculations (in which
$\bar{\phi}_\text{max}$ is chosen large enough) at a range of digitization scales, the leading polynomial dependence
on the small parameter, $\delta_{\tilde{\phi}}$, may be calculated and removed through the introduction of additional Hamiltonian terms.
While the form of such terms may be systematically informed perturbatively or by the simple availability of independent higher-dimension operators~\cite{WEINBERG1979327}, their choice is not unique as the necessity is only to provide polynomial dependence at the correct order for cancellation.
This follows the procedure of Symanzik improvement~\cite{Symanzik:1983dc,Sheikholeslami:1985ij,Luscher:1996sc}
as discussed at the end of Sec.~\ref{sec:SFT} in the context of lattice QCD.
Such improvement procedures are broadly applicable and have been crucial for calculating observables in lattice gauge theories---modestly increasing the complexity of the action rather than calculating closer to the continuum.

\subsubsection{The Impact of Noise}
\label{subsec:HONoise}
\noindent
In the previous sections, we have considered a full non-perturbative improvement of the field conjugate-momentum
operator implemented in field space through a QuFoTr,
and  a perturbative improvement that
systematically eliminates increasing orders in the  digitization  introduced by  finite-difference approximations of
derivatives in field space.
These correspond to different matrices for $\bar \Pi^2$
acting on the basis states in field conjugate-momentum space.
The exact $\bar \Pi^2$  provides exponential precision in the low-lying eigenstates of the system,
but deviations from this matrix may lead to only polynomial precision---as evidenced from the behavior of the perturbatively
improved Hamiltonians and Fig.~\ref{fig:HOa}.
Imperfect gates and decoherence will result in an imperfect application of  $\bar \Pi^2$---introducing errors into  calculations of observables and potentially making superfluous, at a practical level,
the exponentially small improvements in digitization errors below the threshold of quantum noise.

In the presence of noisy gates and decoherence, it remains preferable
to work with the exact  $\bar \Pi^2$ operator,
but the precision of its application is limited.
For a given level of desired precision, the digitization and extent
of the field basis required to ensure that the precision matches that of the noise can be determined.
This would require an iterative tuning procedure in which multiple measurements are performed,
systematically increasing $\bar\phi_{\rm max}$ and decreasing $\delta_{\tilde\phi}$ until the
results of calculations become  stable.
These may or may not correspond to a situation that satisfies the NS sampling bound, depending upon the magnitude of the noise.
In Fig.~\ref{fig:HOa}, the results of calculations are shown with the use of the unimproved, improved and
exact conjugate-momentum operator through QuFoTr with the inclusion of different levels of gate-noise.
The noise is included as an offset to each diagonal element of $\bar \Pi^2$ after QuFoTr from a
Gaussian distribution of width $\sigma$ in conjugate-momentum space.
The value of $\bar\phi_{\rm max} = 5.5$ is chosen to allow for a precision of $\sim 10^{-12}$ for an ideal quantum computer
for digitizations below a critical value of $\delta_{\tilde\phi}$.  For a given gate-noise level, there is a value of
$\delta_{\tilde\phi}$ below which smaller digitizations  do not improve the precision of the calculation.
The  conclusion is that the error associated with digitization can be reduced below errors from
other sources for an arbitrary number of low-lying energy eigenstates with only a small number of qubits.

\begin{figure}
	\centering
\includegraphics[width=0.45\columnwidth]{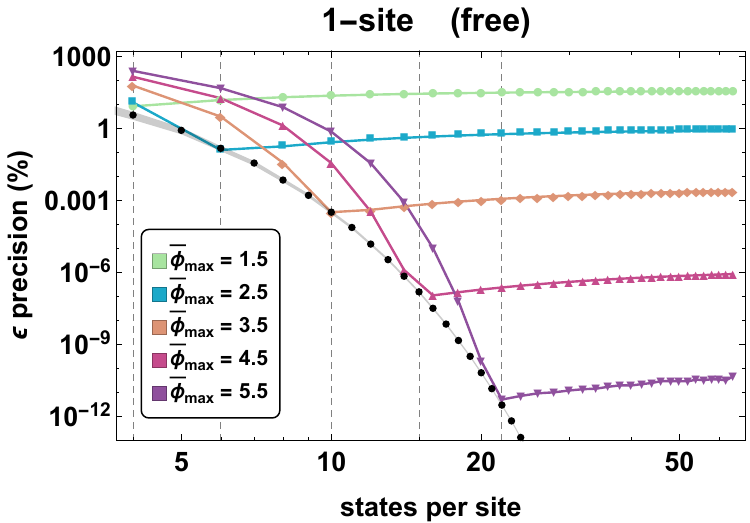}\ \
	\includegraphics[width=0.45\columnwidth]{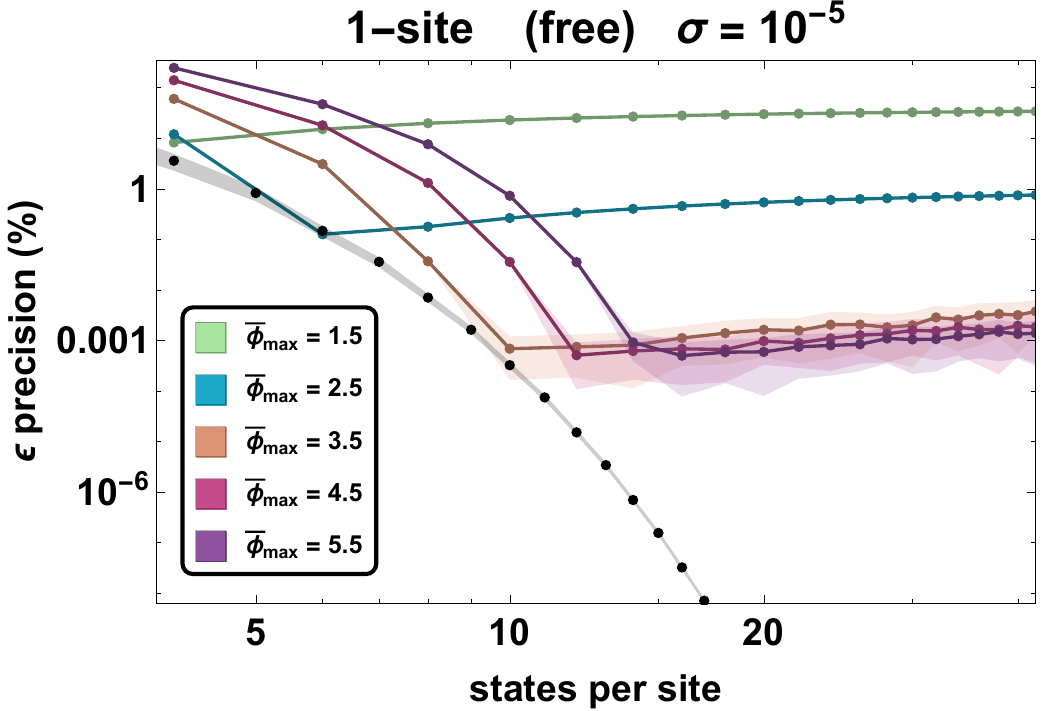}
	\caption{
	The precision of calculations of the ground state energy of the HO in Eq.~(\ref{eq:Hhatlatt01free}).
	The left panel shows
	expectations for
	an ideal quantum computer for different values of $\bar\phi_{\rm max}$ as a function of the number of states.
	The vertical gray-dashed lines correspond to the location of inflection points predicted by the
	NS sampling theorem for the indicated values of $\bar\phi_{\rm max}$.
	A fit to the grey points calculated at the
	NS saturation point indicates  $\epsilon \sim (1.8(2)\times10^3)\ 2^{-2.234(4) n_s}$,
	quantifying the double-exponential scaling between $\epsilon$ and $n_Q$.
	The right panel shows the expected precision from a device with noise at the level of $\sigma=10^{-5}$
	in the application of the field conjugate-momentum operator (as described in the text).  In both panels, the NS saturation points arise left-to-right in the top-to-bottom order of the legend.
		}
		\label{fig:HOphiM}
\end{figure}
The impact of different sampling ranges in $\bar\phi$ space upon the precision of calculations with an
ideal quantum computer (perfect gates), is shown in the left panel of Fig.~\ref{fig:HOphiM}.
The employed value of $\bar\phi_{\rm max}$ limits the overall precision of calculations as $\delta_{\tilde\phi}\rightarrow 0$ (states per site $n_s\rightarrow \infty$)
due to under sampling of the field at large $\bar{\phi}$, which is suppressed by $\sim e^{-\phi^2/2}$ for a HO wavefunction.
The field truncation also limits the precision of calculations
for large values of $\delta_{\tilde\phi}$ due to under sampling of the field in momentum space.  In between these regimes, the NS saturation point is found---perceived as a simple discontinuity in the first derivative---where the position-space sample rate becomes sufficient to capture the structure of momentum-space.
Tracking this saturation point with the gray band of Fig.~\ref{fig:HOphiM} shows a precision that increases exponentially in the number of states and thus double-exponentially in the number of qubits ($n_s = 2^{n_Q}$).  The coefficients of this precision scaling are calculated to be~$\epsilon \sim (1.8(2)\times10^3)\ 2^{-2.234(4) n_s}$, which serves as a general estimate for qubit requirements to capture the low-energy Hilbert space of localized scalar fields.

An interesting observation that can be drawn from
Fig.~\ref{fig:HOa} is that, for the parameters of the calculations explored,
reducing the amount of noise in the application of the field conjugate-momentum operator
below $\sim 10^{-13}$ will have little impact on the precision of the extracted final result.
The demonstration is made more concrete in the right panel of Fig.~\ref{fig:HOphiM},
where the noise level is fixed and the precision of  calculations are determined over a range of
$\bar\phi_{\rm max}$.
For this noise level, there is no improvement in precision as $\bar\phi_{\rm max}$ is increased beyond
$\sim 3.5$.
These are simple special cases of a general conclusion, that for a given
calculation designed with a set of digitization parameters,
there is a level of noise in the quantum device(s)
below which the precision of the results will  be only minimally impacted.  This general conclusion works in both directions and emphasizes the importance of matching precision in the qubit representation to that available from the NISQ hardware.  Exceeding precision in either direction would result in a wasteful use of quantum resources---using extra qubits and gates to represent the physical system with a precision beyond the quantum hardware's capability to resolve or using a noise-resilient quantum device   to probe physics beyond that represented in the qubit representation of the system.
\par One plausible scenario in which it may be beneficial to exceed the precision of the quantum hardware with the qubit mapping is in the presence of post-measurement noise-mitigation techniques as shown for implementations of variational quantum eigensolvers in \cite{McClean,Kandala2017,PhysRevA.98.032331}.  By extrapolating in a parameter scaling with the noise of the system (in the NISQ-era, this is conventionally a number increasing with the number of two-qubit interactions), the precision of a calculation can be improved beyond the precision capable for any ensemble measurement with the device at a fixed noise parameter.  In this case, it is the extrapolated precision of the quantum hardware that needs to be balanced with the theoretical precision of the qubit mapping in order to optimize the use of quantum resources.

\par It can be seen from Figs.~\ref{fig:HOa} and~\ref{fig:HOphiM} that the simple~\footnote{While this structure of quantum noise is acknowledged to be quite primitive, it is a simple model of issues expected in real quantum devices---in this case, a gaussian-distributed over- or under-rotation in the application of phases in conjugate-momentum space---leading to substantial theoretical considerations.  It is expected that current research in error correction on small quantum devices \cite{Kelly2015,PhysRevA.90.062320,Nigg1253742,ChaoReichardt,1367-2630-20-4-043038,PhysRevA.96.032338} will allow quantum noise and decoherence to be modeled in a more accurate, architecture-specific way when designing calculations for quantum hardware.}, yet physically-motivated, noise model implemented here does not significantly modify the results of calculations above the effective noise level.  As has been shown for the use of momentum-space phases associated with finite difference field-space $\tilde{\Pi}^2$ operators in section~\ref{sec:PertImprovHam}, there exist simple modifications to the conjugate-momentum space phases that modify the precision convergence by introducing polynomial sources of error.  Having now determined that gaussian random noise on conjugate-momentum space phase gates does not result in such a dramatic degradation of the calculation's precision above the noise tolerance, we proceed with noiseless calculations---remembering that this property must be monitored as noise models become more accurate and relevant to specific hardware implementations.

\subsection{Harmonic Oscillator Basis}
\noindent
As we have discussed previously, any set of basis states can be used to digitize the field, $\bar\phi$,
in $\bar H$ in Eq.~(\ref{eq:Hhatlatt01free}).
If the basis spans the $\bar\phi$-space and $\bar\Pi$-space of the lowest-lying eigenstates, the NS sampling theorem ensures exponential convergence to those eigenstates and associated eigenvalues.
A basis that is commonly used, beyond the eigenstates of the $\bar\phi$ operator,
is formed by a finite set of  eigenstates of a HO
with angular frequency $\omega_\phi$ that is tuned to optimize convergence in the number of states.
If $\omega_\phi$ is tuned to $\omega_\phi=1$, the basis states are the eigenstates of
$\bar H$  in Eq.~(\ref{eq:Hhatlatt01free})
and the evolution matrix is diagonal in the basis,
and the number of basis states required to converge to the lowest $N$ eigenstates is obviously equal to $N$.
For $\omega_\phi\ne 1$, the basis states are not eigenstates, and the evolution matrix is not diagonal.

It should be emphasized that bases formed from HO eigenstates, that are explored in this section,
are different in nature to those formed from  digitized HO eigenstates,
that have been considered previously~\cite{Somma:2016:QSO:3179430.3179434,PhysRevLett.121.110504,Macridin:2018oli}.
In those works, the eigenstates of the HO were digitized onto the eigenstates of the field operator, e.g.
$\langle \bar\phi |\psi_n\rangle\rightarrow \langle \bar\phi_i|\psi_n^d\rangle$, reducing each field-space
eigenstate from a continuous function to a discrete set.
It was the properties and time-evolution of the
$|\psi_n^d\rangle \sim \sum\limits_i  \psi_n(\phi_i) |\phi_i\rangle$
using the JLP framework that were examined in
Refs.~\cite{Somma:2016:QSO:3179430.3179434,PhysRevLett.121.110504,Macridin:2018oli}.
A HO basis was also used in the pioneering calculations of the deuteron ground state energy using the IBM and Rigetti quantum hardware by an ORNL team~\cite{PhysRevLett.120.210501}.  The mapping of the system onto qubits was accomplished using a 2nd quantization framework, where occupancy of quantum states is encoded in the orientation of the qubit.  In contrast, we consider a first quantized mapping with HO basis states mapped directly onto states of the quantum register.

Unlike the situation found with the JLP digitization of $\bar\phi$ in terms of eigenstates of the $\bar\phi$ operator,
where it is valuable to QuFoTr into conjugate-momentum space to evaluate the exact action of $\bar\Pi^2$,
digitization of the field space is accomplished explicitly by the HO basis
with the coverage in field and conjugate-momentum spaces
determined by the maximum number of basis states and the value of $\omega_\phi$.
As such, quantum circuits implementing the action of the Hamiltonian in the HO basis can be constructed in $\bar\phi$ space only.
The Hamiltonian and ladder operators defining the basis states are,
\begin{eqnarray}
H_{\rm basis} & = &  {1\over 2} \bar\Pi^2 + {1\over 2}  \omega_\phi^2\  \bar \phi^2
\ =\ \omega_\phi \left( a_{\omega_\phi}^\dagger a_{\omega_\phi} + {1\over 2} \right)
\nonumber\\
a_{\omega_\phi} & = & \sqrt{\omega_\phi\over 2} \ \bar \phi + i \sqrt{1\over 2\omega_\phi} \bar \Pi
\ \ ,\ \
a_{\omega_\phi}^\dagger \ = \  \sqrt{\omega_\phi\over 2} \ \bar \phi - i \sqrt{1\over 2\omega_\phi} \bar \Pi
\ \ \ ,
\label{eq:Hhatlatt01basis}
\end{eqnarray}
and the Hamiltonian in Eq.~(\ref{eq:Hhatlatt01free}) can be conveniently written in terms of the basis operators,
\begin{eqnarray}
\bar H & = & {1\over 2} \bar\Pi^2 \ +\ {1\over 2} \omega_\phi^2 \bar\phi^2 + {1\over 2} \left(1-\omega_\phi^2\right) \bar\phi^2
\ =\ H_{\rm basis}\ +\ \delta H_{\omega_\phi}
\ \ \ .
\end{eqnarray}

The eigenvalues and eigenstates of $\bar H$, in Eq.~(\ref{eq:Hhatlatt01free}),
are determined by diagonalizing the Hamiltonian matrix formed from
matrix elements of $\bar H$ in a truncated basis of eigenstates of $H_{\rm basis}$,  in Eq.~(\ref{eq:Hhatlatt01basis}).
An explicit example of the HO basis for three qubits-per-site may be found in the Appendix Sec.~\ref{sec:HObasis3qubitexample}.
Figure~\ref{fig:HOphiHOa} shows the precision of calculations of
the ground state energy of the HO Hamiltonian in Eq.~(\ref{eq:Hhatlatt01free})
expected on an ideal quantum computer as a function of the size of the HO basis for
different values of $\omega_\phi$.
\begin{figure}[!ht]
	\centering
	\includegraphics[width=0.85\columnwidth]{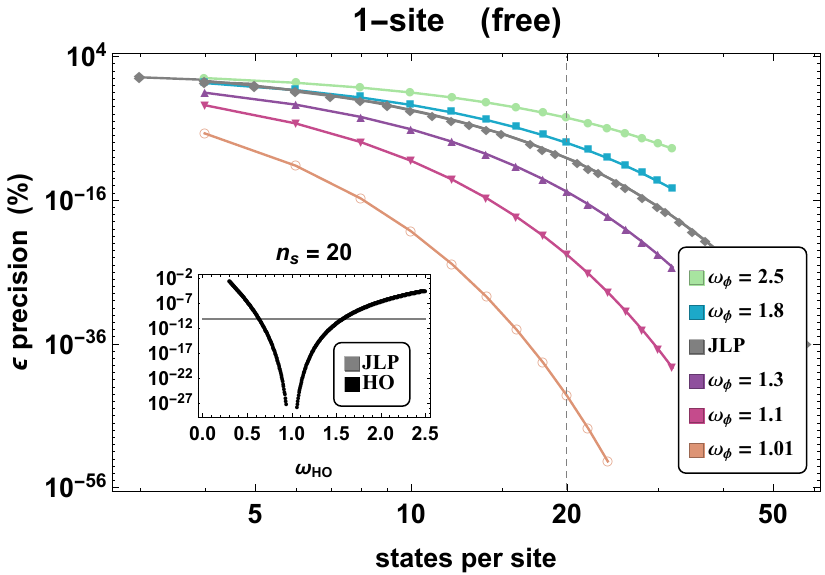}
	\caption{
	The expected precision of the ground state energy of the HO Hamiltonian in Eq.~(\ref{eq:Hhatlatt01free})
	on an ideal quantum computer using a HO basis defined by $\omega_\phi$ in Eq.~(\ref{eq:Hhatlatt01basis})
	as a function of the number of basis states.
	The inset figure shows the scaling for a HO basis of 20 states, which is a slice of the main figure indicated by a vertical gray dashed line (the calculations cross this line top-to-bottom in the order of the legend).
	In the limit of $\omega_\phi=1$ the lowest-lying basis state is an eigenstate, and $\epsilon=0$.
	Tuning $\omega_\phi$ to be in the vicinity of the optimal value $\omega_\phi=1$ outperforms field-space digitization,
	shown by the gray JLP curve.
			}
		\label{fig:HOphiHOa}
\end{figure}
Obviously, when $\omega_\phi=1$ the error vanishes.
For $\omega_\phi$ tuned to be in the vicinity of $\omega_\phi=1$ the precision obtained with the HO basis is better than that obtained with field-space digitization discussed in the previous sections.
However, poor choices of $\omega_\phi$ lead to inferior precision compared with field-space digitization.

The time-evolution induced by $H_{\rm basis}$ is simple, involving only single-phases,
and the quantum circuit to implement it corresponds to only phases applied to each qubit.
Since there are no interactions in this basis, all operators commute and there is no need for a Trotter decomposition, as the total phase can be determined and applied in one application.
When detuned away from $\omega_\phi=1$, the size of the Trotter step required to time-evolve the system
will be determined by the detuning.  In such a detuned scenario, the operator structure  from
$\delta H_{\omega_\phi}$ involves interactions between all qubits, as evidenced from Eq.~(\ref{eq:HOHOnQ3ops}).

\begin{table}[h]
  \centering
  \small
  \begin{tabular}{cc|ccccccc|cc}
  \hline
  \hline
  Basis & $n_Q$ & 0-body & 1-body & 2-body & 3-body & 4-body & 5-body & 6-body & $QFT$ & CNOTs \\
  \hline
   & 2 & 1 & 8  & 2 & & & & & $\checkmark$ & 8\\
   & 3 & 1 &  14 & 6 & & & & & $\checkmark$ & 24 \\
  JLP & 4 & 1 & 20 & 12 & & & & & $\checkmark$ & 48 \\
   & 5 & 1 & 26 & 20 & & & & & $\checkmark$ & 80 \\
   & 6 & 1 & 32 & 30 & & & & &  $\checkmark$ & 120 \\
  \hline
  JLP & $n_Q$ & 1 & $6 n_Q-4$ & $2*\binom{n_Q}{2}$ & & & & & $\checkmark$ & $8 \binom{n_Q}{2} $\\
  \hline
  \hline
     & 2 & 1& 2 & & & & & & & 0\\
     & 3 & 1& 3 & & & & & & & 0\\
  HO$_{\omega \equiv 1}$
     & 4 & 1& 4& & & & & & & 0\\
     & 5 & 1& 5& & & & & & & 0\\
     & 6 & 1& 6& & & & & & & 0\\
  \hline
  HO$_{\omega \equiv 1}$ & $n_Q$ & 1 & $n_Q$ & & & & & & & 0\\
  \hline
  \hline
   & 2& 1& 3 & 1 & & & & & & 2\\
   & 3& 1& 4 & 4 & 3 & & & & & 20\\
  HO$_{\omega \neq 1} $
   & 4& 1 & 5 & 5 & 11 & 7& & & & 96\\
   & 5& 1 & 6 & 6 & 16 & 26 & 15& & & 352\\
   & 6& 1 & 7 & 7 & 22  & 42 & 57 & 31 & & 1120\\
   \hline
   \hline
  \end{tabular}
  \caption{
  Resource requirements for one step in the Trotterized time evolution of a HO in the field-digitization JLP basis, a tuned HO basis, and a
  detuned HO basis. CNOT counts are based upon a standard multi-Pauli implementation requiring
  $2(k-1)$ CNOTs for each $k$-body operator.
  When a QuFoTr is required (JLP), the standard CNOT counts of $2\binom{n_Q}{2}$ for this operation
  (and its inverse) are included.   With the expected limitations in the number of gate operations applied in NISQ-era devices, only systems with $n_Q \leq 4$ may be practical.
  }
  \label{tab:HOcircuitBasisCompare}
\end{table}
In Table~\ref{tab:HOcircuitBasisCompare},
comparisons
in the types and numbers of operations and gates
required to time-evolve the HO described by $\bar H$ in Eq.~(\ref{eq:Hhatlatt01free})
between the field-digitization basis and a tuned/detuned HO basis are presented.
The 2-qubit, CNOT gate requirements are distinguished separately as their presence often represents the largest source of noise on NISQ-era quantum hardware.
The numbers in Table~\ref{tab:HOcircuitBasisCompare} are accumulated for a standard implementation of multi-Pauli gates~\cite{NielsenChuang}
and do not represent expected reductions of the HO basis operations through parity calculation or cancellations that may occur for
particular choices of the operator ordering~\cite{Hastings:2015}.
From Table~\ref{tab:HOcircuitBasisCompare}, it is clear that a tuned HO basis requires significantly fewer operations to evolve a free HO than does the field-digitization basis.   This is because the eigenstates of the system correspond exactly to the basis states.
However, a detuned HO basis involves an exponentially-growing number of multi-qubit operations,
leading to significantly more operations than the field-digitization basis.
Even when the eigenbasis is unknown, JLP has resource requirements limited to 2-body operators.
As a detuned HO basis shares features of a self-interacting system (detailed subsequently),
we conclude that, for this very simple system, the field-digitization basis examined in detail
in the works of JLP is more robust than a generic HO basis.
By this, we mean that for the evolution of an arbitrary, apriori unknown system, the field-digitization basis
will typically require fewer
quantum computational resources while possibly requiring fewer qubits,
as seen from Fig.~\ref{fig:HOphiHOa}.

It is interesting to consider whether the tuned HO could be used as a \enquote{standard candle} for the calibration of quantum hardware.  Its eigenstates and eigenenergies are known to infinite precision and thus could be considered not only as a calibration source but also as a calculation to distinguish the computational precision capable on classical and quantum hardware.  Using the details above and specifically the information of Table~\ref{tab:HOcircuitBasisCompare}, it can be seen that the tuned HO requires 0 two-qubit gates to implement. As such, it contains no entanglement and thus no unique signal that could not be generated with other pre-determined rotation gates to quantify and explore noise in NISQ-era hardware.

\subsection{$\lambda\phi^4$ Scalar Field Theory: Comparing Bases}

After the field and Hamiltonian redefinition of Eq.~(\ref{eq:Hhatlatt01free}), the interacting 0+1 scalar field is described by,
\begin{eqnarray}
\bar H & = &  {1\over 2} \bar\Pi^2 + {1\over 2}  \bar \phi^2 + \frac{\bar{\lambda}_0}{4!} \bar{\phi}^4
\ \ \ ,
\label{eq:Hbar01lam}
\end{eqnarray}
where $\hat \phi = {1\over \sqrt{\hat m_0}} \bar\phi$,
$\hat \Pi = \sqrt{\hat m_0} \bar\Pi$,
$\hat H=\hat m_0  \bar H$, and $\hat{\lambda}_0 = \hat{m}_0\bar{\lambda}_0$.
This system has been numerically studied previously by
Somma~\cite{Somma:2016:QSO:3179430.3179434}.
A value of $\bar{\lambda}_0=32$ will be chosen as a representative case of strong coupling,
where the system is no longer  a HO (nor perturbatively close)
and the basis selection for the description of the wavefunction between JLP digitization and
HO basis functions is relevant within the multi-dimensional space of precision,
qubits, gate decompositions, and tuning requirements.

When using the digitization techniques of JLP, introducing additional interactions does not introduce new challenges.
The only necessary modifications to the method are rescalings of the sampling distributions
(applying considerations for both field and conjugate-momentum space coverage).
\begin{figure}
  \centering
  \includegraphics[width = 0.5\textwidth]{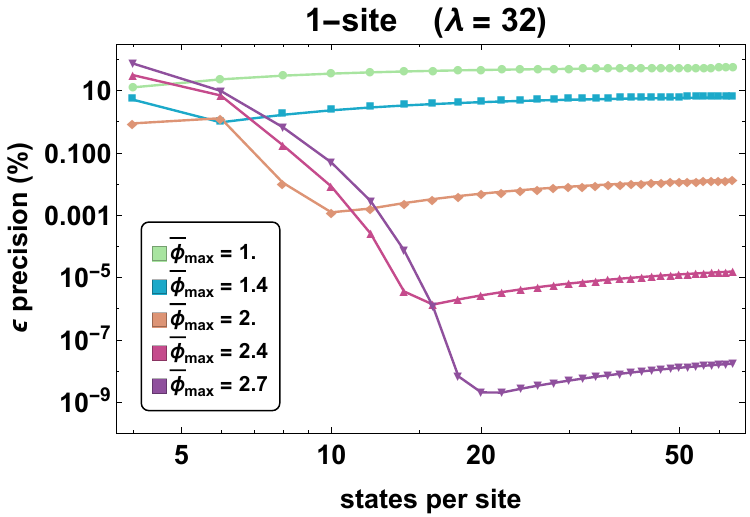}\\
  \includegraphics[width=0.46\textwidth]{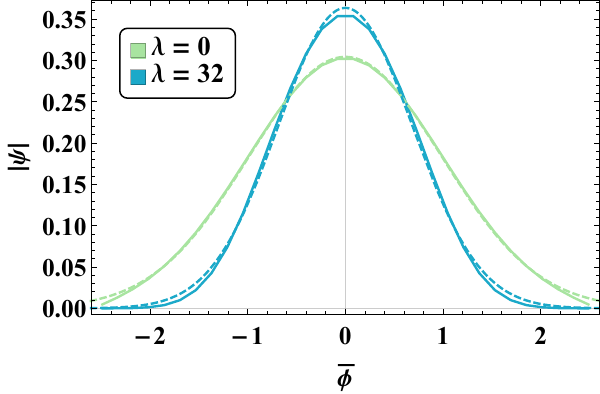} \
  \includegraphics[width=0.46\textwidth]{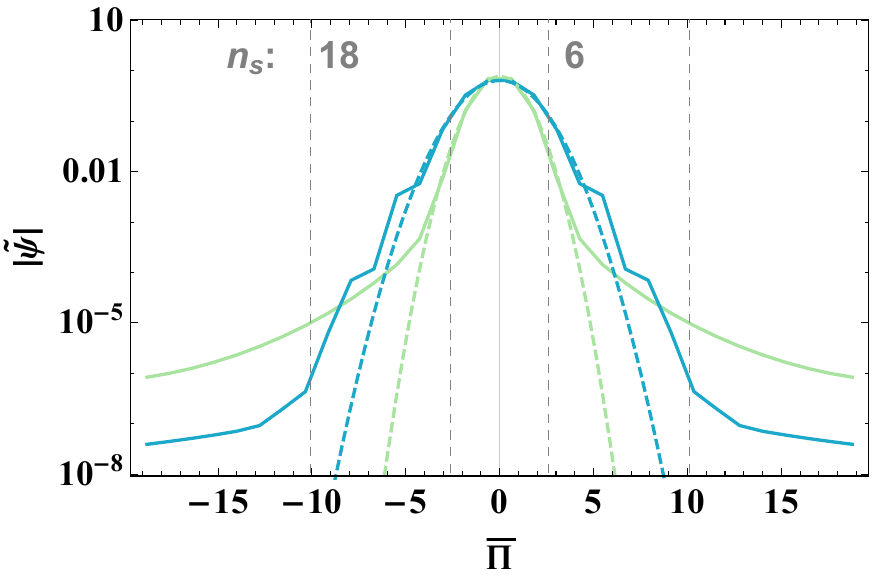}
  \caption{
 The panel in the upper row shows the expected precision of calculations of the ground-state
 energy of a 1-site $\lambda \phi^4$ scalar field theory
	performed with an ideal quantum computer for different values of $\bar\phi_{\rm max}$
	as a function of the number of states (the NS saturation points arise left-to-right in the top-to-bottom order of the legend).
 The lower row shows the field and conjugate-momentum space wavefunctions
 (left and right, respectively) for the free(green or lightly-shaded) and interacting(blue or darkly-shaded), 1-site $\lambda \phi^4$ shown at constant $\bar{\phi}_{\rm max} = 2.5$.
 The introduction of non-zero $\lambda$ reduces the spatial support of the wavefunction while increasing its support in momentum space.
 The small-dashed green(lightly-shaded) and blue(darkly-shaded) lines in the right panel are Fourier transforms of
 Gaussian fits to the wavefunctions in the left panel.
 The vertical, gray dashed lines in the right panel show the truncations in $\bar{\Pi}$ for 6 and 18 states---the location of NS saturation for $\bar{\phi}_{\rm max} = 2.5$ as seen in Figs.~\ref{fig:HOphiM}~and~the first row here.
  }
  \label{fig:wavefunction1site}
\end{figure}
In the case of $\lambda \phi^4$ with $\lambda = 32$,  introduction of the self-interaction shrinks the domain over which the wavefunction has support as shown in the lower left panel of  Fig.~\ref{fig:wavefunction1site}.
As a result, smaller values of $\bar{\phi}_{\rm max}$ may be used for precise calculations.
This can be seen in a comparison between Fig.~\ref{fig:HOphiM} and the upper panel of Fig.~\ref{fig:wavefunction1site}.
For a $\bar{\phi}_{\rm max}$ of 2.5, the highest precision attainable with $\lambda = 0$ and $\lambda = 32$
differs by $\sim 5$ orders of magnitude.
The precision with $\bar{\phi}_{\rm max}$ of 2.5 saturates with 18 states for $\lambda = 32$,
but saturates with only 6 states for $\lambda = 0$, indicating that the value of $\bar{\Pi}_{\rm max}$
has also increased with the introduction of the self-interaction, requiring a smaller value of $\delta_{\tilde{\phi}}$ in order to accurately represent the enlarged Fourier space.
This trade-off can be seen in the lower-right panel of Fig.~\ref{fig:wavefunction1site}.
To capture the Gaussian structure of the free HO requires only the inclusion of a small region of $\bar{\Pi}$ around zero.
For 6 states, the maximum value of the momentum can be determined by
Eq.~(\ref{eq:kphi}) to be $\pm 2.62$.
This value is indicated by the vertical, gray dashed lines in Fig.~\ref{fig:wavefunction1site}.  Outside of this region, the exponential behavior
turns power-law and inclusion of this portion of the wavefunction no-longer informs the sampling about the
physical momentum space, only about artifacts of the truncation.
By fitting a continuous Gaussian of infinite spatial extent to the wavefunction at left and plotting its
Fourier transform on the right (small-dashed curves),
6 states are found to lead to a $\delta_{\tilde{\phi}}$, and thus a maximum $k_{\tilde{\phi}}$,
that captures the Gaussian central region of the wave function.
For $\lambda = 32$, this maximum value in momentum space is no longer sufficient to saturate the NS sampling limit.
There is a significantly larger domain in momentum space before the wavefunction transitions to power-law behavior,
not appearing until $\bar{\Pi}$ values of $\sim \pm 10$.
Again, with Eq.~(\ref{eq:kphi}),   18 states per site are seen to be required for this truncation in momentum space,
a value in agreement with the location of the NS saturation point seen in the upper panel of
Fig.~\ref{fig:wavefunction1site}.

\begin{figure}
\centering
  \includegraphics[width=0.8\textwidth]{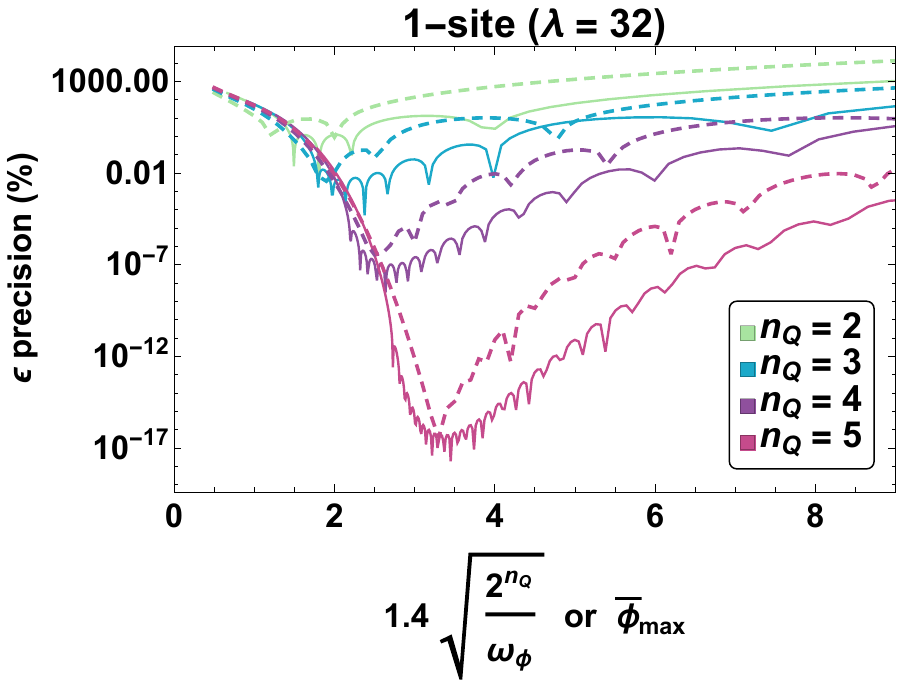}
  \caption{
  Exploration of sensitivity in JLP field digitization (dashed lines) and
  the HO basis (solid lines) to tuning of digitization parameters determining the low-energy states in momentum space.
  For JLP, the relevant parameter is $\bar{\phi}_{\rm max}$,
  while for the HO basis it is a combination of the frequency defining the  basis, $\omega_\phi$ and the number of states,
  $\sim \sqrt{2^{n_Q}/\omega_\phi}$.
  The horizontal axes of the HO curves have been rescaled to $1.4 \sqrt{2^{n_Q}/\omega_\phi}$
  to align them with the JLP curves.  These tuning curve pairs are minimized with smaller values of $\epsilon$ in the top-to-bottom order of the legend.
  }
  \label{fig:tuning1sitelam32}
\end{figure}

By comparing the gray band in Fig.~\ref{fig:HOphiM}, the scaling of the NS saturation for the free 1-site HO,
with the highest precisions attained in the upper panel of Fig.~\ref{fig:wavefunction1site},
it can be seen that the number of states (or qubits) required to achieve a particular precision is relatively stable for this self-interaction.  The values of $\bar{\phi}_{\rm max}$ along this band are skewed from those in the free theory, but the maximum precision attained through distribution of a fixed number of wavefunction sample points is not.  As this self-interaction causes a smooth deformation of the wavefunction, trading extent in field space for that in conjugate-momentum space, it is not surprising that the interacting ground state wavefunction achieves similar precision given similar quantum resources.

When using a basis of HO wavefunctions, the main consideration is, again, assuring that the chosen representation of the wavefunction sufficiently spans both field and conjugate-momentum space.  With JLP, $\bar{\phi}_{\rm max}$ is used to control the domain of support in field space while $\delta_{\tilde{\phi}}$ (or equivalently the number of states per site) is used to control the domain of support in momentum space.
With the HO basis functions, the parameters to be tuned are $\omega_\phi$ and the number of states.
Unlike the lattice-parameters of JLP, these parameters give correlated modifications to field and conjugate-momentum space.  Increasing $\omega_\phi$ creates basis functions that are more
localized in field space while exploring higher momentum-space truncations.
Increasing the number of states  also increases the momentum-space truncation,
but expands the field-space region of support.
Because of these correlations, it is meaningful to compare JLP's dependence of
$\bar{\phi}_{\rm{max}}$ with a combination of $n_Q$ and $\omega_\phi$
dictating the extent of the HO wavefunction basis,
$\sqrt{\frac{2^{n_Q}}{\omega_\phi}}$, reflecting the  fact that
$\sqrt{ \langle \phi^2 \rangle} \sim \sqrt{2^{n_Q}}$
and
$\sim 1/\sqrt{\omega_\phi}$.

\begin{table}[h]
  \centering
  \footnotesize
  \begin{tabular}{cc|ccccccc|ccc}
  \hline
  \hline
  Basis & $n_Q$ & 0-body & 1-body & 2-body & 3-body & 4-body & 5-body & 6-body & $QFT$ & CNOT$_\text{uncompiled}$ & CNOT$_\text{compiled}$ \\
  \hline
   & 2 & 1 & 8  & 2 & & & & & $\checkmark$ & 8 & 8\\
   & 3 & 1 &  14 & 6 & & & & & $\checkmark$ & 24 & 24\\
  JLP & 4 & 1 & 20 & 12 & & 1 & & & $\checkmark$ & 54 & 52 \\
   & 5 & 1 & 26 & 20 & & 5& & & $\checkmark$ & 110 & 96\\
   & 6 & 1 & 32 & 30 & & 15& & &  $\checkmark$ & 210 & 164\\
  \hline
  JLP & $n_Q$ & 1 & $4 n_Q-6$ & $2*\binom{n_Q}{2}$ & & $\binom{n_Q}{4}$ & & & $\checkmark$ & $8 \binom{n_Q}{2} + 6 \binom{n_Q}{4}$ & \\
  \hline
  \hline
     & 2 & 1& 3 & 2 & & & & & & 4 & \\
     & 3 & 1& 5 & 9 & 4 & & & & & 34 & \\
  HO & 4 & 1& 6& 16 & 18 & 10 & & & & 164 & \\
     & 5 & 1& 7& 22 & 32 & 44 & 22 & & & 612 & \\
     & 6 & 1& 8& 29 & 44 & 84 & 98 & 46 & & 1982 & \\
   \hline
   \hline
  \end{tabular}
  \caption{
  Resource requirements for
  one first-order-Trotterized step of
  time evolution for 1-site $\lambda\phi^4$ scalar  field theory in the field-digitization JLP basis and HO basis.
  CNOT counts are based upon a standard multi-Pauli implementation requiring $2(k-1)$ CNOTs for each $k$-body operator. When the QFT is required (JLP), the standard CNOT counts of $2\binom{n_Q}{2}$
  for this operation (and its inverse) are included in the second-to-last column.  The last column contains the required CNOT gates after manual compilation (see Appendix~\ref{sec:circcomp}).  With the expected limitations in the number of gate operations applied in NISQ-era devices, only systems with $n_Q \leq 4$ may be practical.
  }
  \label{tab:HOcircuitBasisComparelam32}
\end{table}

In Fig.~\ref{fig:tuning1sitelam32}, the expected precision of the ground-state energy is shown
as a function of $\bar{\phi}_{\rm max}$ and $\sqrt{\frac{2^{n_Q}}{\omega_\phi}}$ for JLP (dashed) and HO (solid) bases, respectively.
Values on the left of the minimum of each curve have reduced precision due to insufficient sampling in field space, while
to the right of the minimum, the precision is  reduced due to insufficient sampling in momentum space.
Only at the minimum is the sampling in both spaces optimal.
It can be concluded that for these parameters, values of $\bar{\phi}_{\rm max}$ or $\omega_\phi$ can be selected (for $n_Q\geq3$)  such that the errors introduced by digitization are significantly smaller than those expected from computations on NISQ-era hardware.  As such, the digitization of the scalar field is not expected to limit the accuracy of NISQ-era computations.  For $n_Q < 3$, the field digitization is expected to provide a limit to the accuracy of NISQ-era computations.
Comparing the basis choices given a fixed number of qubits, there is a value of the HO basis parameters that produce a higher-precision result in this system than a $\bar{\phi}_{\rm max}$-tuned JLP wavefunction digitization.
For a desired precision, the HO basis offers a larger acceptable window in the basis tuning parameters
than does the JLP field digitization basis.  This translates, through the circuit descriptions of
Figs.~\ref{fig:HOphiC3}~and~\ref{fig:HOHOnQ3circuit},
to reduced sensitivity on the exact angles applied in the Z-axis rotation gates.
This sensitivity will be relevant in the NISQ era with imperfect gate fidelities,
and will continue to be relevant once fault-tolerant quantum computing is available
(where the precision determines the number of $T$ gates~\footnote{The T gate,
$\begin{pmatrix}
  1 & 0 \\
  0 & e^{i \pi/8}
\end{pmatrix}$,
is the gate commonly added to the Clifford group to create universal quantum computation.  Its proliferation is considered a meaningful cost model for many plausible implementations considered for future fault-tolerant quantum computing.} needed to decompose any Z-axis rotation with expected scaling of $\big|\log_2|\epsilon_\theta|\big|$ \cite{Selinger:2015:ECA:2685188.2685198,Ross2016OptimalAC,Kliuchnikov2013}.
)

While Fig.~\ref{fig:tuning1sitelam32} shows desirable qualities when using HO basis functions to digitally
describe the wavefunction, quantum simulations of quantum systems have
many resource requirements to consider beyond qubit number and necessary precision of rotation angles.
Specifically, a large consideration in the feasibility of successfully implementing a quantum calculation in the NISQ era is the number and type of gates required to implement a single Trotter step of the time-evolution operator.  These gate counts are detailed in Table~\ref{tab:HOcircuitBasisComparelam32}.
For JLP, the 1-body operators from the QFT and $\binom{n}{2}$ 2-body operators from the terms quadratic in the field and its conjugate momentum are still present.  The $\lambda \phi^4$ interaction term introduces only $\binom{n}{4}$ 4-body operators and additional contributions to the identity and 2-body operators.
The latter can be consolidated with the operators previously identified
and thus does not contribute to the gate cost
(it does, however necessitate separate operator coefficient structures in field and conjugate-momentum space,
e.g., $\mathcal{O}_0$ from Eq.\eqref{eq:HOdec} can be written as $\mathcal{O}_{\tilde{\Pi}}$ and $\mathcal{O}_{\tilde{\phi}}$ which contain the same operators but with different relative coefficients).
The fact that operators are limited to interacting between a number of qubits equal to the highest power of field interaction included in the Hamiltonian is a feature of JLP not shared by the HO basis.
Here, the additional 2-qubit CNOT gates required to implement the QuFoTr for JLP field digitization
are quickly outnumbered by the CNOT gates required to implement the $k$-body operators for $k$ limited by the number of qubits in the site-register.

The fact that the scaling of CNOTs in the JLP basis is limited to $n_Q^4$  is advantageous
when considering the noise landscape of NISQ-era hardware dominated by 2-qubit interactions.
In Table~\ref{tab:HOcircuitBasisCompare}
and
Table~\ref{tab:HOcircuitBasisComparelam32},
the CNOT gate counts generally
do not include cancellations that may occur for particular operator orderings in the
Trotterization~\cite{Hastings:2015}.
In the JLP basis, we have performed a manual circuit compilation
of the $\lambda\phi^4$ scalar field theory,
eliminating pairs of adjacent CNOT gates, resulting in the CNOT gate counts shown in the right-most column of Table~\ref{tab:HOcircuitBasisComparelam32}.
While the $\bar{\phi}^2$ operator set by itself does not permit a reduction of the number of gates, in combination with the
$\bar{\phi}^4$ operator set, and also among the $\bar{\phi}^4$ operators, redundant CNOT operations in the leading
Trotter expansion can be removed.
While a similar reduction can be applied to the circuits of the HO basis, many changes of Pauli bases between operations make systematic cancellation difficult.  As was the case with the JLP basis, it is not expected that carrying out this elimination in the HO basis will change the scaling of the CNOT-operator accumulation.
A discussion of this manual compilation is given in Appendix~\ref{sec:circcomp}.

\subsubsection{Delocalized Wavefunctions: $m^2 < 0$}
\label{subsubsec:delocalized}

As mentioned in the introduction, $\lambda {\bm \phi}^4$ scalar field theory in $3+1$ dimensions is a cornerstone of the standard model of electroweak interactions~\cite{Glashow:1961tr,Weinberg:1967tq,Salam:1968rm},
where ${\bm \phi}$ is an electroweak doublet of complex real scalar fields.
At low energies, its potential is such that the vev of $\langle {\bm \phi} \rangle \ne 0$, breaking the electroweak gauge group
SU(2)$_L\otimes \text{U(1)}_Y \rightarrow \text{U(1)}_Q$ down to that of quantum electrodynamics.
This minimal symmetry-breaking mechanism,
the Higgs mechanism,
generates masses for the weak gauge bosons and the fermions, and gives rise to a single physical
scalar particle, the Higgs boson~\cite{Higgs:1964pj,Englert:1964et,Aad:2012tfa,Chatrchyan:2012xdj}.
\begin{figure}
  \centering
  \includegraphics[width = 0.45\textwidth]{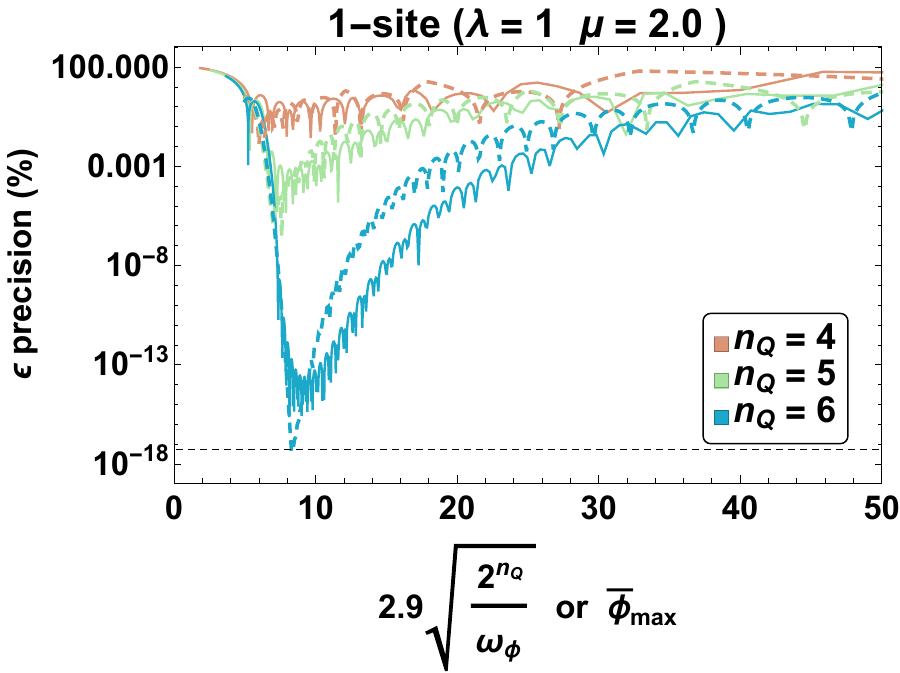}
  \includegraphics[width = 0.45\textwidth]{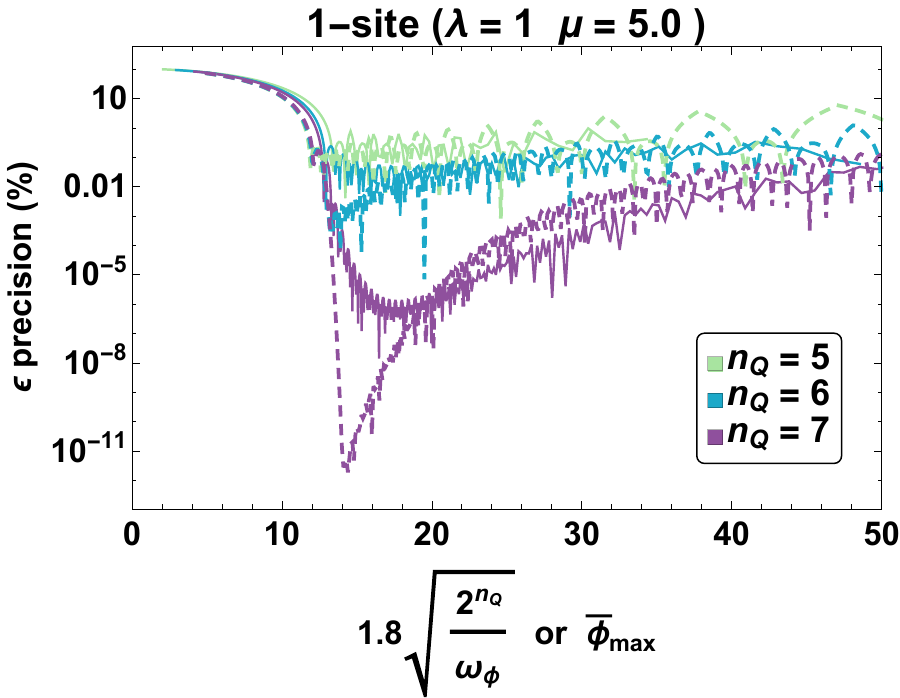}
  \caption{
  Exploration of the sensitivity in JLP field digitization (dashed lines) and
  the HO basis (solid lines) to tuning of digitization parameters determining
  the low-energy states for a 0+1-dimensional scalar field theory with $m^2 < 0$,
  with $\mu=2$ (left panel) and $\mu=5$ (right panel), resulting in delocalized wavefunctions.
 For JLP, a relevant parameter is $\bar{\phi}_{\rm max}$,
  while for the
  HO basis it is a combination of the frequency defining the HO basis, $\omega_\phi$, and the number of states per site.
    The horizontal axes of the HO curves have been rescaled to
        $2.9 \sqrt{2^{n_Q}/\omega_\phi}$ in the left panel and
    $1.8 \sqrt{2^{n_Q}/\omega_\phi}$ in the right panel
  to align them with the JLP curves.
  The black-dashed horizontal line in the left panel
  corresponds to the precision required to distinguish between the ground state and
  $1^{st}$ excited state for $\mu=2$. The corresponding line for $\mu = 5$ in the right panel lies many orders of magnitude beyond the range of the figure.  In both panels, the tuning curve pairs are minimized with smaller values of $\epsilon$ in the top-to-bottom order of the legend.
  }
  \label{fig:tuningSSB}
\end{figure}
In a 0+1 dimensional theory, the parameter regime $-\mu^2 = m^2 < 0$ produces a potential that
contains two minima located at $\phi = \pm \frac{\sqrt{3!} \mu}{\sqrt{\lambda}}$.
For any physical value of $\mu$,
the ground state wavefunction of the Hamiltonian is symmetric
under $\phi\rightarrow -\phi$
and non-degenerate and, as such, respects
the discrete global $Z_2$ symmetry of the Hamiltonian, with a vev of $\langle \phi\rangle = 0$.
However, it is delocalized with maxima near the two minima of the potential.
The wavefunction of the $1^{st}$-excited-state  of the system is similar to that of the ground state,
but it is antisymmetric under $\phi\rightarrow -\phi$.
As $\mu$ becomes large,
and the components of both wavefunctions become increasingly localized around the minima of the potential,
the energy difference between the ground state and the
$1^{st}$-excited-state becomes exponentially small, determined by the barrier-penetration amplitude
for transitioning from $+\phi$ to $-\phi$.
\begin{figure}
  \centering
  \includegraphics[width = 0.3\textwidth]{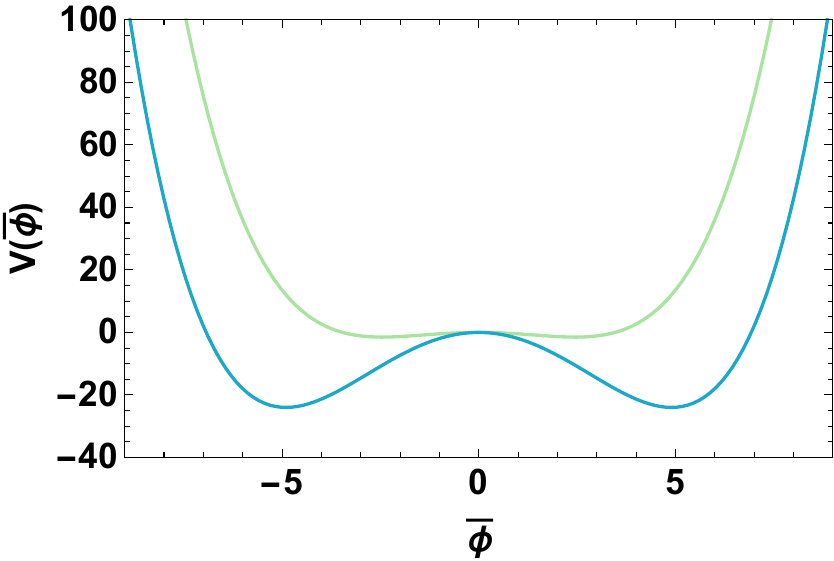}\
  \includegraphics[width = 0.3\textwidth]{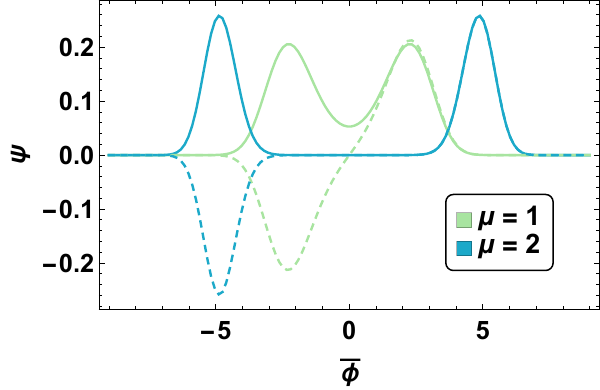}\
  \includegraphics[width = 0.3\textwidth]{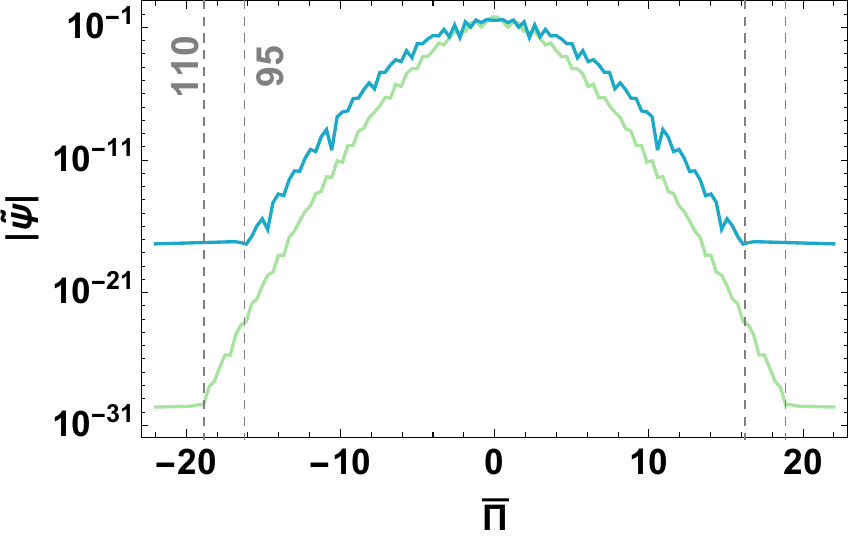}\
  \caption{
  The potentials (left panel) and wavefunctions (center and right panels) of the ground states (solid curves) and $1^{st}$ excited states (dashed curves)
  for systems with $m^2 < 0$.
  The center panel shows the spatial wavefunctions for $\lambda=1$ and $\mu=1, 2$,
  while the right panel shows the corresponding momentum-space wavefunctions.
  These wavefunctions result from using the JLP basis with $\overline{\phi}_{\rm max}=9$
  and $n_Q=7$ qubits.
  (The vertical dashed-grey lines in the right panel indicate
  the number of states at which the NS bound is saturated and thus an increase in $\bar{\phi}_{\rm max}$ would be profitable over an increase in quantum states).
    }
  \label{fig:SSBpsis}
\end{figure}

It is again relevant to consider alternate digitizations for representing the distributions in field
and conjugate-momentum space.
For large $\mu^2>0$,
(where the quantity $\frac{\mu}{\sqrt{\lambda}}$ is large with respect to the wavefunction's natural spatial extent),
the field space wavefunction expands toward two localized and distinct regions of support.
This is the case for the parameter values of
$\mu = 2, 5$ and $\lambda = 1$ chosen in Fig.~\ref{fig:tuningSSB}
and in Fig.~\ref{fig:SSBpsis}.
This enlarged field-space coverage demands similarly-large values of $\bar{\phi}_{\rm max}$ when working in the JLP digitization, or smaller values of $\omega_\phi$ in defining the HO basis.
Achieving these requirements can be accomplished in either basis when they are
tuned, as shown in Fig.~\ref{fig:tuningSSB}.
An additional consideration in considering the configuration of quantum simulations is that
the $1^{st}$-excited-state is becoming very close in energy with the ground state, a feature that is not
present in the previously considered situations.
A low-precision calculation, resulting from the use of a small number of qubits,
will be unable to resolve the ground state from the $1^{st}$-excited-state, and the wavefunctions
emerging from such calculations will likely be  arbitrary  combinations of the two.
Higher-precision calculations, requiring a larger number of qubits, will be required to resolve the low-lying states
in such systems.
For such delocalized states, in contrast to the results obtained from a potential
with $m^2>0$ in Fig.~\ref{fig:tuning1sitelam32},
the JLP basis can be tuned to produce higher precision in the ground-state energy
than the HO basis with the same number of qubits.
This outcome is not surprising---if the wavefunction is deformed into a distribution that is far from Gaussian,
as seen in Fig.~\ref{fig:SSBpsis},
a set of HO basis functions is no longer expected to offer superior coverage in the digital sampling.
An interesting result of this demonstration is the degree to which the formulation of JLP,
in which the basis is a periodic collection of delta functions agnostic to the structure of the wavefunction,
is capable of exceeding the precision of a basis specialized for an alternate symmetry of the low-lying wavefunctions.
The ability of JLP to perform with precision when applied to a range of systems,
and thus require little knowledge of the structure of the low-lying states,
will be a desirable feature of quantum simulations of more sophisticated, strongly-interacting field theories.

In these types of systems, and others, with near-degenerate low-lying states, the impact of noise in the quantum device
upon correctly identifying the ground state wavefunction is expected to be significant.
As discussed in Appendix~\ref{app:noisySimulation},
the noise levels (from either the propagator approximation of step (2) or the intrinsic gate implementation noise of step (3) in Fig.~\ref{fig:errorsourcediagram}) present in calculations with multiple degenerate extrema in the potential producing  delocalized low-lying states
will limit the systems that can be reliably explored as energy splittings are buried below the software (2) and hardware (3) noise levels.

\FloatBarrier
\section{1+1 Dimensional $\lambda\phi^4$ Scalar Field Theory}
\label{sec:1p1}
\noindent
The detailed analysis of $0+1$ scalar field theory presented in the previous sections provides a solid
foundation with which to consider scalar field theory in higher spatial dimensions with NISQ-era quantum computers.
In section~\ref{sec:SFT},
the Hamiltonian for scalar field theory in $d+1$ dimensions was presented, along with
its na\"ive layout on a spatial lattice.
The operator structure for multiple spatial sites is the same as for one spatial site except for the presence of the
$\phi \nabla^2 \phi$ operator, which includes contributions from particle  motion into the Hamiltonian.
The na\"ive representation of this operator as $\phi \nabla_a^2 \phi$ introduces terms that couples
the fields at two adjacent spatial sites.
In general,  smearing the fields to tame high-energy quantum fluctuations,
while preserving low-energy observables, will introduce couplings beyond adjacent spatial sites, but these can be
implemented with operations on two sites also.

In the situation with $d>0$,
the text-book way to construct field theory calculations is to work with HO's for each spatial-momentum mode,
i.e. define fields in terms of quanta with well-defined spatial momentum.
In perturbative calculations that can be performed by hand, this  method is extremely efficient.
In numerical computations of non-perturbative field theories,
such as LQCD, the system is typically defined with regard to fields in position space, while components of calculations
involve determining eigenvectors of the Dirac operator in the presence of a particular configuration of gauge fields.
In the study of systems with few sites in each spatial direction, it is likely the case that calculating with the
momentum-space modes is efficient~\cite{Yeter-Aydeniz:2017ubh}.
First implementation of this quantization procedure on quantum devices has been completed by an ORNL team~\cite{PhysRevA.99.032306}.
However, as argued by JLP~\cite{Jordan:2011ci,Jordan:2011ne,Jordan:2014tma,Jordan:2017lea},
as the interactions that are local in position space, such as
$\lambda\phi^4$, become non-local\footnote{For discussions of the implementation of non-local quantum interactions dominating the cost of quantum chemistry systems, see Refs.~\cite{Jordan1928,BRAVYI2002210,PhysRevA.95.032332,doi:10.1021/jz501649m,PhysRevX.8.011044,PhysRevLett.120.110501}
where alternate choices of qubit mappings or quantum simulation methods are explored to increase the locality of quantum operations.} in momentum space (distant momentum oscillators are capable of producing momentum-conserving contributions to the Hamiltonian), time evolving the system to a given state
defined in momentum-space
will become increasingly inefficient with increasing system size relative to a state defined in
position space~\cite{Jordan:2011ci,Jordan:2011ne,Jordan:2014tma,Jordan:2017lea,doi:10.1021/jz501649m}.
For the remaining discussion, we will limit ourselves to states and operations defined in
position space.

\begin{figure}
\includegraphics[width = 0.75\textwidth]{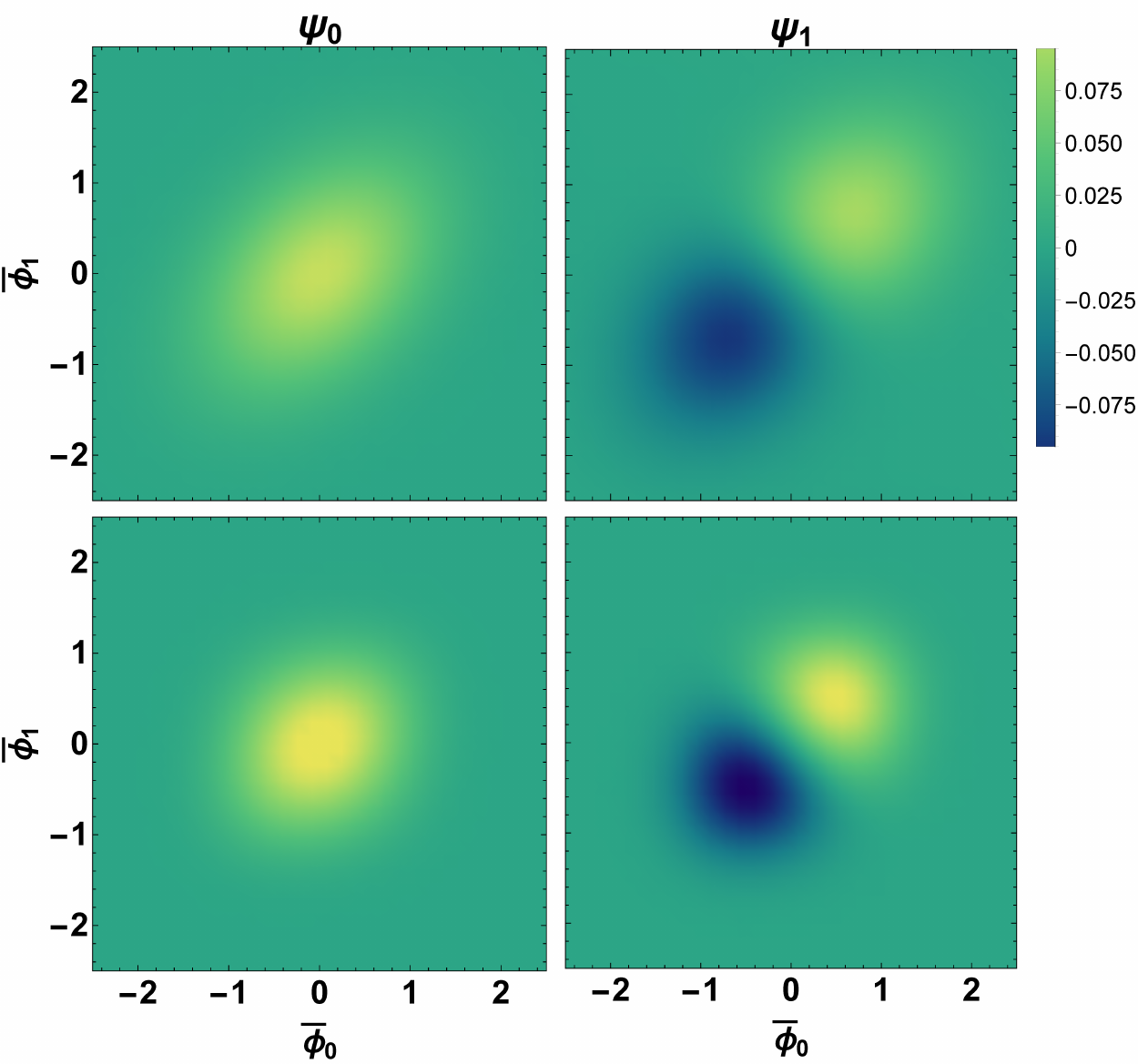}
  \caption{
  The ground and first-excited state wavefunctions for 2-site
  $\lambda \phi^4$ scalar field theory with $\lambda = 0$ and $\lambda = 32$ (top and bottom, respectively).
  The first excited state shows positive correlation between the oscillations at the two spatial
  sites $\left( \bar\phi_0, \bar\phi_1\right)$.
  }
  \label{fig:wavefunction2site}
\end{figure}

Application of the $d+1$-dimensional $\lambda\phi^4$-Hamiltonian time evolution operator to a position-space state can be accomplished site-by-site,
and involve at most $d$ neighboring two-site interactions at each site.
Therefore,  for a system with $\left(L/a\right)^d$ spatial lattice sites, this will require $\left(L/a\right)^d$ such applications.
This being the case, study of the 2-site $1+1$ dimensional $\lambda\phi^4$ theory provides a complete inventory of
the operations and gate counts required to perform a $d+1$-dimensional $\lambda\phi^4$ calculation,
and we have performed such estimates and numerical calculations in this 2-site theory.  Given this 2-site locality and quantum hardware capable of parallelizing the implementation of gates acting in separate tensor product spaces,  application of the Hamiltonian to a position-space state can be accomplished with a circuit of constant depth with increasing lattice size \cite{Lloyd1996}.
The field-space wavefunctions associated with the ground state and first-excited state of the 2-site $1+1$
dimensional theory are shown in Fig.~\ref{fig:wavefunction2site},
with the wavefunction at site-0 denoted by $\bar\phi_0$ and at site-1 by  $\bar\phi_1$.
 A large value of the self-interaction coupling, $\lambda$, focuses this correlation in $\left( \bar\phi_0, \bar\phi_1\right)$.

\begin{figure}
  \includegraphics[width = 0.75\textwidth]{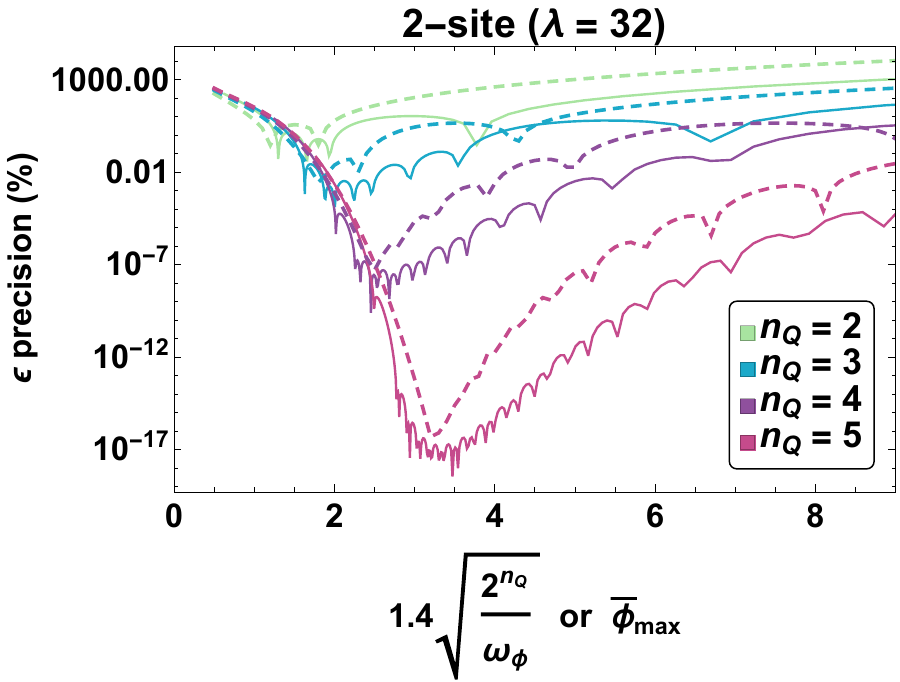}
  \caption{
  The precision of the calculated ground-state energy for the 2-site  lattice
  $\lambda\phi^4$ scalar field theory
  with $\lambda = 32$
  performed with an ideal quantum computer for different numbers of qubits as a function of support in field space.
  For JLP, the relevant parameter is $\bar{\phi}_{\rm max}$,
  while for the HO basis it is a combination of the frequency defining the HO basis, $\omega_\phi$,
  and the number of states per site.
  The shown precision does not include deviations of this 2-site $1+1$ dimensional theory
  from the continuum limit of the $1+1$ dimensional theory for which the number of spatial sites approaches infinity for a fixed spatial extent.
  The horizontal axes of the HO curves have been rescaled to $1.4 \sqrt{2^{n_Q}/\omega_\phi}$
  to align them with the JLP curves.  These tuning curve pairs are minimized with smaller values of $\epsilon$ in the top-to-bottom order of the legend.
  }
  \label{fig:epsilonscaling2site}
\end{figure}

As seen in Fig.~\ref{fig:epsilonscaling2site},
the 2-site $\lambda\phi^4$  theory experiences double-exponential convergence in $n_Q$ to the un-digitized value.
However, just as the use of a finite-difference operator in the field-space implementation of $\bar{\Pi}^2$ introduced polynomial deviations in $\delta_{\bar{\phi}}$ (see results from local and improved operators in Fig.~\ref{fig:HOa}),
the finite-difference implementation of $\phi \nabla_a^2 \phi$
in position space
introduces analogous  polynomial deviations in $a$ from the continuum limit.
These lattice-spacing errors are not shown in Fig.~\ref{fig:epsilonscaling2site}.
Thus, this method converges to the continuum value with lattice-spacing errors that scale as $\epsilon \sim 2^{N_Q}$.
Of course, with a large
quantum computer, it could become possible to
remove these polynomial lattice-spacing artifacts through use of the QuFoTr
and subsequent implementation of the exact lattice phases in
Fourier space to create a smeared, non-local gradient operator (exactly as was done in field space).
Rather than requiring a QuFoTr to be applied on each of the modest-sized qubit registers associated
with individual lattice sites, this proposal would require a QuFoTr across the entire lattice---an
entangling operation amongst all $N_Q$ qubits.
At least in the NISQ era, it is expected that such global operations will be prohibitive both in gate fidelity
as well as coherence time.
For this reason, the finite-difference form of the gradient operator,
demanding only local interactions between the qubit registers at neighboring sites, appears to be
optimal~\cite{Jordan:2011ci,Jordan:2011ne,Jordan:2014tma,Jordan:2017lea}.

\begin{table}
\footnotesize
\begin{tabular}{cc|cccccccccccc}
\hline
\hline
  Basis &$n_Q$ & 2-body & 3-body & 4-body & 5-body & 6-body & 7-body & 8-body & 9-body & 10-body & 11-body & 12-body & CNOT \\
  \hline
 &2 & 4 &  &  &  &   &  &  &  &  &  & & 8 \\
 &3 & 9 &  &  &   &  &  &  &  &  &  &  & 18 \\
 JLP& 4 & 16 &  &  &  &  &  &  &  &  &  &  & 32 \\
 &5 & 25 &  &  &  &  &  &  &  &  &  &  & 50 \\
 &6 & 36 &  &  &  &  &  &  &  &  &  &  & 72 \\
 \hline
 & $n_Q$ & $n_Q^2$ & & & & & & & & & & & $2 n_Q^2$\\
 \hline
 \hline
 &2 & 1 & 6 & 9 &  &  &  &  & &  &  &  & 80 \\
 &3 & 1 & 8 & 30 & 56 & 49 &  &  & &  &  &  & 1,152 \\
 HO&4 & 1 & 10 & 47 & 140 & 271 & 330 & 225 &  &  &  &  & 11,264 \\
 &5 & 1 & 12 & 68 & 244 & 630 & 1204 & 1668 & 1612 & 961 &  &  & 89,600 \\
 &6 & 1 & 14 & 93 & 392 & 1186 & 2772 & 5154 & 7560 & 8541 & 7182 & 3969 & 626,688 \\
 \hline
 \hline
\end{tabular}
  \caption{
  Operators associated with the additional $\bar\phi(x) \bar\phi(x+1)$ operator
  resulting from the finite-difference spatial gradient operator $\phi \nabla_a^2 \phi$
  for time evolution of 2-site  lattice $\lambda\phi^4$ scalar field theory
 in the JLP  and HO field-digitization bases.  CNOT counts are based upon a standard multi-Pauli implementation requiring
  $2(k-1)$ CNOTs for each $k$-body operator.
  }
  \label{tab:phiphip1circuitBasisComparelam32}
\end{table}

\par When implementing the gradient operator as a finite difference,
there is only one set of operators $\bar{\phi}(x) \bar{\phi}(x+1)$ acting between the spatial sites that need be additionally considered.
Table~\ref{tab:phiphip1circuitBasisComparelam32} shows the nature and number of pauli terms associated with
this additional operator in the $1+1$-dimensional
Hamiltonian\footnote{These gate counts are in addition to those resulting from action on the individual sites that have been determined in previous sections of this paper.}.
In this 1+1-dimensional system, the coefficient of the mass term in field space is modified by two of the terms in the
$\phi \nabla_a^2 \phi$ operator, but the operator structure is unaltered.
As mentioned above, the quantum resources calculated in this paper may be easily combined to determine the requirements for larger lattices in $d$-dimensions, for example
\begin{equation}
  {\rm CNOT}_{lattice}(n_Q,d,L/a) = L^d a^{-d}\  {\rm CNOT}_{\text{1-site}}(n_Q) + d L^d a^{-d} \ {\rm CNOT}_{\bar{\phi}(x) \bar{\phi}(x+1)}(n_Q)
\end{equation}
expresses the total number of CNOT gates required to evolve the field across a lattice with CNOT$_{\text{1-site}}(n_Q)$ extracted from Tables~\ref{tab:HOcircuitBasisCompare} or~\ref{tab:HOcircuitBasisComparelam32} for the free or self-interacting fields, respectively, CNOT$_{\bar{\phi}(x) \bar{\phi}(x+1)}(n_Q)$ extracted from Table~\ref{tab:phiphip1circuitBasisComparelam32},
$n_Q$ the number of qubits used to digitize the field at each site, $d$ the dimensionality of space, $L$ the spatial extent in each dimension, and $a$ the lattice spacing in each dimension.
The nearest-neighbor interactions
between sites in the JLP digitization requires all $n_Q^2$ 2-body operators that can be created
between the two site registers.
This is contrasted by the HO basis where operators between the two site registers are not limited to
2-body qubit interactions, but require tensor product Pauli operators acting on up to all $2n_Q$ qubits.
Because of this dramatic difference in the structure of necessary operators, even for the smallest number of qubits per spatial site, the JLP basis requires fewer resources to implement the
$\phi \nabla_a^2 \phi$ operator---emphasizing the importance of an application's physical
representation onto qubit degrees of freedom in quantification of its required quantum resources.

\section{Summary and Outlook}
\noindent
Quantum computing and quantum information science is anticipated to provide disruptive changes to scientific computing and to the ways that we think about addressing  scientific challenges.
The prospect of being able to explore quantities in quantum many-body systems, including quantum gauge field theories such as quantum chromodynamics, that
require exponentially-large classical computing resources, such as for
dense matter or in the time-evolution of non-equilibrium systems, is truly exciting.
In this work, we have built upon foundational works by
Jordan, Lee and Preskill~\cite{Jordan:2011ci,Jordan:2011ne,Jordan:2014tma,Jordan:2017lea}
on how to formulate scalar field theory on quantum computers to determine
properties of the scalar particle and interactions, both elastic and inelastic, between particles.
In an attempt to understand the magnitude of resources required for even modest quantum
computations in a simple field theory,
our work has focused on the digitization of scalar field theories with only a small number
of qubits per spatial lattice site.
The recent work by Macridin, Spentzouris, Amundson and Harnik~\cite{PhysRevLett.121.110504,Macridin:2018oli},
which,
building upon the work of Somma~\cite{Somma:2016:QSO:3179430.3179434},
emphasized the utility of the Nyquist-Shannon (NS) sampling theorem,
is a theme for our work as it provides an important guide for tuning digitization
parameters in quantum field theories for the accurate representation of field and conjugate-momentum spaces
on quantum devices (and  may also have  implications for classical calculations).

In addition to an in-depth exploration of the requirements for a basis of eigenstates of the field operator,
as introduced by JLP, we have introduced and explored the resources required for, and the utility of, a basis
of harmonic oscillator eigenstates.
We have performed operator decompositions of the Hamiltonians for a small number of qubits in
$0+1$ and $1+1$ dimensional systems.
As tunings are required in both bases for an optimal computation on an ideal quantum device,
we find that both bases are effective, but that the JLP basis appears to be more robust
for systems that are delocalized in field space or not smooth in either field or conjugate-momentum space.
We considered the impact of noise on calculations in such systems and found that
parameters defining the field theory should be tuned given the limits in
precision imposed by the quantum device in order to optimize the scientific productivity of the calculation.
In either basis, when tuned,
a quantum device with $n_Q=3$ or $n_Q=4$ qubits used to define the field at each
spatial lattice site is found to be
able to provide a precision of better than $\sim 10^{-6}$ for a given lattice spacing
for a potential with $m^2>0$.
Separating the spatial lattice-spacing systematic error from the digitization systematic error in field space,
the digitization error in the space of low-lying energy eigenstates, $\epsilon_{\rm dig}$,  is found to scale as
$|\log|\log |\epsilon_{\rm dig}|||\sim n_Q$ for $n_Q$ qubits per site, while the lattice spacing error, $\epsilon_{\rm latt}$,
scales as
$|\log |\epsilon_{\rm latt}||\sim N_Q$ where $N_Q$ is the total number of qubits in the simulation.

The lessons learned from studying the digitization of a scalar field onto qubit degrees of freedom have been numerous.
The following is an itemized summarization of those lessons appearing in 0+1 dimensional field theory---before the additional complications of a lattice spacing and spatial momentum are introduced in section~\ref{sec:1p1}.

\begin{enumerate}
  \item
  The scalar field digitization techniques of
  Jordan-Preskill-Lee~\cite{Jordan:2011ci,Jordan:2011ne,Jordan:2014tma,Jordan:2017lea,Somma:2016:QSO:3179430.3179434,PhysRevLett.121.110504,Macridin:2018oli},
  a momentum-space mode expansion~\cite{Yeter-Aydeniz:2017ubh}
  and a harmonic oscillator basis
  are relevant for NISQ-era hardware implementations.
  The number of qubits per site needed to reduce the digitization and discretization systematic errors below near-term hardware noise levels are
  $n_Q \sim 4$ for potentials with $m^2>0$ and
  $n_Q \gsim 6$  for potentials with $m^2<0$.
  These qubit requirements are consistent with those of Refs.~\cite{Somma:2016:QSO:3179430.3179434,PhysRevLett.121.110504,Macridin:2018oli} and extend these modest requirements to delocalized field-space wavefunctions.
  \label{itm:relevant}
  \item
  When the Nyquist-Shannon sampling bound,
  introduced in this context by Macridin, Spentzouris, Amundson and Harnik~\cite{PhysRevLett.121.110504,Macridin:2018oli},
  building on work by Somma~\cite{Somma:2016:QSO:3179430.3179434},
   is saturated, field and conjugate-momentum space are
  described to comparable accuracy and the ground-state energy can be reproduced with a precision scaling
  with the number of qubits in the site register as $|\log | \log | \epsilon ||| \sim n_Q$.
  For a free theory in 0+1 dimensions, the coefficients of this relationship are calculated to be $\epsilon_{(\%)} = (1.8(2)\times10^3)\ 2^{-2.234(4) n_s}$.
  This rapid convergence is responsible for item~\ref{itm:relevant}. \label{itm:NS}
  \item
  In order to enjoy the double-exponential convergence of item~\ref{itm:NS},
  the conjugate-momentum operator must be constructed with exact  phases in momentum space,
  leading to a highly-non-local operator in field space.
  This is possible through use of the quantum Fourier transform as an efficient entangling
  operation among all qubits in the register~\cite{Somma:2016:QSO:3179430.3179434,PhysRevLett.121.110504,Macridin:2018oli}.
  Note that for spatial dimensions greater than zero,
  the size of the qubit register that undergoes QuFoTr grows with precision of the scalar field digitization,
  and not the size of the spatial lattice.
  Given the qubit estimates of item~\ref{itm:relevant},
  global entanglement within this register is a reasonable  goal for  NISQ-era hardware. \label{itm:exactphases}
  \item
  The implementation of \emph{exact}  phases required in item~\ref{itm:exactphases} does not supersede the effects of noise.
  Under a generic noise model on phases in conjugate-momentum space, the double-exponential convergence
  stated in item~\ref{itm:NS} is only seen up to a precision barrier set by the magnitude of the noise.
  In spite of this physical limitation, the use of exact phases is still
  recommended as it minimizes the number of states needed in the quantum system and is no more costly than implementation of conjugate-momentum operators local in field
  space.
  As an additional feature, using exact phases in momentum space yields symmetry between the gates required in field and conjugate-momentum space.  This analysis, shown in Figs.~\ref{fig:HOa}~and~\ref{fig:HOphiM}, informs a balancing between the noise level of the quantum system and the precision with which the quantum field theory is mapped onto qubits.
  It is naturally expected that there will be  advantages in ``matching'' the precision of a calculation to the
  noise level in a given quantum device or vice-versa.
  \label{itm:noise}
  \item While the relative precision attainable with the JLP and HO bases depends on the structure of the low-energy wavefunctions,  the comparatively-burdensome operator structure of the HO basis may be a deciding factor in the NISQ era.  While JLP Pauli operators are limited to 1-, 2-, and $m$-body qubit interactions for $\lambda \phi^m$, the HO basis includes all $k$-body operators up to $k = n_Q$.
 These operators necessitate larger numbers of CNOT gates, the two-qubit entangling gates dominating the error contribution for many instances of near-term hardware.
  \item Gate decompositions can be  sensitive to symmetries that may be broken through classically-inconsequential
  truncation artifacts.
  When truncating spaces to contain states that have interactions beyond the truncated space
  (e.g., in the HO basis), it is preferable to truncate  \emph{after} construction of the Hamiltonian in an enlarged space to remove edge-effects in the interactions.  When deciding upon boundary conditions (e.g., in the JLP basis), it is beneficial to consider alternatives (such as twisted boundary conditions) that symmetrize the distribution of wavefunction samples in Fourier space.
  \item For the near-term, including and beyond the NISQ-era, digitization and discretization of the field onto qubits need not limit the accuracy obtained in simulations of scalar field theory.  Rather, the software approximation of the propagator (e.g., through Trotterization) and hardware gate-error rates (currently above $10^{-4}$ in the simple model of Appendix~\ref{app:noisySimulation}) are expected to be the dominant limitations to simulations of the scalar field.
\end{enumerate}

The content of this paper has provided information to make a hardware-specific, informed
decision on the parameters chosen to implement a scalar field on quantum devices.
For example, imagine a future in which the application of CNOT gates become relatively inexpensive \footnote{This is the case for many models of fault-tolerant quantum computing}, rotation gates contain small but non-negligible errors in their rotation angle, and a hypothetical goal is to simulate a 0+1-dimensional scalar field with quartic self-interaction to at least $10^{-11} \%$ precision.
Both the JLP and HO bases are capable of achieving this goal, as seen in Fig.~\ref{fig:tuning1sitelam32}.
However, given the wider range of tuning parameters allowable in the HO basis, making the precision more robust to noise in the rotation gates' angles, an informed choice might be to work with a HO basis.
Imagine, as a modification to this scenario, that gates are expensive
(either due to short coherence times or to their imperfect fidelity) but qubits are cheap.
In this case, the contents of Table~\ref{tab:HOcircuitBasisComparelam32} raise concerns over the 612 entangling gates required to implement the Trotterized circuit in the HO basis.
Instead, it may be logical to use JLP digitization, add one qubit  to increase the range of the
tuning parameter $\bar{\phi}_{\rm max}$ capable of satisfying the above precision requirement,
and as a result require only one third of the previous number of CNOT gates for each Trotter step, a number also more amenable to the NISQ era.
It is further found that small calculations of $\lambda\phi^4$ scalar field theories
can be performed with a modest number of qubits.
For example, an ideal $\sim 60$-qubit device could be used to describe such a system with up to $\sim 20$ spatial lattice sites (with three qubits per site defining the field digitization),
arranged in a number of dimensions, at the $10^{-6}$ level.
Observing that this error is below that expected for digital gate implementation on NISQ-era devices of this size indicates that properties of a scalar field independent of digitization artifacts will be accessible to quantum devices with fewer than 100 qubits.
Beyond the digitization errors (step (1) of Fig.~\ref{fig:errorsourcediagram}) that have here been demonstrated to be controllable with qubit requirements appropriate for the NISQ era, the quantum simulation errors arising from necessary approximation of the time evolution operator (step (2) of Fig.~\ref{fig:errorsourcediagram}) and imperfect implementation on noisy hardware (step (3) of Fig.~\ref{fig:errorsourcediagram}) now remain as the dominant sources of uncontrolled error in calculations of the scalar field implemented on quantum hardware.

Analyses such as we have presented in this work,
are expected to play a role in optimizing the output of quantum devices in any scientific application domain.
Their use in tuning digitization parameters to \enquote{match} the precision of calculations
to specific hardware will become increasingly important to make the best use of available hardware at any given time,
as is the case in classical high-performance computing.
With rapid development of quantum hardware in the NISQ era, it is likely that the optimal layout
of a quantum system onto qubit degrees of freedom will have significant variability,
both with choice of quantum architecture and with time.  Having a detailed map of the resource landscape is thus critical for creating informed decisions for implementing calculations across a range of quantum architectures.

\appendix

\section{Jordan-Lee-Preskill Basis Example: Three Qubits}
\label{sec:HOnQ3}
\noindent

In order to provide explicit examples to reinforce the generalities described in the main text,
quantum computations of the HO in Eq.~(\ref{eq:Hhatlatt01free}) performed with 3 qubits, $n_Q=3$
(with $n_s=8$ quantum states), are considered in detail.
With 8 states, $\bar\phi$ is sampled at the field  and  conjugate-momentum values
\begin{eqnarray}
\tilde\phi_i & = &
 \{
 \pm 1 , \pm {5\over 7} , \pm {3\over 7} , \pm {1\over 7}
\}\ \bar\phi_{\rm max}
\ \ , \ \
k_{\tilde\phi }\ =\
 \{
 \pm {7\over 8} , \pm {5\over 8} , \pm {3\over 8} , \pm {1\over 8}
\}\ {\pi\over\delta_{\tilde\phi}}
\ \ , \ \
\delta_{\tilde\phi}={2 \bar\phi_{\rm max} \over 7}
 \  \ \  ,
\label{eq:phinQ3}
\end{eqnarray}
where we have dropped the ``$\Delta$''  superscript on $k_{\tilde\phi }^\Delta$ in Eq.~(\ref{eq:kDeldef}).

The operator decomposition of the Hamiltonian for this system is straightforward.
It is useful to extend the basis of Pauli operators to include the identity matrix,
$\bar\sigma = \left( \sigma^x, \sigma^y, \sigma^z, I_2\right)$,
and  to define the general tensor product of $n_Q$ operators
$T^{ijk}=\bar\sigma^i\otimes\bar\sigma^j\otimes\bar\sigma^k$
\begin{eqnarray}
T^{ijk} & = & \bar\sigma^i\otimes\bar\sigma^j\otimes\bar\sigma^k
\ \ ,\ \
{\rm Tr}\left[ \ T^{ijk} \ T^{i^\prime j^\prime k^\prime}\ \right] \ =\ 8 \ \delta^{ii^\prime}\delta^{jj^\prime}\delta^{kk^\prime}
 \ ,
\label{eq:Tdef}
\end{eqnarray}
where the orthogonality of the $T^{ijk}$ is helpful in decomposing the Hamiltonian into qubit operators.
Projecting against the $T^{ijk}$, it is straightforward to show that
\begin{eqnarray}
\tilde\phi^2
& = &
{4\over 49}\ \bar\phi_{\rm max}^2
\ {\cal O}_0^{(n_Q=3)}
\ \ ,\ \
\tilde\Pi^2 \ =\
{49 \pi^2\over 64\ \bar\phi_{\rm max}^2}
\ {\cal O}_0^{(n_Q=3)}
\ \ \ \ ,
\nonumber\\
{\cal O}_0^{(n_Q=3)}
& = &
4\   \sigma^z\otimes\sigma^z\otimes I_2
\ +\ 2\  \sigma^z\otimes I_2\otimes\sigma^z
\ +\  I_2\otimes\sigma^z\otimes \sigma^z
\ +\ {21\over 4}\  I
\ \ \ \ ,
\nonumber\\
& = & {\cal O}_{03}^{(n_Q=3)} \ + \ {21\over 4}\  I
\ \ \ ,
\label{eq:HOdec}
\end{eqnarray}
where $I_2$ is the identity operator acting on a single qubit, and
where the operator has been  split into an overall identity and non-identity terms.
As expected from the JLP explicit construction,  the decomposition of the digitized HO Hamiltonian
into Pauli operators acting on individual qubits is quite simple, and easily extended to larger numbers of
qubits~\footnote{
The analogous decomposition for a 4-qubit system is
\begin{eqnarray}
\tilde\phi^2_{n_Q=4}
& = &
{4\over 225}\ \bar\phi_{\rm max}^2
\ {\cal O}_0^{(n_Q=4)}
\ \ ,\ \
\tilde\Pi^2_{n_Q=4}  \ =\
{225 \pi^2\over 256 \  \bar\phi_{\rm max}^2}
\ {\cal O}_0^{(n_Q=4)}
\ \ ,
\nonumber\\
{\cal O}_0^{(n_Q=4)}
& = &
16\  \sigma^z\otimes\sigma^z\otimes I_2\otimes I_2
\ +\
8\  \sigma^z\otimes I_2\otimes\sigma^z\otimes I_2
\ +\
4\ \sigma^z\otimes I_2\otimes I_2\otimes\sigma^z
\ +\
4\   I_2\otimes\sigma^z\otimes\sigma^z\otimes I_2
 \nonumber\\
&&
\ +\
2\  I_2\otimes\sigma^z\otimes I_2\otimes\sigma^z
\ +\
 I_2\otimes I_2\otimes\sigma^z\otimes\sigma^z
\ +\
{85\over 4}\   I
\ \ \ ,
\label{eq:HOnQ4dec}
\end{eqnarray}
while for the $n_Q=5$ system,
\begin{eqnarray}
\tilde\phi^2_{n_Q=5}
& = &
{4\over 961} \ \bar\phi_{\rm max}^2
\ {\cal O}_0^{(n_Q=5)}
\ \ ,\ \
\tilde\Pi^2_{n_Q=5}  \ =\
{961 \pi^2\over 1024\  \bar\phi_{\rm max}^2}
\ {\cal O}_0^{(n_Q=5)}
\ \ ,
\nonumber\\
{\cal O}_0^{(n_Q=5)}
& = &
64\  \sigma^z\otimes\sigma^z\otimes I_2\otimes I_2\otimes I_2
\ +\
32\  \sigma^z\otimes I_2\otimes\sigma^z\otimes I_2\otimes I_2
\ +\
16\  \sigma^z\otimes I_2\otimes I_2\otimes\sigma^z\otimes I_2
 \nonumber\\
&&
\ +\
8\  \sigma^z\otimes I_2\otimes I_2\otimes I_2\otimes\sigma^z
\ +\
16\  I_2\otimes\sigma^z\otimes  \sigma^z\otimes   I_2\otimes I_2
\ +\
8\  I_2\otimes\sigma^z\otimes I_2\otimes  \sigma^z\otimes   I_2
  \nonumber\\
&&
\ +\
4\  I_2\otimes \sigma^z\otimes  I_2\otimes I_2\otimes\sigma^z
\ +\
4\  I_2\otimes  I_2\otimes \sigma^z\otimes\sigma^z\otimes  I_2
\ +\
2\  I_2\otimes  I_2\otimes \sigma^z\otimes  I_2\otimes\sigma^z
  \nonumber\\
&&
\ +\
 I_2\otimes  I_2\otimes I_2\otimes\sigma^z\otimes \sigma^z
\ +\
{341\over 4}\   I
\ \ \ .
\label{eq:HOnQ5dec}
\end{eqnarray}
}.
The structure of the operators, and their extensions to larger systems, is interesting.
The only nontrivial operators that appear involve operations on 2 qubits only,
without the appearance of higher-qubit operators, such as those involving 3-, 4- or 5-qubits.
This simple operator structure extends to  larger systems.
If instead of applying the exact conjugate-momentum operator, the finite-difference
conjugate-momentum operator is applied,
the resulting operator structure is more complicated,
involving higher-qubit operators beyond  2-qubits.
For instance, in the case of $n_Q=4$ there is a contribution to $\tilde\Pi^2$ of the form
$ \sigma^z\otimes\sigma^z\otimes\sigma^z\otimes\sigma^z$ from the operator
in
Eq.~(\ref{eq:PiFD}),
which is absent in the operator decomposition of $k_{\tilde\phi}$ in Eq.~(\ref{eq:kDeldef}).
For the $n_Q=6$ qubit system, there are all
combinations of operators involving two $\sigma^z$'s, four $\sigma^z$'s,  and one
six-qubit operator of the form $^{\otimes 6}\sigma^z$.
As the  resource costs  of applying  circuits to implement  higher-qubit operators
are significantly more than those for two-qubit operators,
significantly more resources are required
to simulate the finite-difference Hamiltonian (with
power-law deviations from exact results) than it is to simulate the exact
Hamiltonian (that provides results that are exponentially close to the exact result on an ideal quantum computer).
It is amusing to note that most of the resources required to simulate the finite-difference Hamiltonian
would be expended to  determine  polynomial deviations from the exact result.

\begin{figure}[!ht]
	\centering
	\includegraphics[width=0.85\columnwidth]{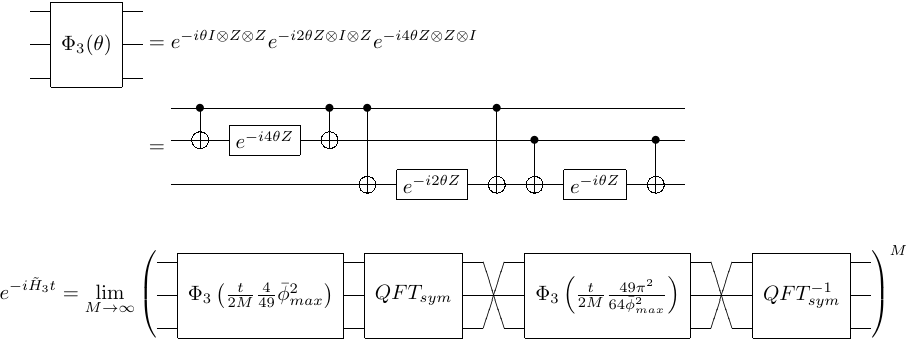}
	\caption{
	A quantum circuit required to perform Trotterized time evolution of the HO Hamiltonian in Eq.~(\ref{eq:Hhatlatt01free}),
	digitized with $n_Q=3$ qubits in JLP digitization, denoted by $\tilde H_3$.
	The upper circuit implements the operator $\Phi_3(\theta) $ defined in Eq.~(\ref{eq:Phi3def}),
	with an arbitrary angle, $\theta$, while the lower circuit implements
	that circuit in both $\tilde\phi$ and $\tilde\Pi$ space,
	making use of a symmetric QuFoTr and its inverse, to achieve Trotterized Hamiltonian evolution of the system
	defined by $\tilde H_3$.
	These circuits are equivalent to controlled-rotation gate circuits
	appearing in MSAH~\cite{PhysRevLett.121.110504,Macridin:2018oli}.
		}
		\label{fig:HOphiC3}
\end{figure}

Quantum circuits to implement the action of the operator(s) in Eq.~(\ref{eq:HOdec}),
in particular for the action of the evolution operator, $e^{-i \tilde H t}$,
for an arbitrary number of qubits have been presented by
MSAH~\cite{PhysRevLett.121.110504,Macridin:2018oli} in terms of controlled-rotation gates.
In terms of CNOT gates and single qubit phase rotations, the quantum circuit implementing  the exponentiated
action of the non-identity
operators in Eq.~(\ref{eq:HOdec}), ${\cal O}_{03}^{(n_Q=3)}$,
\begin{eqnarray}
\Phi_3(\theta) & = & e^{-i\theta {\cal O}_{03}^{(n_Q=3)}}
\ \ \ ,
\label{eq:Phi3def}
\end{eqnarray}
is given in the upper panel in Fig.~\ref{fig:HOphiC3}.
Because the three operators contributing to ${\cal O}_{03}^{(n_Q=3)}$ commute, the  operations
can be performed in any order.
One application of $\Phi_3(\theta)$ to the $n_Q=3$ qubit system requires 6 CNOT gates and 3 single-qubit phase operations.
One application of the (simplest-) Trotterized time-evolution operator associated with $\tilde H$ in Eq.~\eqref{eq:Hhatlatt01free},
over a time-step $\Delta t = \frac{t}{M}$, is accomplished by  acting with
$\Phi_3(\theta)$ with
$\theta= {2\over 49} \bar\phi_{\rm max}^2 \Delta t $
to evolve with $e^{-i \tilde\phi^2/2}$, followed by a symmetric QuFoTr, followed by acting with
$\Phi_3(\theta^\prime)$ with
$\theta^\prime= {49\  \pi^2\over 128\  \bar\phi_{\rm max}^2} \Delta t $
to evolve with $e^{-i \tilde\Pi^2/2}$,  followed by the inverse symmetric QuFoTr.
This sequence is shown in the lower panel of Fig.~\ref{fig:HOphiC3}.
Appendix~\ref{sec:SQuFoTr} provides  circuits and associated discussions for a symmetric QuFoTr
(which is similar to the permuted QuFoTr introduced by Somma~\cite{Somma:2016:QSO:3179430.3179434}
for the same purpose on a different  conjugate-momentum-space basis).
The total gate-counts for one application of the (unimproved-) Trotterized evolution-operator associated with
$\tilde H$ in Eq.~(\ref{eq:Hhatlatt01free}) on $n_Q=3$ qubits,
including  those from the symmetric QuFoTr(s), are
24 CNOT gates, 6 Hadamard gates and 24 single-qubit phase rotations.

\section{Jordan-Lee-Preskill Circuit Compilation}
\label{sec:circcomp}
Constructing optimal quantum circuits for implementing desired quantum operations is an optimization problem often inhibited by large dimensionality \cite{Hastings:2015,Selinger:2015:ECA:2685188.2685198,Ross2016OptimalAC,Kliuchnikov2013,1367-2630-20-11-113022,2018arXiv180710781A}.
Similar to the work of Ref.~\cite{Hastings:2015}, it is possible to eliminate neighboring CNOT operators from the na\"ive, uncompiled estimates in the tables in the main text.  Because all operators implemented between QuFoTr operations are diagonal in the JLP basis, this cancellation may be accomplished systematically.  Without the $\lambda \bar{\phi}^4$ operators, the mass term is introduced with all possible two-body operators, which contain identical CNOTs only separated by single-qubit rotations.  Thus, there is no cancellation of CNOT operators before the field self-interaction term is introduced.
Note that a similar argument indicates an absence of CNOT cancellations for implementation of the finite-difference gradient operator as $\bar{\phi}(x) \bar{\phi}(x+1)$ introduces only two-body operators, extended between qubit registers for neighboring sites of the lattice.
Thus, for the JLP basis, the only regime in which CNOT cancellation is expected to occur is in the position-space implementation of an interacting field.
\begin{figure}
  \includegraphics[width = 0.9\textwidth]{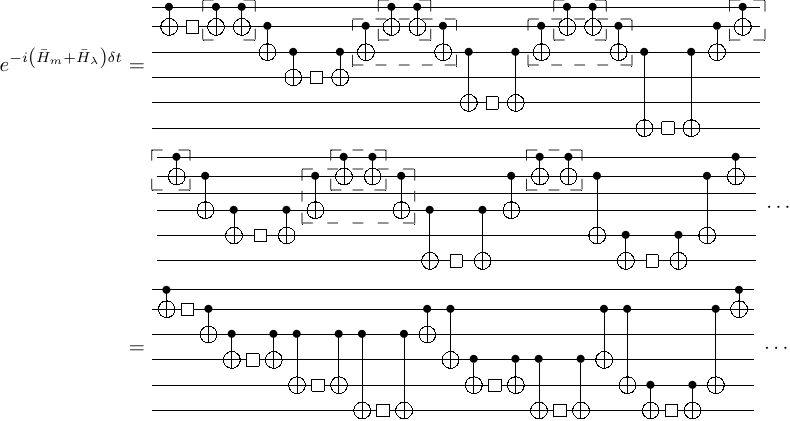}
  \caption{Partial circuit for one first-order-Trotterized step of time evolution for a single site of the interacting scalar field $\bar{\phi}^2$ and $\bar{\phi}^4$ terms with $n_Q=6$ in the JLP basis.  Boxes indicate CNOT pairs that can be eliminated, resulting in the reduced circuit in second expression.}
  \label{fig:CNOTelimination}
\end{figure}
\par Determining the possible eliminations of CNOT pairs requires understanding the combinatorics of available operators and ordering them to maximize the number of qubits shared by neighboring operators in leading-order Trotterization.
Upon implementation, a convenient way to organize these orderings is through connected \enquote{strings}---maximal sets of operators that may be connected through CNOT cancellation.  An example of such a string may be seen in Fig.~\ref{fig:CNOTelimination} where the leading-order Trotterization of the first six four-body operators is implemented.
The string begins with a two-body operator that, due to cancellation, is implemented without increase in the CNOT cost.  Dashed boxes around pairs of CNOT gates indicate where cancellations occur and are removed in the reduced circuit of the third line.  This string implements all six four-body operators that contain both the first and second qubits.  As a result, the entanglement between these two qubits needs only to be established once at the beginning and finally removed at the end.  A similar argument is made for the second pair of qubits in the operators, matching with the second and third qubits for the first three operators and the second and fourth qubits for the next two operators.  To complete the $15$ four-body operators for $n_Q = 6$, it can be shown that two other strings are necessary---beginning with a two-body operator either between qubits (3,4) or (4,5) containing 5 and 4 operators, respectively.
\par The resulting gate counts after elimination of redundant CNOTs may be found in the last column of Table~\ref{tab:HOcircuitBasisComparelam32} in the main text.  It is important to note that the cancellation demonstrated here does not reduce the na\"ive scaling of CNOT gate cost as was found in Ref.~\cite{Hastings:2015}, where the CNOT cost for implementing four-orbital operators in quantum chemistry is reduced by one polynomial power in the number of orbitals used to describe a molecule.  The scaling improvement of Ref.~\cite{Hastings:2015} was found by reducing the cost of a Jordan-Wigner (JW) string (the collections of sequential CNOT gates used to enforce fermionic statistics on a register of qubits) from linear in the number of orbitals to constant.  Because the scalar field is bosonic, there are no JW strings and it remains expected that gate costs will scale with the fourth power (for the four-qubit operators of $\lambda \bar{\phi}^4$) of the number of qubits per lattice site.  While the gate compilation of the HO basis is significantly more cumbersome due to abundant changes in bases truncating the possible operator strings, a similar lack of modified scaling is expected.

\section{Harmonic Oscillator Basis Example: Three Qubits}
\label{sec:HObasis3qubitexample}

\begin{figure}
  \includegraphics[width=0.8\textwidth]{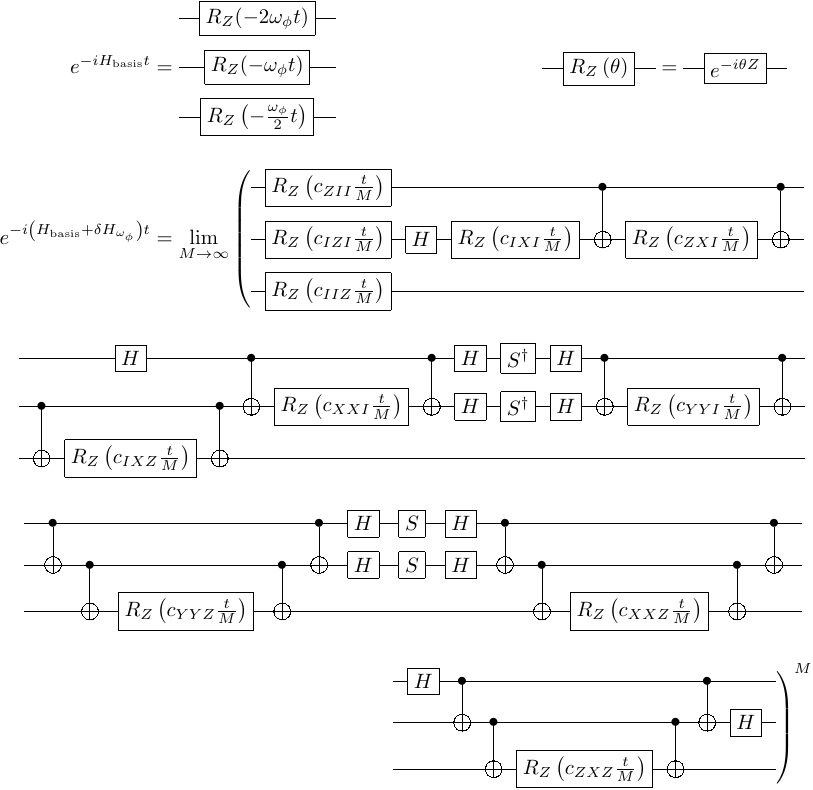}
  \caption{
  Quantum circuits required to perform time evolution of the HO Hamiltonian in Eq.~(\ref{eq:Hhatlatt01free})
  using a HO basis with $n_Q = 3$. The upper circuit is applied when the HO basis can be precisely tuned to the HO being simulated.  When detuned, the lower circuit is necessary with coefficients, $c_{ijk}$, attained from the corresponding operators in
  Eq.~(\ref{eq:HOHOnQ3ops}).
  }
  \label{fig:HOHOnQ3circuit}
\end{figure}

The gate decomposition for a basis of 8 states distributed on $n_Q=3$, of the
matrices,~\footnote{
There is one  subtly in forming the operator decomposition that is to do with defining the truncated
operator matrices.
If the operator basis is restricted to $2^{n_Q}$ from the outset, prior to performing all matrix multiplies,
operations that move the states out of and then back into the truncated space are absent.
In contrast, such operations are included if they are performed in a larger space
with the truncation imposed as the final step.
Initial truncations produce matrix structures with reduced symmetry compared to those where the
truncation is performed last, and as a result give rise to more complex operator structures than those
given in, for example, Eq.~(\ref{eq:HOHOnQ3ops}).
}
$H_{\rm basis}={1\over 2}\omega_\phi\ {\rm diag}\left(1,3,5,7,9,11,13,15\right)$ and
\begin{eqnarray}
\delta H_{\omega_\phi} & = &
{1-\omega_\phi^2\over \omega_\phi}\
\left(
\begin{array}{cccccccc}
 \frac{1}{4} & 0 & \frac{1}{2 \sqrt{2}} & 0 & 0 & 0 & 0 & 0 \\
 0 & \frac{3}{4} & 0 & \frac{\sqrt{\frac{3}{2}}}{2} & 0 & 0 & 0 & 0 \\
 \frac{1}{2 \sqrt{2}} & 0 & \frac{5}{4} & 0 & \frac{\sqrt{3}}{2} & 0 & 0 & 0 \\
 0 & \frac{\sqrt{\frac{3}{2}}}{2} & 0 & \frac{7}{4} & 0 & \frac{\sqrt{5}}{2} & 0 & 0 \\
 0 & 0 & \frac{\sqrt{3}}{2} & 0 & \frac{9}{4} & 0 & \frac{\sqrt{\frac{15}{2}}}{2} & 0 \\
 0 & 0 & 0 & \frac{\sqrt{5}}{2} & 0 & \frac{11}{4} & 0 & \frac{\sqrt{\frac{21}{2}}}{2} \\
 0 & 0 & 0 & 0 & \frac{\sqrt{\frac{15}{2}}}{2} & 0 & \frac{13}{4} & 0 \\
 0 & 0 & 0 & 0 & 0 & \frac{\sqrt{\frac{21}{2}}}{2} & 0 & \frac{15}{4} \\
\end{array}
\right)
\ \ \ .
\label{eq:HOHOnQ3mats}
\end{eqnarray}
is
\begin{eqnarray}
H_{\rm basis} & = &
\omega_\phi \ \left(
4 \ I \ -\  2\  \sigma^z\otimes I_2\otimes I_2 \ -\  I_2\otimes \sigma^z\otimes I_2 \ -\  {1\over 2}\  I_2\otimes I_2 \otimes \sigma^z
\right)
\ \ \ ,
\nonumber\\
\delta H_{\omega_\phi} & = &
{1-\omega_\phi^2\over \omega_\phi}\ \left(
{\sqrt{3}-\sqrt{5}\over 8} \sigma^x\otimes\sigma^x\otimes\sigma^z
\ +\
{\sqrt{3}+\sqrt{5}\over 8} \sigma^x\otimes\sigma^x\otimes I_2
\right. \nonumber\\
&& \left.
\ +\
{\sqrt{3}-\sqrt{5}\over 8} \sigma^y\otimes\sigma^y\otimes\sigma^z
\ +\
{\sqrt{3}+\sqrt{5}\over 8} \sigma^y\otimes\sigma^y\otimes I_2
\right. \nonumber\\
&& \left.
\ +\
{1-\sqrt{3} +\sqrt{21}-\sqrt{15} \over 8\sqrt{2}} \sigma^z\otimes\sigma^x\otimes \sigma^z
\ +\
{1+\sqrt{3} -\sqrt{21}-\sqrt{15} \over 8\sqrt{2}} \sigma^z\otimes\sigma^x\otimes  I_2
\right. \nonumber\\
&& \left.
\ +\
{1-\sqrt{3} -\sqrt{21}+\sqrt{15} \over 8\sqrt{2}} \ I_2\otimes\sigma^x\otimes \sigma^z
\ +\
{1+\sqrt{3} +\sqrt{21}+\sqrt{15} \over 8\sqrt{2}} \ I_2\otimes\sigma^x\otimes  I_2
\right. \nonumber\\
&& \left.
\ -\
 \sigma^z\otimes I_2\otimes  I_2
\ -\
{1\over 2}\ I_2 \otimes\sigma^z\otimes I_2
\ -\
{1\over 4}\ I_2\otimes I_2 \otimes\sigma^z
\ +\
2 I
\right)
\ \ \ .
\label{eq:HOHOnQ3ops}
\end{eqnarray}
For $H_{\rm basis}$,  there are only
3 nontrivial commuting single-qubit operators and
an identity operator.
However, for $\delta H_\phi$, there are
3 three-qubit operators,
4 two-qubit operators, and
4 single-qubit operators. Circuit-representations of the propagators for both the tuned and detuned HO can be seen in Fig.~\ref{fig:HOHOnQ3circuit}.

As an example, in a situation considered by MSAH, electron-phonon interactions can be described via
a linear coupling to the field-space coordinate of a HO.
For such a system, mapped onto 3 qubits, the operator decomposition of the $\bar\phi$ operator contains
7 3-qubit operators,
4 2-qubit operators, and
1 single-qubit operator.
\begin{align}
 \bar{\phi} &=  \frac{1}{\sqrt{\omega_\phi}} \left( \frac{\sqrt{2} - \sqrt{6} - \sqrt{10} + \sqrt{14}}{8} \sigma^z \otimes \sigma^z \otimes \sigma^x \ +\ \frac{2 - 2 \sqrt{3}}{8} \sigma^z \otimes \sigma^y \otimes \sigma^y \right. \nonumber\\ & \qquad \qquad  + \ \frac{2 - 2 \sqrt{3}}{8} \sigma^z \otimes \sigma^x \otimes \sigma^x \ +\ \frac{1}{2 \sqrt{2}}\sigma^y \otimes \sigma^y \otimes \sigma^x  \ +\ \frac{1}{2\sqrt{2}} \sigma^y \otimes \sigma^x \otimes \sigma^y \nonumber \\ & \qquad \qquad - \frac{1}{2\sqrt{2}} \sigma^x \otimes \sigma^y \otimes \sigma^y + \frac{1}{2 \sqrt{2}} \sigma^x \otimes \sigma^x \otimes \sigma^x   + \frac{\sqrt{2} + \sqrt{6} - \sqrt{10} - \sqrt{14}}{8} \sigma^z \otimes I_2 \otimes \sigma^x \nonumber \\ & \qquad \qquad + \frac{\sqrt{2} - \sqrt{6} + \sqrt{10}-\sqrt{14}}{8} I_2 \otimes \sigma^z \otimes \sigma^x  \ +\ \frac{2 + 2 \sqrt{3}}{8} I_2 \otimes \sigma^y \otimes \sigma^y  \nonumber \\ & \qquad \qquad +\left. \frac{2 + 2 \sqrt{3}}{8} I_2 \otimes \sigma^x \otimes \sigma^x \ +\ \frac{\sqrt{2} + \sqrt{6} + \sqrt{10} + \sqrt{14}}{8} I_2 \otimes I_2 \otimes \sigma^x \right)
  \ \ \ .
  \label{eq:phibarop}
\end{align}
Notice that the 3-body operators present in this interaction term are not repeated versions of the 3-body operators already present and thus  increase gate requirements.
Decompositions of
the $\bar\phi$ interactions
in Eq.~(\ref{eq:phibarop}) for larger $n_Q$ look  similar, in terms of the quantum resources of
Table~\ref{tab:HOcircuitBasisCompare}, to those of the detuned HO with only a single one-qubit operator and double the number of $n_Q$-qubit operators.
Therefore,  while the free HO evolution is computationally inexpensive when using a basis of tuned HO eigenstates, an interaction requires many multi-qubit interactions.

\section{Symmetric Quantum Fourier Transform}
\label{sec:SQuFoTr}

Small modifications to the symmetry properties of operators can impact the
gate decomposition necessary for implementation on quantum hardware.
As discussed by Somma~\cite{Somma:2016:QSO:3179430.3179434}
who presented a circuit for the permuted QuFoTr,
a different choice of conjugate-momentum eigenstates requires a different circuit to accomplish a QuFoTr.
In this Appendix,  a circuit  is provided for the symmetric QuFoTr that is used in the time-evolution of the
systems considered in this work.
By acting with a set of single-qubit phase gates before the standard QuFoTr, the states in
momentum space may be distributed symmetrically between $\pm \pi$, avoiding both edges of the first Brillouin zone.
Because this structure now resembles that of field space, distributed around zero between $\pm \bar\phi_{\rm max}$,
the gate decompositions within these two conjugate spaces are identical (differing only in rotation angles) for a free theory.

For a position-space register written in binary,
\begin{equation}
 |x\rangle = |x_{n-1} x_{n-2} \cdots x_0\rangle = \Big\lvert \sum_{i = 0}^{n-1} x_i 2^i \Big\rangle
 \ \ \ ,
\end{equation}
the symmetric QuFoTr implements the following transformation
\begin{equation}
  |x\rangle = \frac{1}{\sqrt{2^n}} \sum_{k = - \frac{2^n -1}{2}}^{\frac{2^n-1}{2}} e^{\frac{2\pi i x \cdot k}{2^n}}|k\rangle
  \ \ \ .
\end{equation}
This differs from the standard QuFoTr only in the introduction of an additional $k$-independent phase,
\begin{equation}
  \exp \left[ 2 \pi i \frac{1}{2^{n+1}} \left( - \sum_{j = 0}^{n-1} 2^j \right) \left( \sum_{i = 0}^{n-1} x_i 2^i \right) \right] \ \ \ ,
\end{equation}
determined only by the value of the $x$ register.
This dependence suggests that it should be applied prior to the transformation to Fourier space.
Indeed, this phase can be implemented with a single layer of single-qubit phase gates.
With the usual definition of
$R(\theta) = \begin{pmatrix}
  1 & 0 \\
  0 & e^{i \pi \theta}
\end{pmatrix}$
and defining $M = \sum\limits_{j = 0}^{n-1} 2^j$, the symmetric QuFoTr may be written as:
\begin{align}
QFT^{sym}_{ij} &= \frac{1}{\sqrt{n_s}} e^{i \frac{2 \pi x_i k_j}{n_s}}  \quad x = \left\{ 0, ..., n_s-1\right\} \ \ k = \left\{ -\frac{n_s-1}{2} , ..., \frac{n_s-1}{2} \right\}
\\
\nonumber \\
QFT^{sym}_{n_q = 3}&=
\begin{gathered}
  \Qcircuit @C=1em @R = 0.7em{
  & \gate{R\left(-\frac{M}{2}\right)} & \gate{H} & \gate{R\left( \frac{1}{2}\right)} & \gate{R\left( \frac{1}{4}\right)} & \qw & \qw & \qw &  \qw \\
  & \gate{R\left(-\frac{M}{4}\right)} & \qw & \ctrl{-1} & \qw & \gate{H} & \gate{R\left( \frac{1}{2} \right)} & \qw & \qw  \\
  & \gate{R\left(-\frac{M}{8}\right)} & \qw & \qw & \ctrl{-2} & \qw & \ctrl{-1} & \gate{H}  & \qw \\
  }
\end{gathered} \\ \nonumber \\
QFT^{sym} &=
\begin{gathered}
  \Qcircuit @C=1em @R = 0.7em{
    & \gate{R\left(-\frac{M}{2}\right)} & \multigate{3}{QFT} & \qw \\
    & \gate{R\left(-\frac{M}{4}\right)}  & \ghost{QFT} & \qw  \\
    & \gate{\vdots}  & \ghost{QFT} & \qw \\
    & \gate{R\left(-\frac{M}{2^{n_Q}}\right)} & \ghost{QFT} & \qw
  }
\end{gathered}
\end{align}
Note that the swap network conventionally required to reverse the qubit orderings in
Fourier space is neglected as written here.
This reversal (and the one appearing in the inverse symmetric QuFoTr returning the calculation to position space)
will be implemented instead by simply reading the qubits backwards when applying the $\bar{\Pi}^2$ operator
in momentum space.
This reading inversion is notated by crossing qubit lines so that e.g., the first qubit is associated with the last input to the momentum phase gate as shown in Fig.~\ref{fig:HOphiC3}.
In this way, two depth-$n_Q$ swap networks (per lattice site and per Trotter step) each containing $\lfloor{ \frac{n_Q}{2} }\rfloor \lceil{\frac{n_Q}{2} }\rceil +  \lfloor{\frac{n_Q-1}{2}}\rfloor \lfloor{\frac{n_Q}{2}}\rfloor$ swap gates can be removed from the quantum circuit with an addition of negligible classical preprocessing.

In application to scalar field theory, where the $\tilde{\phi}$ and $\tilde{\Pi}$ operators (in position and conjugate-momentum space, respectively) can be written as tensor products of single-qubit operators leading to only 2-qubit operators in the free Hamiltonian, the advantage of the symmetric QuFoTr is dominantly aesthetic (and potentially experimentally-convenient) as the operator structure applied in position and momentum space is identical,
as shown in Sec.~\ref{subsec:HONoise} for a free HO.
Had the standard QuFoTr been used, single-qubit diagonal gates would also be present in the Fourier-space
implementation of $\tilde{\Pi}^2$, a factor of 2 fewer single-qubit rotations than needed to symmetrize the QuFoTr and its inverse.
However, when all $k$-body operators with $k\leq n$ are required to implement phases in
Fourier space (as is the case when a finite-difference or polynomially-corrected operator is chosen for $\tilde{\Pi}$),
use of the symmetric QuFoTr results in removal of all operators with odd values of $k$, or roughly a  factor of 2 reduction in the exponential (in $n_Q$) number of operators required for a Pauli decomposition of the necessary phases.

\section{Field Conjugate-Momentum Operators}
In this Appendix, we show explicitly the finite-difference, $\delta_{\bar{\phi}}^2$-corrected, and the
exact conjugate-momentum operators in position space, and show that the structure of the
finite-difference operator is increasingly smeared to form the exact lattice $\bar{\Pi}^2$ operator.
Having the capability of implementing these operators directly as diagonal operators in Fourier space is an advantage of working in a qubit formulation.

\begin{figure}
\centering
  \includegraphics[width=0.3\textwidth]{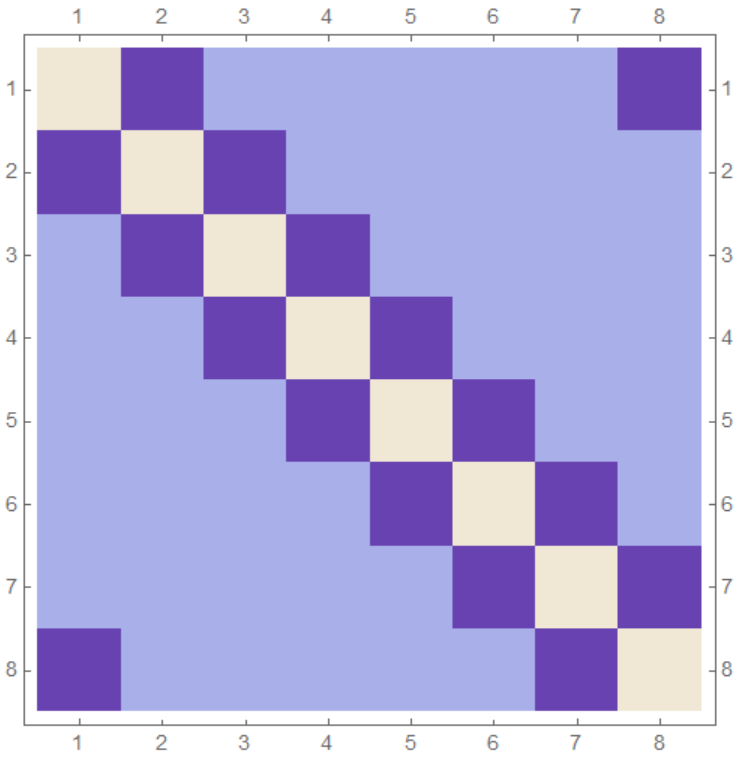}
  \includegraphics[width=0.3\textwidth]{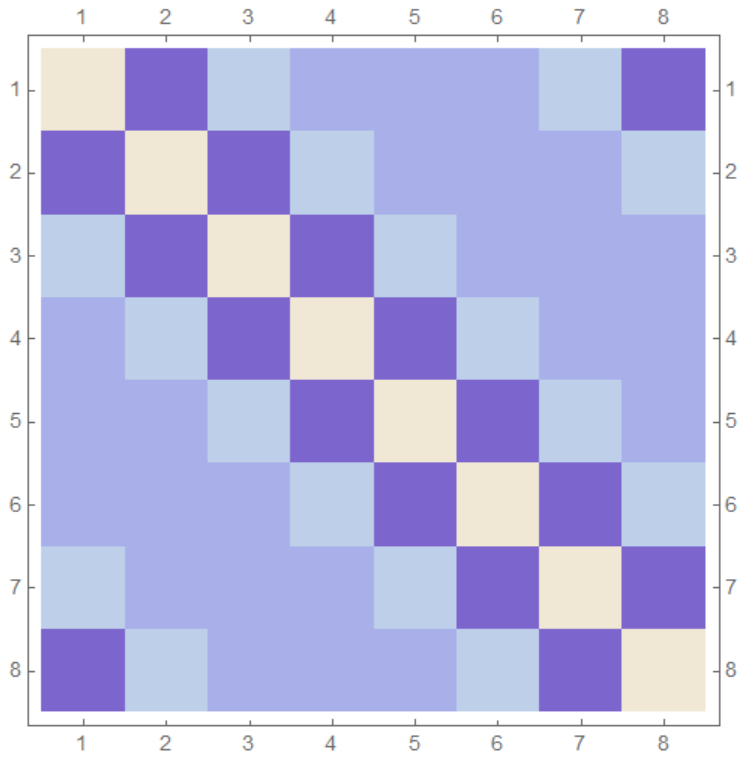}
  \includegraphics[width=0.3\textwidth]{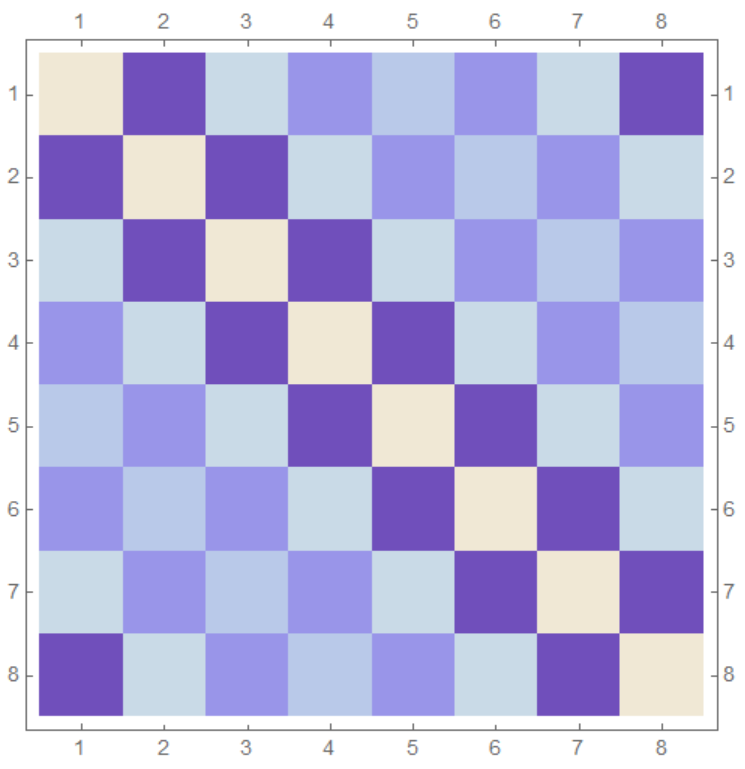}
  \caption{
  Visual representations of Eqs.~\eqref{eq:Pi2Local}-\eqref{eq:Pi2Exact}.
  From left to right, the finite-difference, $\delta_{\bar{\phi}}^2$-improved, and exact field conjugate-momentum operators
  obtained from PBCs
  show increasing non-locality in field space.
  }
  \label{fig:PBCheatmap}
\end{figure}

In the case of PBCs imposed on the field space spanned by $n_Q=3$ qubits,
with momentum eigenvalues
$k={\pi\over \delta_{\bar{\phi}} }\left( -{3\over 4}, -{1\over 2}, -{1\over 4}, 0, {1\over 4}, {1\over 2}, {3\over 4}, 1 \right)$,
the finite-difference, $\delta_{\bar{\phi}}^2$-corrected, and the exact conjugate-momentum operators in position space
are
\setlength{\jot}{15pt}
\begin{align}
&\tilde{\Pi}^2_{\rm finite-difference} =
 \frac{1}{\delta_{\bar{\phi}}^2}
 \left(
\begin{array}{cccccccc}
 2 & -1 & 0 & 0 & 0 & 0 & 0 & -1 \\
 -1 & 2 & -1 & 0 & 0 & 0 & 0 & 0 \\
 0 & -1 & 2 & -1 & 0 & 0 & 0 & 0 \\
 0 & 0 & -1 & 2 & -1 & 0 & 0 & 0 \\
 0 & 0 & 0 & -1 & 2 & -1 & 0 & 0 \\
 0 & 0 & 0 & 0 & -1 & 2 & -1 & 0 \\
 0 & 0 & 0 & 0 & 0 & -1 & 2 & -1 \\
 -1 & 0 & 0 & 0 & 0 & 0 & -1 & 2 \\
\end{array}
\right)\ \ \ , \label{eq:Pi2Local} \\
&\tilde{\Pi}^2_{\delta_{\bar\phi}^2-\rm{improved}} =
 \frac{1}{\delta_{\bar{\phi}}^2}
\left(
\begin{array}{cccccccc}
2.5 & -1.3 & 0.083 & 0 & 0 & 0 & 0.083 & -1.3 \\
-1.3 & 2.5 & -1.3 & 0.083 & 0 & 0 & 0 & 0.083 \\
0.083 & -1.3 & 2.5 & -1.3 & 0.083 & 0 & 0 & 0 \\
0 & 0.083 & -1.3 & 2.5 & -1.3 & 0.083 & 0 & 0 \\
0 & 0 & 0.083 & -1.3 & 2.5 & -1.3 & 0.083 & 0 \\
0 & 0 & 0 & 0.083 & -1.3 & 2.5 & -1.3 & 0.083 \\
0.083 & 0 & 0 & 0 & 0.083 & -1.3 & 2.5 & -1.3 \\
-1.3 & 0.083 & 0 & 0 & 0 & 0.083 & -1.3 & 2.5 \\
\end{array}
\right) \ \ \ ,
\label{eq:Pi2Improved}
\\
&\bar{\Pi}^2_{\rm exact} =
 \frac{1}{\delta_{\bar{\phi}}^2}
\left(
\begin{array}{cccccccc}
3.39 & -2.11 & 0.617 & -0.361 & 0.308 & -0.361 & 0.617 & -2.11 \\
-2.11 & 3.39 & -2.11 & 0.617 & -0.361 & 0.308 & -0.361 & 0.617 \\
0.617 & -2.11 & 3.39 & -2.11 & 0.617 & -0.361 & 0.308 & -0.361 \\
-0.361 & 0.617 & -2.11 & 3.39 & -2.11 & 0.617 & -0.361 & 0.308 \\
0.308 & -0.361 & 0.617 & -2.11 & 3.39 & -2.11 & 0.617 & -0.361 \\
-0.361 & 0.308 & -0.361 & 0.617 & -2.11 & 3.39 & -2.11 & 0.617 \\
0.617 & -0.361 & 0.308 & -0.361 & 0.617 & -2.11 & 3.39 & -2.11 \\
-2.11 & 0.617 & -0.361 & 0.308 & -0.361 & 0.617 & -2.11 & 3.39 \\
\end{array}
\right) \ \ .
\label{eq:Pi2Exact}
\end{align}
``Heat maps'' of the entries in each of the previous operators are shown in Fig.~\ref{fig:PBCheatmap}.

\begin{figure}
\centering
  \includegraphics[width=0.3\textwidth]{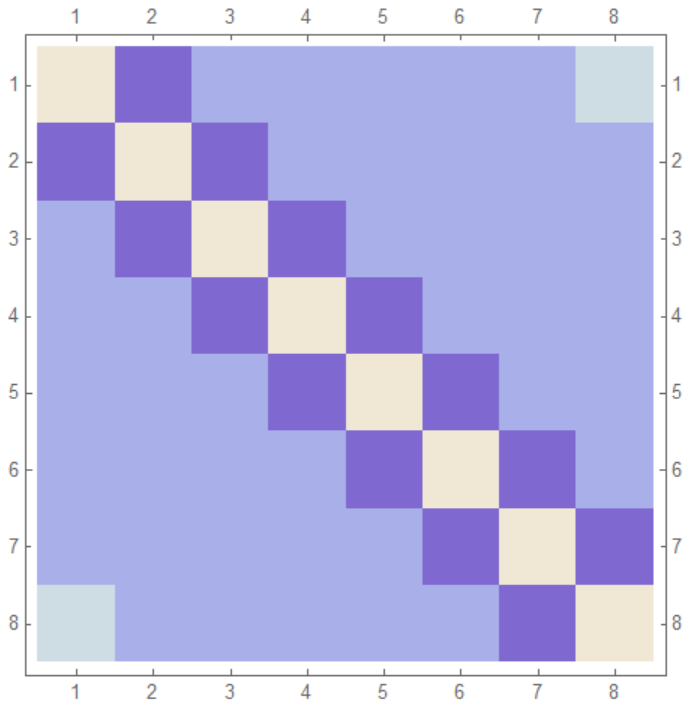}
  \includegraphics[width=0.3\textwidth]{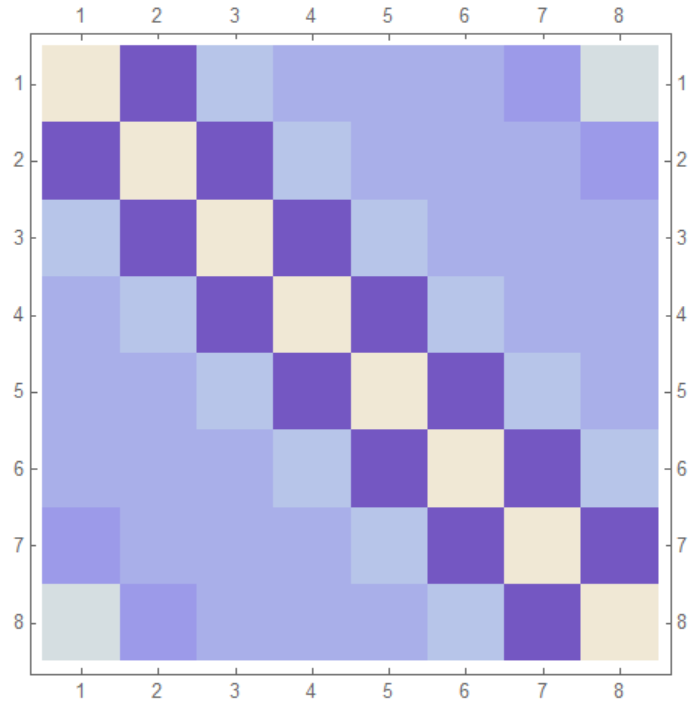}
  \includegraphics[width=0.3\textwidth]{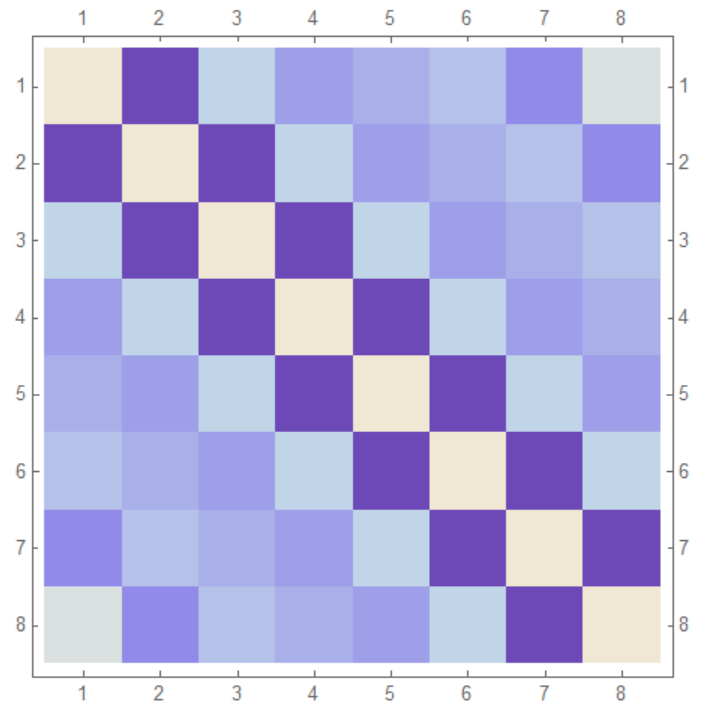}
  \caption{
  Visual representations of Eqs.~\eqref{eq:Pi2Localtwist}-\eqref{eq:Pi2Exacttwist}.
  From left to right, the finite-difference, $\delta_{\bar{\phi}}^2$-improved, and exact field conjugate-momentum operators
    obtained from twisted boundary conditions
  show increasing non-locality in field space.
  }
  \label{fig:TBCheatmaptwist}
\end{figure}

Twisted BCs are used for the calculations performed in this work,
For the $n_Q=3$ system, as defined in Eq.~(\ref{eq:kDeldef}).
with momentum eigenvalues
$k={\pi\over \delta_{\bar{\phi}} }\left( -{7\over 8}, -{5\over 8}, -{3\over 8}, -{1\over 8},  {1\over 8}, {3\over 8}, {5\over 8}, {7\over 8} \right)$,
the finite-difference, $\delta_{\bar{\phi}}^2$-corrected, and the exact conjugate-momentum operators in field space
are
\setlength{\jot}{15pt}
\begin{align}
&\tilde{\Pi}^2_{\rm finite-difference} =
 \frac{1}{\delta_{\bar{\phi}}^2}
\left(
\begin{array}{cccccccc}
2 & -1 & 0 & 0 & 0 & 0 & 0 & 1 \\
-1 & 2 & -1 & 0 & 0 & 0 & 0 & 0 \\
0 & -1 & 2 & -1 & 0 & 0 & 0 & 0 \\
0 & 0 & -1 & 2 & -1 & 0 & 0 & 0 \\
0 & 0 & 0 & -1 & 2 & -1 & 0 & 0 \\
0 & 0 & 0 & 0 & -1 & 2 & -1 & 0 \\
0 & 0 & 0 & 0 & 0 & -1 & 2 & -1 \\
1 & 0 & 0 & 0 & 0 & 0 & -1 & 2 \\
\end{array}
\right) \ \ \ ,
 \label{eq:Pi2Localtwist} \\
&\tilde{\Pi}^2_{\delta_{\bar\phi}^2-\rm{improved}} =
 \frac{1}{\delta_{\bar{\phi}}^2}
\left(
\begin{array}{cccccccc}
2.5 & -1.3 & 0.083 & 0 & 0 & 0 & -0.083 & 1.3 \\
-1.3 & 2.5 & -1.3 & 0.083 & 0 & 0 & 0 & -0.083 \\
0.083 & -1.3 & 2.5 & -1.3 & 0.083 & 0 & 0 & 0 \\
0 & 0.083 & -1.3 & 2.5 & -1.3 & 0.083 & 0 & 0 \\
0 & 0 & 0.083 & -1.3 & 2.5 & -1.3 & 0.083 & 0 \\
0 & 0 & 0 & 0.083 & -1.3 & 2.5 & -1.3 & 0.083 \\
-0.083 & 0 & 0 & 0 & 0.083 & -1.3 & 2.5 & -1.3 \\
1.3 & -0.083 & 0 & 0 & 0 & 0.083 & -1.3 & 2.5 \\
\end{array}
\right)  \ \ \ ,
\label{eq:Pi2Improvedtwist}
\\
&\bar{\Pi}^2_{\rm exact} =
 \frac{1}{\delta_{\bar{\phi}}^2}
\left(
\begin{array}{cccccccc}
3.24 & -1.95 & 0.436 & -0.138 & 0 & 0.138 & -0.436 & 1.95 \\
-1.95 & 3.24 & -1.95 & 0.436 & -0.138 & 0 & 0.138 & -0.436 \\
0.436 & -1.95 & 3.24 & -1.95 & 0.436 & -0.138 & 0 & 0.138 \\
-0.138 & 0.436 & -1.95 & 3.24 & -1.95 & 0.436 & -0.138 & 0 \\
0 & -0.138 & 0.436 & -1.95 & 3.24 & -1.95 & 0.436 & -0.138 \\
0.138 & 0 & -0.138 & 0.436 & -1.95 & 3.24 & -1.95 & 0.436 \\
-0.436 & 0.138 & 0 & -0.138 & 0.436 & -1.95 & 3.24 & -1.95 \\
1.95 & -0.436 & 0.138 & 0 & -0.138 & 0.436 & -1.95 & 3.24 \\
\end{array}
\right) \ \ .
\label{eq:Pi2Exacttwist}
\end{align}
\noindent The corresponding  ``heat maps'' of the entries in each of the twisted operators are shown in Fig.~\ref{fig:TBCheatmaptwist}.
%

\section{Basic Circuit Construction}
In the art and science of quantum circuit development, improvements to generic circuits can often be found when considering the structure of the problem of interest.  The techniques presented in this Appendix are well known and exist in standard literature e.g., Ref.~\cite{NielsenChuang}.  Because the intended audience of this paper is diverse and a number of considerations for the digitization of the scalar field are made with the following circuit construction in mind, it will be useful to explicitly describe the basic methods for applying unitary operators of the form $e^{i \theta \sigma^j\otimes \sigma^k \otimes  \cdots}$ on NISQ-era quantum hardware capable of implementing $z$-axis rotations $e^{i\theta \sigma^z}$.   For the Trotterized time evolution of a Pauli-decomposed Hamiltonian, there are two degrees of freedom needed to modify this operation: increasing the number of qubits in the exponentiated tensor product of Paulis and changing the $z$-axis rotation to $x$- or $y$-axis rotations.  In order to increase the number of qubits in the tensor product, a string of CNOT operators on either side of a single-qubit rotation may be applied.  This computes (and subsequently un-computes) the parity of the of the qubit register down into the last qubit.
\begin{equation}
  e^{i \theta\  \sigma^z \otimes \sigma^z \otimes \sigma^z \otimes \sigma^z} = \begin{gathered}
    \Qcircuit @R 0.7em @C 0.7em {
    & \ctrl{1} & \qw & \qw & \qw & \qw & \qw & \ctrl{1} & \qw\\
    & \targ & \ctrl{1} & \qw & \qw & \qw & \ctrl{1} & \targ & \qw\\
    & \qw & \targ & \ctrl{1} & \qw & \ctrl{1} & \targ & \qw & \qw\\
    & \qw & \qw & \targ & \gate{e^{i \theta \sigma^z}} & \targ & \qw & \qw & \qw
    }
  \end{gathered}
\end{equation}
To change the axis of rotation for any qubit, the following unitary transformations may be used:
\begin{equation}
  \begin{gathered}
  \Qcircuit @R 0.7em @C 0.7em{
  &\gate{X} & \qw
  }
  \end{gathered}
  =
  \begin{gathered}
  \Qcircuit @R 0.7em @C 0.7em{
  &\gate{H} & \gate{Z} & \gate{H} & \qw
  }
  \end{gathered}
  \qquad
  \begin{gathered}
  \Qcircuit @R 0.7em @C 0.7em{
  &\gate{Y} & \qw
  }
  \end{gathered}
  =
  \begin{gathered}
  \Qcircuit @R 0.7em @C 0.7em{
  &\gate{S^\dagger} & \gate{H} & \gate{Z} & \gate{H} & \gate{S}&  \qw
  }
  \end{gathered}
\end{equation}
Because applying these transformations in the exponential is equivalent to applying them to the unitary operator itself, these basis-change operations can be implemented as multiplicative unitaries.
\begin{equation}
e^{i \theta \sigma^x} = \begin{gathered}
\Qcircuit @R 0.7em @C 0.7em {
& \gate{H} & \gate{e^{i \theta \sigma^z}} & \gate{H} & \qw
}
\end{gathered}
\qquad
e^{i \theta \sigma^y} = \begin{gathered}
\Qcircuit @R 0.7em @C 0.7em {
&\gate{S^\dagger} & \gate{H} & \gate{e^{i \theta \sigma^z}} & \gate{H} & \gate{S} &\qw
}
\end{gathered}
\end{equation}
Combining these two degrees of freedom, the exponent of any tensor product of Pauli operators can be created from the single-qubit $z$-axis rotation through use of a CNOT-distributed parity calculation and a change of Pauli bases at the beginning and end of the circuit.
\begin{equation}
  e^{i \theta \ \sigma^x \otimes \sigma^y \otimes \sigma^x \otimes \sigma^z \otimes \sigma^y} =
  \begin{gathered}
    \Qcircuit @R 0.7em @C 0.7em {
    & \qw & \gate{H} & \ctrl{1} & \qw & \qw & \qw & \qw & \qw & \qw & \qw & \ctrl{1} & \gate{H} & \qw & \qw \\
    & \gate{S^\dagger} & \gate{H} & \targ & \ctrl{1} & \qw & \qw & \qw & \qw & \qw & \ctrl{1} & \targ & \gate{H} & \gate{S} & \qw \\
    & \qw & \gate{H} & \qw & \targ & \ctrl{1} & \qw & \qw & \qw & \ctrl{1} & \targ & \qw & \gate{H} & \qw & \qw\\
    &\qw & \qw & \qw & \qw & \targ & \ctrl{1} & \qw & \ctrl{1} & \targ & \qw & \qw & \qw & \qw & \qw\\
    & \gate{S^\dagger} & \gate{H} & \qw & \qw & \qw & \targ & \gate{e^{i \theta \sigma^z}} & \targ & \qw & \qw &\qw & \gate{H} & \gate{S} & \qw
    }
  \end{gathered}
\end{equation}
Before considering cancellations that usually occur when sequentially implementing operators for Trotterization in this way \cite{Hastings:2015}, these basic circuits lead to a CNOT contribution of $2(k-1)$ for the implementation of each unitary with a $k$-body Pauli operator in the exponent.  This is the counting used for the resource estimates shown in Tables~\ref{tab:HOcircuitBasisCompare},~\ref{tab:HOcircuitBasisComparelam32},~and~\ref{tab:phiphip1circuitBasisComparelam32}.

\section{Lowest-Lying Energy Eigenvalues}
The ground state and $1^{st}$ excited state energies of the systems studied in this work are given in Table~\ref{tab:energies}.
\begin{table}[ht!]
\centering
  \begin{tabular}{|l|c|c|c}
  \hline
  \hline
  System & GS  & $1^{st}$ \\
  \hline
  $0+1 \ \ \lambda = 0 \ \ m = 1$
  & ${1\over 2}$
  & ${3\over 2}$ \\
  $0+1 \ \ \lambda = 32 \ m = 1$
  & 0.859\ 742\ 690\ 445\ 509\ 019\ 355\ 96
  & 2.949\ 363\ 767\ 009\ 968\ 902\ 29  \\
  $0+1 \ \ \lambda = 1 \ \ \mu = 2$
  & -22.596\ 382\ 373\ 935\ 095\ 119\ 775\ 874
  & -22.596\ 382\ 373\ 935\ 095\ 118\ 634\ 895    \\
  $0+1 \ \ \lambda = 1 \ \ \mu = 5$
  &  -933.966\ 134\ 532\ 634\ 985\ 047\ 797\ 739
  &  -933.966\ 134\ 532\ 634\ 985\ 047\ 797\ 739   \\
  $1+1 \ \ \lambda = 32 \ m = 1$
  &  2.124\ 233\ 123\ 438\ 790\ 185\ 081\ 206\ 39
  &  4.141\ 788\ 964\ 874\ 434\ 527\ 967\ 370\ 80
  \\
  \hline
  \hline
  \end{tabular}
  \caption{
  Values of the ground-state and $1^{st}$-excited-state
  energies for the 1-site and 2-site systems studied in this work.
  The eigenvalues of the 1-site theory with $\lambda = 0$ and $m = 1$ (HO) are known exactly.
  All other ground state values are displayed with a number of digits sufficient to produce the
  tuning and precision plots that appear in the main text.
  }
  \label{tab:energies}
\end{table}

\section{Noisy Simulations}
\label{app:noisySimulation}

In this Appendix, we show that
in a simple model of quantum noise (step (3) of Fig.~\ref{fig:errorsourcediagram})
representative of near-term quantum hardware,
and
using first-order Trotterized time evolution (step (2) of Fig.~\ref{fig:errorsourcediagram}),
simulation errors exceed the theoretical systematic errors of digitization
(step (1) of Fig.~\ref{fig:errorsourcediagram}).
This identifies the simulation errors (steps 2,3) as the dominant source of uncertainty.
This focuses future improvements on bolstering the system against simulation errors to have the greatest impact on the exploration of scalar fields on near-term quantum devices.

\begin{figure}
\centering
\includegraphics[width=0.8\textwidth]{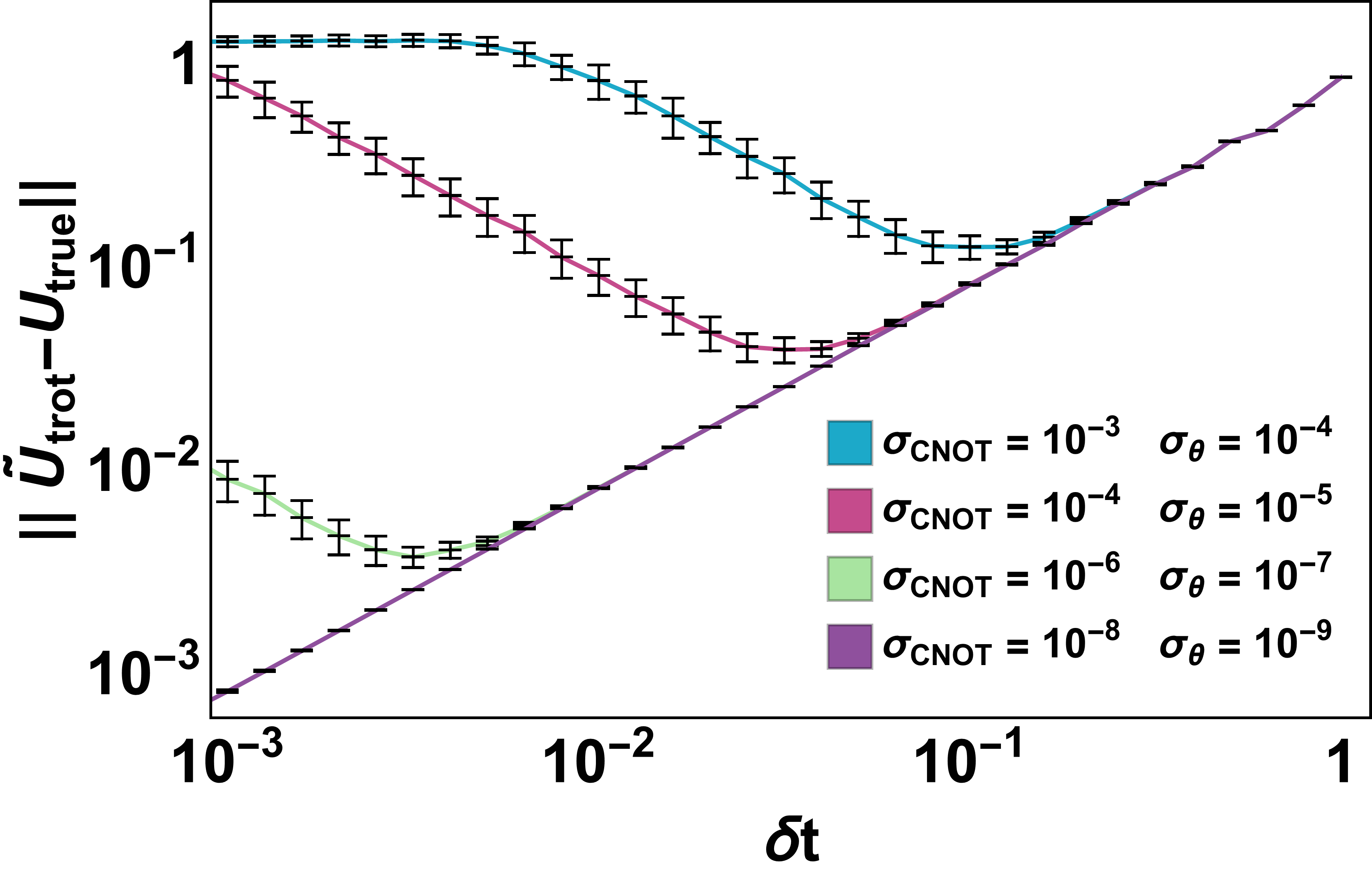}
\caption{Distance measure (Schatten 1-norm) between the noisily-Trotterized and exactly-digitized time evolution operator (after step (3) and (1) of Fig.~\ref{fig:errorsourcediagram}, respectively) as a function of the Trotter step size, $\delta t$, for an integrated evolution time of $T_f = 1$.
The noisy, first-order Trotterized propagator, $\tilde{U}_{\rm trot}$ is implemented with sampled error rates on 1- and 2- qubit gates set by $\sigma_\theta$ and $\sigma_{\rm CNOT}$, respectively.  The digitization scheme is defined by $n_Q = 3$ and $\bar{\phi}_{\rm max} = 3.0$.  From right-to-left, the calculations deviate from the from the ideal result in the top-to-bottom order of the legend with the last calculation at $\sigma_{\rm CNOT} = 10^{-8}$ maintaining visual agreement for the entire plotted domain.  }
\label{fig:onenorm}
\end{figure}

The first quantity to examine in assessing the error landscape specific to the scalar field is the Schatten 1-norm of the time evolution operator evolved to final time $T_f = 1$ as shown in Fig.~\ref{fig:onenorm}.
The system studied in this and the following figures is the one-site, free scalar field digitized onto three qubits (the minimum number identified in the main text necessary to achieve $\sim 1\%$ errors on the low-energy eigenvalues).
Referencing Fig.~\ref{fig:HOphiMnq3}, the tuning of the JLP basis that places the calculation at the NS saturation point and thus optimizes the balance between field-space and momentum-space representations of the wavefunction may be chosen leading to a field-space truncation of $\bar{\phi}_{\rm max} = 3.0$.
In Fig.~\ref{fig:onenorm}, $U_{\rm true}$ is the exact, eight-dimensional, digitized propagator described in the qubit system after step (1) in Fig.~\ref{fig:errorsourcediagram}.
The Schatten 1-norm of this propagator with the Trotterized propagator $U_{\rm trot}$ (after step (2) in Fig.~\ref{fig:errorsourcediagram}) scales linearly with the Trotter step size, $\delta t$, and is degenerate in this figure with the minimum-error data defined by $\sigma_{\rm CNOT} = 10^{-8}$.
The noisy, Trotterized propagator $\tilde{U}_{\rm trot}$ (after step (3)
in Fig.~\ref{fig:errorsourcediagram}) deviates from this linear scaling at low values of $\delta t $ where the errors of $T_f/\delta t$ noisy Trotter steps have accumulated, effectively smearing the evolution.
This type of accumulation of fluctuations in gate implementations may be easily visualized in the single qubit case where the uncertainty in the final quantum state eventually wraps a significant portion of the Bloch sphere's surface.
For long evolution, the accumulation of these errors leads to a constant, $\mathcal{O}(1)$ operator norm
(as can be seen at $\delta t \sim  3\times 10^{-3}$ with $\sigma_{\rm CNOT} = 10^{-3}$).
As the noise model implemented here is unitary, containing no effects of decoherence or amplitude damping
modeling quantum decoherence of the qubit hardware, this saturation indicates that the system has surpassed
a \emph{software coherence time}, a limit encountered due to imperfect gates
(even if implemented on ideal qubits isolated from their environment).
Of course, when the gate error rate is reduced, the \emph{software coherence time} is increased and a longer circuit (smaller $\delta t$ for a fixed $T_f = 1$) may be implemented before saturation of the propagator's Schatten 1-norm with the ideal Trotterized propagator.
It is a hardware-specific question for future investigation whether NISQ-era digital quantum simulations will be
limited by hardware or software coherence times.

The noise model simulated in this Appendix is that of local, unitary errors correlated with the presence of computational operations.
For each $m$-qubit gate, an $SU(2)^{\otimes m}$ operator is applied before and after its application with Euler angles sampled from a normal distribution centered at zero with a standard deviation of
$\sigma_\theta$ and $\sigma_{\rm CNOT}$ for $m = 1$ and $m = 2$, respectively.
All operators with $m > 2$ are decomposed into 1- and 2-qubit gates as demonstrated in the main text before implementing this noise model.
While this procedure neglects the possible non-unitary, non-Markovian, non-local quantum fluctuations that may plague the calculations implemented on NISQ-era hardware,
examining its effects proves enlightening to the limitations imposed by even this simple form of quantum noise.
Note that this formulation includes, and is more general than, the noise model used in the creation of
Fig.~\ref{fig:HOa} where Gaussian noise was implemented only on the phases applied in conjugate-momentum space.
For the simulations of this Appendix, $\sigma_{\rm CNOT}$ is chosen to be one order of magnitude larger than $\sigma_\theta$ to express the dominance of 2-qubit errors anticipated in the NISQ-era.
The error shown in Fig.~\ref{fig:onenorm} represent the standard deviation of the Schatten 1-norm propagated through sampling of the noisy, Trotterized time evolution operator.

\begin{figure}
\centering
\includegraphics[width = 0.99\textwidth]{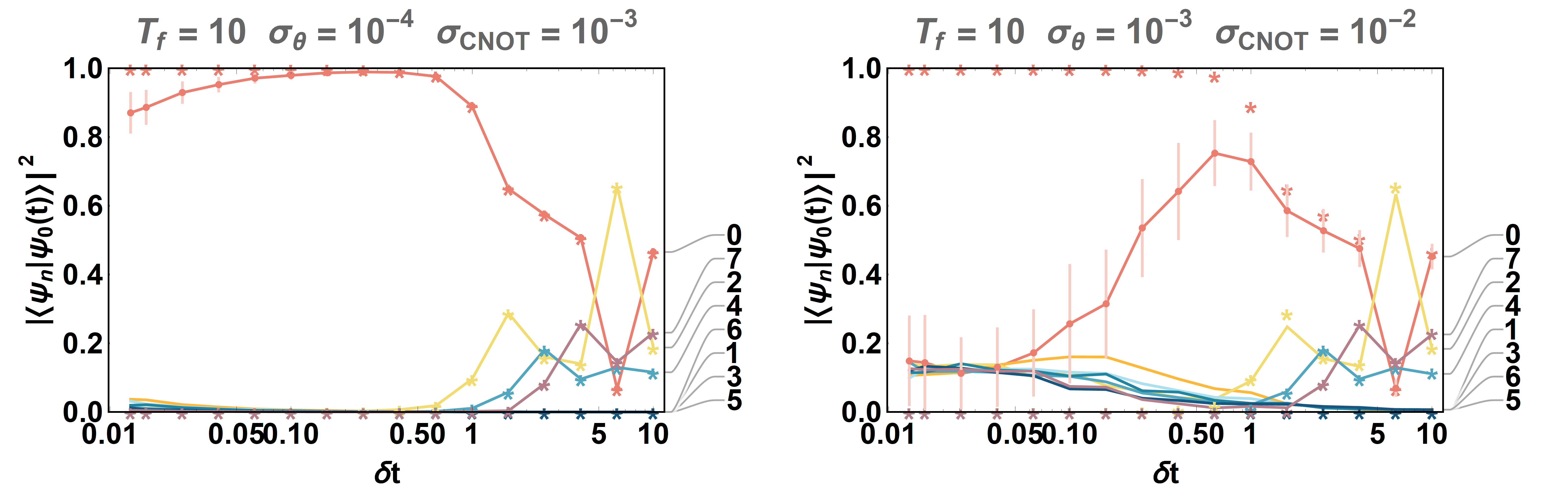}
\caption{
Ground state persistence of a 1-site, free scalar field propagated to time $T_f=10$ that has been
digitized with $n_Q = 3$ with $\bar{\phi}_{\rm max} = 3.0$ as a function of Trotter step size, $\delta t$,
for two error rates.
The star-shaped points are calculated from a first-order Trotterized propagator without quantum noise
and becomes exact in the limit $\delta t \rightarrow 0$.
Points joined by colored lines are sampled with the noise model described in the text at error rates of $\sigma_{CNOT} = 10^{-3}, 10^{-2}$ and $\sigma_\theta$ one order of magnitude smaller in each case.
Eigenvector indices, $n$, are indicated at the right of each panel.
}
\label{fig:GSpersistence}
\end{figure}
While the Schatten 1-norm is a succinct distance measure to quantify the effects of Trotterization and successive noisy implementation of the time evolution operator, it remains unclear, beyond perhaps placing loose bounds, how these errors propagate to physical observables likely to be extracted from the quantum calculation.
To address quantities of direct relevance to the calculation results, two additional properties are examined:  the ground state persistence and the evolution of the expectation value of the field $\langle \phi \rangle(t)$.
Figure~\ref{fig:GSpersistence} shows the decomposition of the time evolved ground state across the t = 0 eigenbasis after a total evolution time of $T_f = 10$ as a function of the Trotter step size, $\delta t$.
The star-shaped points are noiseless Trotterizations of the digitized propagator
(after step (2) in Fig.~\ref{fig:errorsourcediagram}) and show that the ground state persists
for sufficiently small $\delta t$ as $U_{\rm trot}$ becomes exact.
When $\delta t$ is sufficiently large, the ground-state persistence diminishes due to  errors
that are polynomial in $\delta t$
affecting both the energies and eigenvalues of the Trotterized propagator.
Moving through step (3) of Fig.~\ref{fig:errorsourcediagram}, the joined points indicate that the ground state persistence falls also at small $\delta t$ where, as was the case with the Schatten 1-norm, noise in the propagator has accumulated.  The left panel in Fig.~\ref{fig:GSpersistence}
 is calculated with $\sigma_{\rm CNOT} = 10^{-3}$ while the right panel is calculated with larger fluctuations in the gate errors, $\sigma_{\rm CNOT} = 10^{-2}$.
The notable difference in the ground-state persistence between these error rates (both being relevant to NISQ-era hardware) is that the former is capable, for specific but existing choice of $\delta t$, of retaining the ground state content at the 1\% level while the latter is not capable of achieving this for any choice of $\delta t$.
Note that matrix elements to eigenstates 1, 3, 5, and 6 are  not excited in the Trotterization and remain of negligible excitation in the noisy Trotterization.
This feature dramatically limits the Hilbert-space mixing available to the noisy evolution by effectively decoupling half of the Hilbert space.
Even with this structural advantage, the persistence of low-energy eigenstates is seen to be a significant obstacle at noise levels expected in the NISQ-era.

\begin{figure}
  \centering
  \includegraphics[width=0.99\textwidth]{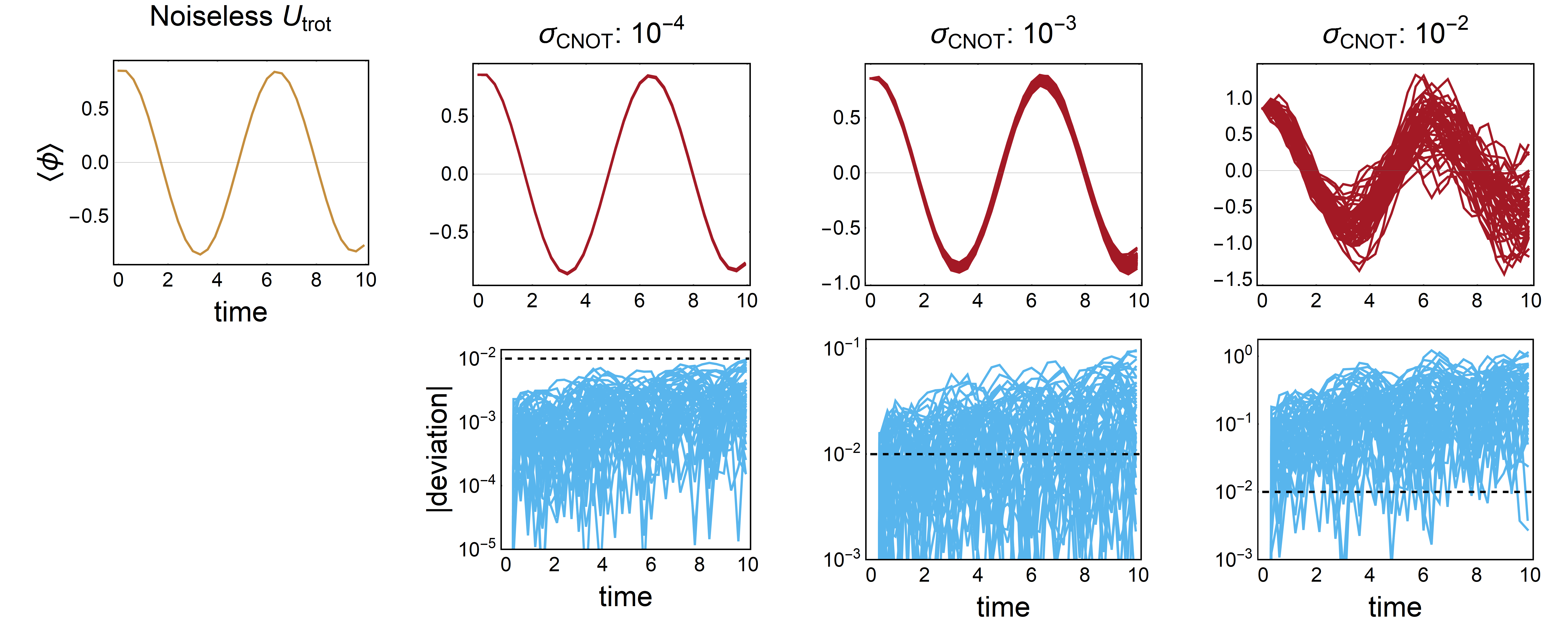}
  \caption{
  The time evolution of the expectation value of the field operator for a state initialized to the ground state rotated by a single site in $\phi$-space.
  The evolution in the upper left (gold curve) is the noiseless first-order Trotterization (after step (2) in Fig.~\ref{fig:errorsourcediagram}) with $\delta t = 0.3$.
  The gate-error rate $\sigma_{\rm CNOT}$ increases to the right with $\sigma_{\theta}$ an order of magnitude smaller
  in each case.
  The deviation shown in the second row is with respect to the noiseless Trotterization
  and represents only the errors arising from step (3) of Fig.~\ref{fig:errorsourcediagram}.
  }
  \label{fig:phiexpectationvalue}
\end{figure}
For studying the time dependence of the expectation value of the field, $\langle \phi \rangle(t)$,
we choose to initialize the system in a state $|\psi_i\rangle$ that is the ground state of the system
rotated by one site in $\phi$-space.
In this state, the expectation value of the field is $\langle \phi \rangle = 0.8567$, a value dictated dominantly by the
$\phi$-space lattice spacing $\delta_\phi = \frac{2 \bar{\phi}_{\rm max}}{2^{n_Q} -1} = \frac{6}{7}$,
with deviations due to the effects of the periodic boundary conditions.
This state is in the low-energy sector of the Hilbert space,
with the projections $|\langle \psi_n|\psi_i\rangle|^2$ into the lowest three eigenstates,  $n=0,1,2$,
of 69\%, 25\%, and 5\%.
Consequently,
this evolution is particularly sensitive to the mixing of low-energy eigenstates.
The inaccuracies in the time evolution of $\langle \phi \rangle$ for the highest error rate of
$\sigma_{\rm CNOT} = 10^{-2}$ are substantial, as can be seen in the right-hand column of Fig.~\ref{fig:phiexpectationvalue}.
To isolate the noise in this observable (step (3) of Fig.~\ref{fig:errorsourcediagram}),
the deviation (light blue curves) between the full noisy expectation value (dark red curves) and that calculated from the noiseless Trotterized propagator (upper left panel, gold curve) is shown in the second row,
with gate errors increasing to the right.
Figure~\ref{fig:phiexpectationvalue} indicates that the errors stemming from step (3)
alone result in $\mathcal{O}(1)$ deviations for this observable with $\sigma_{\rm CNOT} = 10^{-3}$.
It is only when the error rate is decreased to $\sigma_{\rm CNOT} = 10^{-4} $ and $\sigma_\theta = 10^{-5}$
that $\langle \phi \rangle$ can be determined with $\sim 10^{-2}$ precision with respect to the ideal Trotterization,
as emphasized by the black dashed reference line in the second row of Fig.~\ref{fig:phiexpectationvalue}.
This level of precision is
presently unavailable on NISQ-era hardware,  indicating that non-negligible error mitigation is required
for extracting  observables in even small space-time volumes on quantum devices.
This situation is expected to persist for an extended period.

We conclude that the digitization errors depicted in step (1) of Fig.~\ref{fig:errorsourcediagram}
can be made a subdominant error source for NISQ-era applications through the tuning procedures
described in the main text.
The remaining simulation errors of the noisy implementation of
a Trotterized  time-evolution operator depicted in steps (2, 3) of Fig.~\ref{fig:errorsourcediagram} are found, under reasonable assumptions, to be significant barriers to implementing even the smallest representation of scalar field theory.
While higher-order Trotterizations that would reduce the error in step (2) are known,
the implementation of these improved temporal digitizations can require a significantly increased number of quantum gates, necessarily increasing the error in step (3) of Fig.~\ref{fig:errorsourcediagram}.
The results of this Appendix emphasize that further study is needed in the directions of algorithmic improvements and error mitigation strategies before the errors of steps (2,3) can be systematically controlled
in the same way as in the main text of this paper for the digitization step (1).

It is important to remember that desired precision of a calculation need not be achieved with a single set of simulation parameters ($\delta t$, $\sigma_{\theta, \rm CNOT}$, etc.).
While properties of a single parameter set were analysed here, it is also possible (and likely essential in the NISQ-era) to implement a collection of biased or lower-precision calculations and extrapolate to the unbiased or \enquote{zero-error} limit.
Such an extrapolation in the regime of two-qubit gate errors has been shown to be vital to digitally calculate e.g., the deuteron binding energy~\cite{PhysRevLett.120.210501} and the dynamics of pair production in the Schwinger model~\cite{PhysRevA.98.032331}.
In this way, it is possible to achieve precision in an observable beyond the noise levels of available quantum hardware.
A thorough investigation for designing collections of quantum simulations ranging in computational expense allowing extrapolations of the effects of noise in NISQ-era devices is now, in the context of controllable digitization errors, a leading avenue for future algorithmic progress towards implementation of scalar fields on NISQ-era quantum devices.
Of course, such a program of planning the distributions of resources is not unique to quantum computation, and has been essential in optimizing scientific productivity in high-performance (classical) computing projects, such as the lattice QCD production campaigns in high-energy and nuclear physics.

\begin{acknowledgements}
Inspiration for this work emerged from discussions with George Siopsis, Eugene Dumitrescu and Raphael Pooser with the U.S. Department of Energy, Office of Science, Office of Advanced Scientific Computing Research (ASCR) quantum Testbed Pathfinder program under field work proposal number ERKJ335.
We would like to thank Jim Amundson, Silas Beane, Stephen Jordan,  David Kaplan, Pavel Lougovski, Aidan Murran, John Preskill, Kenneth Roche, Alessandro Roggero, Jesse Stryker, and Nathan Wiebe for many interesting discussions.  This work is supported by U.S. Department of Energy grant No. DE-FG02-00ER41132 and by the U.S. Department
of Energy, Office of Science, Office of Advanced Scientific Computing Research (ASCR) quantum algorithm teams
program, under field work proposal number ERKJ333.  NK was supported in part by the Seattle Chapter of the Achievement Rewards for College Scientists (ARCS) foundation.
Calculations were done using \emph{Wolfram Mathematica 11.1} and the quantum circuits appearing in this paper were typeset using the latex package \emph{Qcircuit} originally developed by Bryan Eastin and Steven Flammia.
\end{acknowledgements}

\bibliography{lf4Ibib}
\end{document}